\documentclass[11pt]{article}

\usepackage{amsmath,amssymb}
\usepackage[margin=1in]{geometry}
\usepackage{setspace}
\usepackage{graphicx}

\numberwithin{equation}{section}

\begin{document}

\newcommand{\needref}{\textbf{ [REF] }}
\newcommand{\rn}{\mathbb{R}^n}          
\newcommand{\cn}{\mathbb{C}^n}          
\newcommand{\cpn}{\mathbb{CP}^n}        
\newcommand{\wcp}[2]{\mathbb{WCP}^{#1}_{[#2]}}  
\newcommand{\ci}[1]{\left(#1\right)}        
\newcommand{\cu}[1]{\left\{#1\right\}}      
\newcommand{\sq}[1]{\left[#1\right]}        
\newcommand{\an}[1]{\left<#1\right>}        
\newcommand{\bci}[1]{\Bigl(#1\Bigr)}        
\newcommand{\bcu}[1]{\Bigl\{#1\Bigr\}}      
\newcommand{\bsq}[1]{\Bigl[#1\Bigr]}        
\newcommand{\Bci}[1]{\Biggl(#1\Biggr)}      
\newcommand{\Bcu}[1]{\Biggl\{#1\Biggr\}}    
\newcommand{\Bsq}[1]{\Biggl[#1\Biggr]}      
\newcommand{\dbd}[1]{\frac{\partial}{\partial #1}} 
\newcommand{\hf}{\frac{1}{2}}                   
\newcommand{\ef}[1]{\mathrm{d}#1}           
\newcommand{\textbox}[1]{\begin{center}\fbox{\parbox{14cm}{#1}}\end{center}}  
\newcommand{\der}[2]{\frac{\partial #1}{\partial #2}}   
\newcommand{\diff}[2]{\frac{{\rm d} #1}{{\rm d} #2}}    
\newcommand{\ndiff}[3]{\frac{d^{#3}#1}{d #2^{#3}}}  
\newcommand{\mea}[2]{{\mathrm d}^{#1}#2}        
\newcommand{\cm}{\hspace{1cm}}              
\newcommand{\hcm}{\hspace{0.5cm}}           
\newcommand{\tr}{{\mathrm{Tr}}\:}           
\newcommand{\im}{{\mathrm{Im}}}             
\newcommand{\re}{{\mathrm{re}}}             
\newcommand{\dbar}{\bar{\partial}}          

\newcommand{\cA}{{\cal A}}      \newcommand{\cB}{{\cal B}}
\newcommand{\cC}{{\cal C}}      \newcommand{\cD}{{\cal D}}
\newcommand{\cE}{{\cal E}}      \newcommand{\cF}{{\cal F}}
\newcommand{\cG}{{\cal G}}      \newcommand{\cH}{{\cal H}}
\newcommand{\cI}{{\cal I}}      \newcommand{\cJ}{{\cal J}}
\newcommand{\cK}{{\cal K}}      \newcommand{\cL}{{\cal L}}
\newcommand{\cM}{{\cal M}}      \newcommand{\cN}{{\cal N}}
\newcommand{\cO}{{\cal O}}      \newcommand{\cP}{{\cal P}}
\newcommand{\cQ}{{\cal Q}}      \newcommand{\cR}{{\cal R}}
\newcommand{\cS}{{\cal S}}      \newcommand{\cT}{{\cal T}}
\newcommand{\cU}{{\cal U}}      \newcommand{\cV}{{\cal V}}
\newcommand{\cW}{{\cal W}}      \newcommand{\cX}{{\cal X}}
\newcommand{\cY}{{\cal Y}}      \newcommand{\cZ}{{\cal Z}}

\newcommand{\bA}{{\mathbb A}}   \newcommand{\bN}{{\mathbb N}}
\newcommand{\bB}{{\mathbb B}}   \newcommand{\bO}{{\mathbb O}}
\newcommand{\bC}{{\mathbb C}}   \newcommand{\bP}{{\mathbb P}}
\newcommand{\bD}{{\mathbb D}}   \newcommand{\bQ}{{\mathbb Q}}
\newcommand{\bE}{{\mathbb E}}   \newcommand{\bR}{{\mathbb R}}
\newcommand{\bF}{{\mathbb F}}   \newcommand{\bS}{{\mathbb S}}
\newcommand{\bG}{{\mathbb G}}   \newcommand{\bT}{{\mathbb T}}
\newcommand{\bH}{{\mathbb H}}   \newcommand{\bU}{{\mathbb U}}
\newcommand{\bI}{{\mathbb I}}   \newcommand{\bV}{{\mathbb V}}
\newcommand{\bJ}{{\mathbb J}}   \newcommand{\bW}{{\mathbb W}}
\newcommand{\bK}{{\mathbb K}}   \newcommand{\bX}{{\mathbb X}}
\newcommand{\bL}{{\mathbb L}}   \newcommand{\bY}{{\mathbb Y}}
\newcommand{\bM}{{\mathbb M}}    \newcommand{\bZ}{{\mathbb Z}}

\begin{spacing}{1}

\long\def\symbolfootnote[#1]#2{\begingroup%
\def\thefootnote{\fnsymbol{footnote}}\footnote[#1]{#2}\endgroup}
{\phantom .}
\begin{center}
\textbf {\Large Families of Quintic Calabi--Yau 3--Folds with Discrete Symmetries} \\
\vspace{1.5em}

Charles Doran,$^1$\symbolfootnote[1]{{\bf email:} {\it doran@math.washington.edu}} Brian Greene,$^{2,3}$\symbolfootnote[2]{ {\bf email:} {\it greene@phys.columbia.edu}} and Simon Judes$^2$\symbolfootnote[3]{{\bf email:} {\it judes@phys.columbia.edu}} \\
\vspace{0.75em}
$^1$Department of Mathematics, University of Washington, Seattle, WA 98195. \\
\vspace{0.4em}
$^2$Institute for Strings, Cosmology and Astroparticle Physics, Department of Physics, Columbia University, New York, NY 10027. \\
\vspace{0.4em}
$^3$Department of Mathematics, Columbia University, New York, NY 10027. \\
\end{center}
\normalsize

\vspace{0.5ex}
\begin{abstract}
At special loci in their moduli spaces, Calabi--Yau manifolds are endowed with discrete symmetries. Over the years, such spaces have been intensely studied and have found a variety of important applications. As string compactifications they are phenomenologically favored, and considerably simplify many important calculations. Mathematically, they provided the framework for the first construction of mirror manifolds, and the resulting rational curve counts. Thus, it is of significant interest to investigate such manifolds further. In this paper, we consider several unexplored loci within familiar families of Calabi--Yau hypersurfaces that have large but unexpected discrete symmetry groups. By deriving, correcting, and generalizing a technique similar to that of Candelas, de la Ossa and Rodriguez--Villegas, we find a calculationally tractable means of finding the Picard--Fuchs equations satisfied by the periods of all 3--forms in these families. To provide a modest point of comparison, we then briefly investigate the relation between the size of the symmetry group along these loci and the number of nonzero Yukawa couplings. We include an introductory exposition of the mathematics involved, intended to be accessible to physicists, in order to make the discussion self--contained.
\end{abstract}

\tableofcontents
\normalsize

\section{Introduction --- Mirror Manifolds far from the Fermat Point}
Although a global description of the complex structure moduli space of many Calabi--Yau manifolds is available, it is often very useful to consider special loci with discrete symmetries.

For example, in the context of the $E_8\times E_8$ heterotic string compactified to 4 dimensions, a simple but powerful way to break the gauge symmetry to $SU(3)\times SU(2)\times U(1)^n$ is to allow Wilson lines, which require that the compact 6d manifold is not simply connected\cite{Witten:1985xc}\cite{Greene:1986bm}\cite{Braun:2005ux}. Suitable manifolds with nontrivial fundamental group are most easily constructed by starting out with a simply connected space $\tilde{X}$ with a freely acting discrete symmetry group $G$, and taking the quotient $X=\tilde{X}/G$, which then has $\pi_1 = G$.

A more technical but equally important reason for focusing on models with a discrete symmetry group $G$ (or after quotienting with $\pi_1=G$), is that many interesting calculations are considerably simpler than in the general case with trivial $G$. For example, in heterotic compactifications that pass through an $E_6$ GUT phase,\cite{Witten:1985xc}\cite{Greene:1986bm} phenomenological information is contained in the $27\otimes27\otimes27$ Yukawa couplings:
\begin{align}
    \kappa(\alpha,\beta,\gamma) = \int_{\text{CY}}\ef{^6}x \sqrt{g}\Omega_{ijk}\Omega^{\bar{i}\bar{j}\bar{k}}\alpha^i_{\;\bar{i}}\beta^j_{\;\bar{j}}\gamma^k_{\;\bar{k}}
\end{align}
Here $\Omega$ is the holomorphic 3--form, and $\alpha,\beta,\gamma\in H^1(T)$ correspond to four-dimensional fields that lie in the $27$ of $E_6$. In general, there are a large number of such integrals, and each is burdensome to calculate. If discrete symmetries are present, then there are relations among the couplings, and many vanish \cite{Witten:1985xc}\cite{Candelas:1987se}\cite{Greene:1987xh}.

Other simplifications have been found in calculations of the stabilized values of moduli in type IIB string  backgrounds with nontrivial flux\cite{Giryavets:2003vd}. The vacua are determined by the Gukov--Vafa--Witten superpotential:\cite{Gukov:1999ya}\cite{Giddings:2001yu}\footnote{$G$ is a combination of the 3--form fluxes and the dilaton. So in type IIB string theory, generic nonzero flux creates a potential for the complex structure moduli and the dilaton, but not for the K\"{a}hler moduli.}
\begin{align}
    W = \int_{\text{CY}} G\wedge\Omega(t) \propto \sum_i g_i \varpi_i(t)
\end{align}
where $g_i$ are integers specifying the number of units of flux around the $i$th 3--cycle of the Calabi--Yau, $\varpi_i(t)$ is the integral of $\Omega$ over the $i$th 3--cycle (the $i$th period of $\Omega$), and $t$ denotes the coordinates on the complex structure moduli space. One usually finds the periods by solving certain differential equations that they satisfy, the so--called Picard--Fuchs equations. Unfortunately the order of the equations is in general $b_3(\text{CY})$, which can be very large ($\sim 100$). It was noted in \cite{Candelas:2000fq} that if the manifold has discrete symmetries, then the order of the Picard--Fuchs equations is vastly reduced, greatly facilitating their solution.

From a mathematical point of view as well, Calabi--Yau manifolds with discrete symmetries have been instrumental to key developments. For example, the first construction of a pair of mirror manifolds \cite{Greene:1990ud} involved the prototypical family of 3--folds with discrete symmetry: the Fermat family of quintic hypersurfaces in $\bP_4$, defined in homogeneous coordinates $[x_0,\ldots,x_4]$ by:
\begin{align} \label{FermatPencil}
    Q(t) = (x_0)^5 + (x_1)^5 + (x_2)^5 + (x_3)^5 + (x_4)^5  - 5t x_0x_1x_2x_3x_4 = 0
\end{align}

\begin{figure}[h]
\begin{center} 
\includegraphics[scale=0.9]{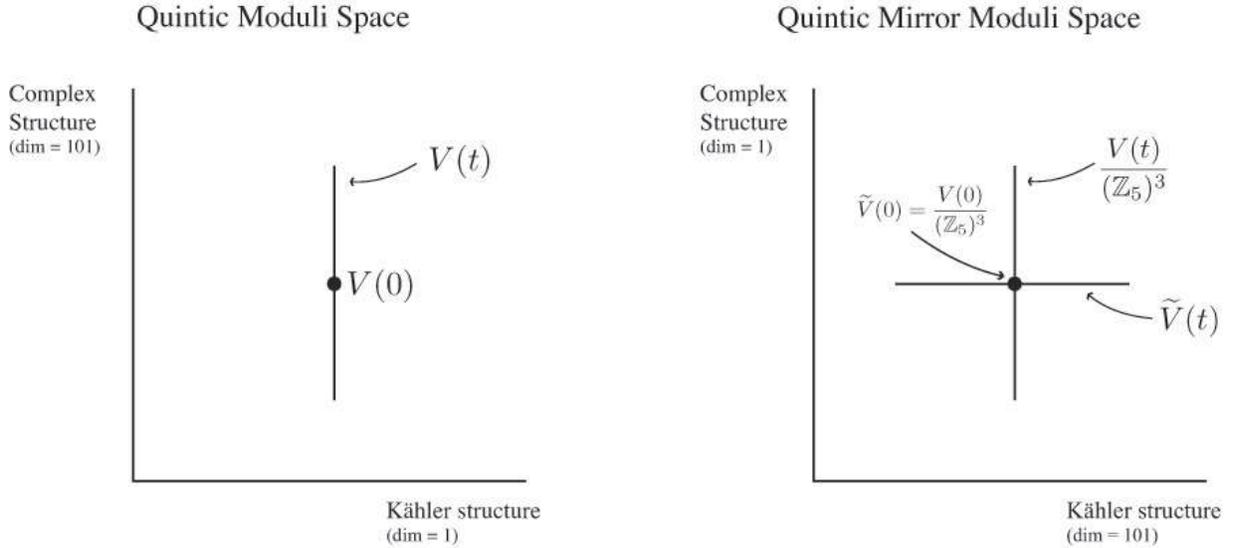}
\vspace{-1em} \begin{minipage}[h]{5in}\caption{\label{mirrorfig} \small A schematic diagram of the moduli spaces of the quintic in $\bP_4$ and its mirror. The line shown in the quintic moduli space is mirror to the horizontal line on the mirror side. But the quotient by $(\bZ_5)^3$ gives the vertical line.}\end{minipage}\end{center}\
\end{figure}

This family (denoted $V(t)$) is invariant under the $S_5$ group of permutations of the $x_i$'s, as well as  4 $\mathbb{Z}_5$ scalings\footnote{We use the notation: $\mathbb{Z}_n = \mathbb{Z}/n\mathbb{Z}$} generated by:
\begin{align} \label{z5gens}
    g_1 &= (1,0,0,0,4) \hcm     g_3 = (1,0,4,0,0)  \\  \nonumber
    g_2 &= (1,0,0,4,0) \hcm g_4 = (1,4,0,0,0)
\end{align}
where $(a,b,c,d,e)$ means $(x_0,x_1,x_2,x_3,x_4)\to(\gamma^a x_0,\gamma^b x_1, \gamma^c x_2, \gamma^d x_3, \gamma^e x_4)$ and $\gamma^5=1\neq\gamma$.
Because we are working in projective space, $g_1,g_2,g_3,g_4$ are not independent symmetries of $V(t)$, since $g_1 g_2 g_3 g_4 = (4,4,4,4,4) = I$, implying that the symmetry group is $[S_5 \ltimes (\mathbb{Z}_5)^4]/\mathbb{Z}_5$. Quotienting $V(0)$ by $(\bZ_5)^3$ (generated by particular combinations of the 4 $g_i$'s) and resolving the orbifold singularities appropriately produces $\widetilde{V}(0)$, the mirror to the Fermat quintic\cite{Greene:1990ud}. The single complex structure parameter of the mirror is then $t$, but somewhat confusingly, $V(t)/(\bZ_5)^3$ is \emph{not} $\widetilde{V}(t)$ (the mirror to $V(t)$) except at $t=0$. This is because $V(t)$ and $V(t)/(\bZ_5)^3$ differ from $V(0)$ and $V(0)/(\bZ_5)^3$ respectively only in their complex structure, but moving in the complex structure moduli space of the mirror corresponds to moving in the \emph{K\"{a}hler} moduli space of the original manifold and vice versa. See figure \ref{mirrorfig}.

One can see that a mirror must exist for any nonsingular quintic in $\bP_4$ by considering deformations of the conformal field theory away from the Fermat point by truly marginal operators. These are interpreted differently on the original and mirror manifolds (complex structure and K\"{a}hler deformations switch roles), but they exist on both sides. It is clear that it is the symmetry of the Fermat quintic that makes this construction possible, by allowing an explicit realization of the mirror at one point in moduli space.

The goal of \cite{Greene:1998yz} was to find other points in moduli space where the mirror can be presented as a resolution of a quotient by a discrete symmetry. One might guess from the simplicity of the form of \eqref{FermatPencil} that moving away from the Fermat family will reduce the symmetry to a subgroup of $[S_5 \ltimes (\mathbb{Z}_5)^4]/\mathbb{Z}_5 $. Indeed this seems to be the case for local deformations, but it turns out to be spectacularly false if one searches the moduli space sufficiently. For example, in \cite{Greene:1998yz} the hypersurface $V_{41}(t)$ defined by:
\begin{align} \label{Z41Quintic}
    Q_{41}(t) = (y_0)^4y_1+(y_1)^4y_2+(y_2)^4y_3+(y_3)^4y_4+(y_4)^4y_0 - 5t y_0y_1y_2y_3y_4=0
\end{align}
was found to possess a $\mathbb{Z}_{41}$ scaling symmetry generated by $(1,37,16,18,10)$ where the entries now indicate nontrivial 41st roots of unity. It is worthwhile to recall the reasoning that leads to the form of \eqref{Z41Quintic}. The idea is to implement the $(\bZ_5)^3$ quotient of the Fermat quintic by making an unusual, apparently ill-defined, change of variables:
\begin{align} \label{fractionaltransform}
    (x_0,x_1,x_2,x_3,x_4) \to (y_0^{4/5}y_1^{1/5},y_1^{4/5}y_2^{1/5},y_2^{4/5}y_3^{1/5},y_3^{4/5}y_4^{1/5},y_4^{4/5}y_0^{1/5})
\end{align}
Generally, the fractional powers would require, at the very least, choices of branch cuts to make the map and its inverse well defined. However, it is easy to see that away from coordinate hyperplanes, appropriate coordinate identifications make this unnecessary.
From \eqref{fractionaltransform} it is immediate that imposing a $(\bZ_5)^3$ group of identifications on the $(x_0,x_1,x_2,x_3,x_4)$--the very same group, in fact, that yields the mirror Calabi--Yau family--the map from the $y$'s to the $x$'s becomes well defined.

However, this is not yet sufficient for \eqref{fractionaltransform} to be one--to--one. By solving for the inverse, we find, for example:
\begin{align} \label{fractionalinverse}
    y_0 = x_0^{\frac{256}{205}}x_1^{\frac{-64}{205}}x_2^{\frac{16}{205}}x_3^{\frac{-4}{205}}x_4^{\frac{1}{205}} \\ \text{and cyclic permutations} \nonumber
\end{align}
which requires further identifications be made on the $y_i$'s. One can check that \eqref{fractionalinverse} is well defined if one identifies $y_i$ with its image under the $\bZ_{41}$ scaling symmetry indicated above. This suggests a relationship:
\begin{align}\label{mirrorrelation}
    \frac{V_{\text{Fermat}}(t)}{(\bZ_5)^3} \sim \frac{V_{41}(t)}{\bZ_{41}}
\end{align}
But since the fractional change of variables is only invertible away from coordinate hyperplanes, the relationship in \eqref{mirrorrelation}  is not a biholomorphism. It was argued in \cite{Greene:1998yz} using the methods of toric geometry, that the two quotients are nevertheless topologically identitical, representing two parametrizations of  the complex structure moduli space of the quintic mirror, at different points in the K\"{a}hler moduli space.  The relationship \eqref{mirrorrelation} suggests that by focusing on points in the quintic moduli space that have a maximal discrete symmetry group in their local neighborhood, and by then quotienting by this maximal group, we generate the mirror partners to these manifolds. Since the initial manifolds differ by deformations of their complex structures, their mirrors would then differ by deformations of their Kahler structures. In particular, this would mean that the Picard--Fuchs equations for the periods of the holomorphic 3--form of the $\bZ_{41}$ quotient, or equivalently for those periods of $V_{41}(t)$ invariant under $\bZ_{41}$, should agree with standard Picard-Fuchs equation on the mirror quintic. In \cite{Greene:1998yz} a tedious calculation using the Griffiths--Dwork technique on $V_{41}(t)/\bZ_{41}$ was shown to yield:
\begin{align}
    \sq{\ci{1- t  ^5}\diff{^4  }{ t  ^4} -10 t  ^4\diff{^3 }{ t  ^3} - 25 t  ^3 \diff{^2 }{ t  ^2} - 15 t  ^2\diff{ }{ t  } - t  }\int\Omega   = 0
\end{align}
which is precisely the Picard--Fuchs equation satisfied by the periods of the holomorphic 3--form of the mirror family $\widetilde{V}(t)$ \cite{Candelas:1990rm}.

It is easily seen that the technique illustrated by \eqref{fractionaltransform} offers numerous variations, providing a rich set of new
enhanced symmetry loci. For example, one can consider:
\begin{align}
    (x_0,x_1,x_2,x_3,x_4) \to (y_0,y_1^{4/5}y_2^{1/5},y_2^{4/5}y_3^{1/5},y_3^{4/5}y_4^{1/5},y_4^{4/5}y_1^{1/5})
\end{align}
which leads to a family with another unfamiliar symmetry group, $\bZ_{51}$.
Several other examples were tabulated in \cite{Greene:1998yz,Greene:1991iv}, which we repeat in table \ref{symmetricfamilies}.

\begin{table}
\begin{centering} \small
\begin{tabular}{|c|c|c|c|}\hline &$Q(t)$ & Scaling Symmetries & Action \\ \hline 1&$  \frac{1}{5}\ci{a^5+b^5+c^5+d^5+e^5} -t abcde  $ & $ (\mathbb{Z}_5)^3  $ & $ (4,1,0,0,0) $, $(4,0,1,0,0)$, $ (4,0,0,1,0) $ \\ 2& $ \frac{1}{5}\ci{a^4b+b^4c+c^4d+d^4e+e^4a} -t abcde  $ & $ \mathbb{Z}_{41}  $ & $ (1,37,16,18,10)$ \\ 3& $ \frac{1}{5}\ci{a^4b+b^4c+c^4d+d^4a+e^5}-t abcde  $ & $ \mathbb{Z}_{51}  $ & $ (1,47,16,38,0) $ \\ 4& $ \frac{1}{5}\ci{a^4b+b^4c+c^4a + d^5+e^5} - t abcde  $ & $\bZ_5\times\bZ_{13} $ & $(0,0,0,4,1)$, $(1,9,3)$  \\ 5& $ \frac{1}{5}\ci{a^4b+b^4c+c^4a+d^4e+e^4d} -t abcde  $ & $\bZ_3\times\bZ_{13}$ &  $(0,0,0,1,2)$, $(1,9,3)$ \\ 6& $ \frac{1}{5}\ci{a^4b+b^4a + c^5+d^5+e^5}-tabcde  $ & $(\bZ_5)^2\times \bZ_3$  & $(0,0,4,1,0)$, $(0,0,4,0,1)$, $(1,2,0,0,0)$ \\ \hline    \end{tabular}
\end{centering}
\caption{\label{symmetricfamilies} Six 1--parameter families of quintic hypersurfaces with discrete symmetries.}
\end{table}

\normalsize
Given how useful loci with discrete symmetry have been in the development of our understanding of Calabi--Yau moduli spaces, new symmetric families are of great interest. They expand the range of examples to which analytic methods can be applied, and provide new testing grounds for mirror symmetry, rational curve counts, moduli stabilization, and phenomenology.

To orient our analysis, it is interesting to ask where in moduli space these new loci reside; for example, where is the $V_{41}(t)$ family in relation to the Fermat locus? We might attempt a linear transformation $x(y)$ on $Q_{41}(t)$ to bring it into the form:
\begin{align}
    Q_{41} = (y_0)^5 + (y_1)^5 + (y_2)^5 + (y_3)^5 + (y_4)^5 + \ldots =0
\end{align}
where the ellipsis indicates a specific combination of quintic monomials at most cubic in any of the $y_i$'s. But this brings a more pressing issue into sharp relief. Notice that such a linear transformation would obscure the presence of the $\bZ_{41}$ symmetry. Similarly, it is clear that an arbitrary linear transformation of the Fermat quintic would make the $(\bZ_5)^3$ symmetry significantly less obvious because the symmetry would no longer act diagonally on the homogeneous coordinates.

This may lead one to wonder whether the $\mathbb{Z}_{41}$ symmetric family and the Fermat family are even distinct loci. Perhaps $\mathbb{Z}_{41}$ acts nondiagonally on \eqref{FermatPencil}, and $(\mathbb{Z}_5)^3$ nondiagonally on $\eqref{Z41Quintic}$. To show that this is not the case, we write the polynomial defining the Fermat family as: $\frac{1}{5}Q_0 - t Q_{\infty} = 0$, where $Q_0= (x_0)^5 + (x_1)^5 + (x_2)^5 + (x_3)^5 + (x_4)^5$ is the Fermat polynomial and $Q_{\infty}=x_0x_1x_2x_3x_4$. If the symmetry groups of $Q_0$ and $Q_{\infty}$ are denoted  $G_0$ and $G_{\infty}$ respectively, then the automorphism group of a generic member of the Fermat family is just $G_0\cap G_{\infty}$. In \cite{fermatsymmetries} it was shown that $G_0=S_5\ltimes (\bZ_5)^4$, i.e the automorphisms of the Fermat quintic are permutations of the homogeneous coordinates, and scalings by 5th roots of unity, excluding an overall scaling which is trivial in projective space.\footnote{The result: lemma 3.2 of \cite{fermatsymmetries} is rather more general. For $d\geq 3$ and $n\geq 2$ and $(d,n)\neq(3,2)$ or $(4,3)$, the symmetries of the degree $d$ Fermat hypersurface in $\bP_n$ are $S_{n+1}\ltimes (\bZ_{d})^n$. In words, a semi--direct product of permutations and scalings by $d$'th roots of unity. } On the other hand $G_{\infty}=S_5\ltimes (\bC^*)^4$, i.e. we can permute the coordinates, and scale them. The full automorphism group of a generic member of the pencil is then $G=S_5\ltimes (\bZ_5)^3$, i.e. the subgroup of $G_0\cap G_{\infty}$ that scales $\frac{1}{5}Q_0 - t Q_{\infty}$ by an overall factor. In other words there are no more symmetries than those found above by inspection. $G$ is a finite group with $5^3\times5!$ elements, so it has no subgroup of order 41 or 51. It follows that the families with $\bZ_{41}$ and $\bZ_{51}$ symmetries cannot be isomorphic to the Fermat family.

The new loci are thus distinct from the Fermat family, and therefore constitute a new probe for enriching our understanding of Calabi--Yau manifolds and their moduli spaces. Utilizing this probe requires that we're able to perform the basic calculations of periods, familiar from studies of Fermat families, which contain essential information about the complex structure and geometric monodromy of the families. The purpose of this paper is to set up the formalism for doing so.

\subsection*{Plan of the Paper}

In section \ref{hypersurfacecohomology}, we review the aspects of the cohomology of families of hypersurfaces required to understand the Picard--Fuchs equations and the Griffiths--Dwork procedure for finding them. The expert reader will find much in this section that is already familiar. However, because our results and, in particular, the way they differ from \cite{Candelas:2000fq}, depend critically on this background, a self-contained summary is essential. We emphasize those aspects which play key roles in the sections that follow. Some examples and further aspects of the formalism are developed in appendices \ref{GDexamples} and \ref{geometricmonodromy}.

In section \ref{CandelasMethodSec} we derive an alternative to the Griffiths--Dwork method for symmetric hypersurfaces in $\bP_n$ which greatly reduces the labor of computation. The technique is similar in its details to that applied to the quintic 3--fold by Candelas, de la Ossa and Rodriguez--Villegas \cite{Candelas:2000fq}, but our results improve on \cite{Candelas:2000fq} in three key respects:
\begin{itemize}
    \item Here (in Section \ref{DiagramMethod}) the technique is derived in a rigorous fashion from fundamental results of Griffiths \cite{Griffiths}.
    \item In \cite{Candelas:2000fq} it is claimed that these methods compute differential equations whose solutions are periods of the holomorphic 3--form. By carefully relating the method to the work of Griffiths \cite{Griffiths} (reviewed in section \ref{hypersurfacecohomology}), we show that this is generally not true. Instead, we show that most of the resulting equations are satisfied by other elements of the period matrix, i.e. integrals of elements of  $H^3(X,\bC)$ not contained in $H^{3,0}(X)$.
    \item The technique involves constructing diagrams which display the relevant relations among periods in a useful way. We find an algorithm for sytematically constructing these diagrams, summarized in section \ref{diagramalgorithm}.
\end{itemize}

We then apply the procedure to calculate the Picard--Fuchs equations for the $\bZ_{41}$--symmetric family of 3--folds. Other families can be treated in the same way, and the results for the $\bZ_{51}$ case are tabulated in appendix \ref{Z51calc}.

To give a feel for the new loci and to see another way in which they differ from the Fermat family, section \ref{yukawas} examines the effect of the discrete symmetries on the Yukawa couplings of the 6 quintics in table \ref{symmetricfamilies}. We find a somewhat surprising relation between the number of nonzero couplings and the size of the symmetry group.

Finally, in appendix \ref{weighted} we indicate how to extend the technique to symmetric Calabi--Yau hypersurfaces in weighted projective spaces, and we compute the Picard--Fuchs equation for the example of a Fermat--type hypersurface in $\bW\bP_{[41,48,51,52,64]}[256]$.

\section{Cohomology of Hypersurfaces} \label{hypersurfacecohomology}

The Calabi--Yau manifolds we will consider are all hypersurfaces in $\bP_n$, i.e. submanifolds of $\bP_n$ defined by the vanishing locus of a single homogeneous polynomial.\footnote{In appendix \ref{weighted} we generalize this slightly to hypersurfaces in weighted projective space.} This is a rather special choice, which can be generalized in many ways, but is particularly convenient to analyze. The goal of this section is to review a useful way to get our hands on elements of the Dolbeault cohomology groups of such a hypersurface.

We consider a smooth hypersurface $V\subset \bP_n$ defined by the zero locus of an irreducible degree $l$ polynomial $Q(x)$ in the homogeneous coordinates $[x_0,\ldots,x_n]$. As might be expected, it is hard to set up coordinates on $V$ and describe differential forms on the hypersurface directly. Instead we make use of a beautiful generalization of the Cauchy integral formula due to Griffiths \cite{Griffiths}.

Starting from Cauchy's theorem
\begin{align}
    \frac{1}{2\pi i}\oint_C \frac{P(z)}{Q(z)}\ef{z} = \sum_i w_i(i\text{th Residue})
\end{align}
where the sum is over the poles enclosed by the contour $C$ that winds around the $i$th pole $w_i$ times, and the residue is the coefficient of the $1/z$ term in a Laurent expansion of $f(z)= \frac{P(z)}{Q(z)}$ about the pole. Griffiths interpreted the right hand side as the integral of a 0--form over a 0--cycle on the Riemann sphere $\bP_1$. The 0--cycle (which we suggestively denote by $V$) is the set of poles of $\frac{P(z)}{Q(z)}\ef{z}$, each weighted by the number of times the contour $C$ winds around, and the 0--form is the value of the residue at each pole. Notice that adding an exact rational 1--form (whose poles are contained in $V$) to the left hand side integrand makes no difference, so we can think of the residue as a map:
\begin{align}
    \text{Res}: \cH(V) \to H^0(V,\bC)
\end{align}
where $\cH(V)$ is like the de Rham cohomology group $H^1(\bP_1-V,\bC)$, but using only rational forms. The purpose of generalizing this story to higher dimensions is to represent $(n-1)$--forms on $V$ (which becomes a hypersurface), as residues of rational $n$--forms on the complement $\bP_n-V$. The latter are considerably easier to work with.

\subsection{Some Results of Griffiths}
Let $A^n(V)$ be the space of rational $n$--forms on $\bP_n$ with polar locus $V$. Then we define $\cH(V)=A^n(V)/\ef{A}^{n-1}(V)$, i.e. the de Rham cohomology of rational $n$--forms on $\bP_n-V$. \footnote{Note that $\partial A^n(V)=0$ since $n$ is the maximal holomorphic degree of a form on $\bP_n$, and $\bar{\partial}A^n(V)=0$ because rational forms are by definition holomorphic. In defining $\cH(V)$ there is therefore no need to restrict the numerator of the quotient to closed forms only --- they are all closed.} The residue map $\text{Res}:\cH(V)\to H^{n-1}(V,\bC)$ is then defined by the property :
\begin{align} \label{residuedef}
    \frac{1}{2\pi i}\int_{T(\gamma)}\varphi = \int_{\gamma}\text{Res}(\varphi)
\end{align}
Here $\varphi\in\cH(V)$, $\gamma$ is an $(n-1)$--cycle in $V$ and $T(\gamma)$ is a tubular neighborhood of $\gamma$ in $\bP_n-V$. More precisely, $T(\gamma)$ is a circle bundle over $\gamma$ with an embedding into $\bP_n-V$ such that it encloses $\gamma$. For small enough radii, any two such bundles are homologous in $H_n(\bP_n-V,\bZ)$, so the construction is unique. A rather abstract way to state the definition \eqref{residuedef} is that $\text{Res}$ is the dual of the so--called \emph{Leray coboundary map}: $H_{n-1}(V,\bZ)\to H_n(\bP_n-V,\bZ)$ which sends $[\gamma]$ to $[T(\gamma)]$.

A more concrete description of the residue map can be given as follows. Let $\varphi$ be a smooth differential form on $\bP_n$, except for possible singularities on $V$. To indicate the order of the singularities, suppose that for some positive integer $k$, $f^k\varphi$ and $f^{k-1}\ef{f}\wedge\varphi$ are smooth everywhere if $f=0$ is a local defining equation for $V$. In terms of local (affine) coordinates $z^1,\ldots,z^n$, we have $\varphi = \varphi(z)\ef{z^1}\wedge\ldots\wedge\ef{z^n}$, but close to the hypersurface (i.e. near $f=0$), we can choose coordinates $(z^1, \ldots,z^{n-1},f)$ and write:
\begin{align} \nonumber
    \varphi &= \frac{\ef{f}\wedge\alpha}{f^k} + \frac{\beta}{f^{k-1}}\\
    &= -\frac{1}{k-1}\ef{}\ci{\frac{\alpha}{f^{k-1}}} + \frac{\beta+\frac{1}{k-1}\ef{\alpha}}{f^{k-1}}
\end{align}
where $\alpha$ and $\beta$ are smooth forms and do not contain $\ef{f}$. This expression is only valid in a single patch, but by making use of a partition of unity, one can show that for $k\neq 1$, $\varphi = \ef{\psi}+\eta$ where $\psi$ and $\eta$ are \emph{globally defined} smooth forms with poles of order $k-1$ along $V$. It follows that up to an exact form, $\varphi$ can be reduced to a form with a pole of order 1 along $V$:
\begin{align} \label{reseqn}
    \varphi - \ef{(\psi_1 + \ldots + \psi_{l-1})} = \frac{\alpha'\wedge\ef{f}}{f} + \beta'
\end{align}
where $\psi_a$ has a pole of order $k-a$ along $V$. The residue is then the coefficient of $\ef{f}/f$ restricted to the hypersurface:
\begin{align}
    \text{Res}(\varphi) = \alpha'|_V
\end{align}
This is precisely analogous to the usual definition of the residue as the coefficient of the $1/z$ term in the Laurent expansion. Note that the residue of a rational $n$--form with $k=1$ is necessarily holomorphic, but since the construction above uses a partition of unity, residues of forms with $k>1$ are only smooth in general.

Having defined the residue map, we now need to explore its properties. In particular, we have the following question: which rational $n$--forms on $\bP_n-V$ map to which cohomology classes on $V$? The answer is provided by another beautiful theorem of Griffiths, in preparation for which we must introduce some further formalism.

Let $A^n_k(V)\subset A^n(V)$ denote the rational $n$--forms on $\bP_n-V$ with poles of order $k$ along $V$. By analogy with $\cH(V)$, we can define the cohomology groups:
\begin{align}
    \cH_k(V) = \frac{A^n_k(V)}{\ef{}A^{n-1}_{k-1}(V)}
\end{align}
It is important to note that two such groups $\cH_k(V)$ and $\cH_{k'}(V)$ generally have a nonzero intersection. If for example we take $k=1$ and $k'=2$, the statement is just that there are rational forms with double poles that differ from rational forms with simple poles only by exact rational forms with simple poles. We will explicitly delineate such intersections shortly, but notice that what we're speaking of here is different from the reduction of pole order in the residue construction. There we were interested in lowering the pole order by adding smooth forms. Here we are only allowed to add rational forms. We will return shortly to the question of when such a reduction is possible.

For the moment, let us follow Griffiths and write the groups $\cH_k(V)$ as a sequence of inclusions:
\begin{align} \label{poleorderfiltration}
    \cH_1(V)\subset\cH_2(V)\subset \cdots\cdots \subset \cH_n(V) = \cH(V)
\end{align}
The nontrivial claim here is the final equality: $\cH_n(V) = \cH(V)$, the proof of which can be found in \cite{Griffiths}. A decomposition like \eqref{poleorderfiltration} with inclusions (as opposed to a direct sum decomposition) is called a \emph{filtration}, and we will refer to \eqref{poleorderfiltration} as the filtration of $\cH(V)$ by \emph{order of pole}.

There is another filtration we are interested in, the so--called \emph{Hodge filtration} of $H^{n-1}(V,\bC)$, given by:
\begin{align}
    \bF^{n-1,n-1}(V)\subset\bF^{n-1,n-2}(V)\subset\cdots\cdots\subset\bF^{n-1,0}(V) = H^{n-1}(V,\bC)
\end{align}
where $\bF^{a,b}(V) = H^{a,0}(V)\oplus H^{a-1,1}(V)\oplus\ldots\oplus H^{b,a-b}$. This time the equality on the right hand side is just the Hodge decomposition of cohomology, which holds for all K\"{a}hler manifolds:
\begin{align}
    H^{n-1}(V,\bC) = H^{n-1,0}(V)\oplus H^{n-2,1}(V)\oplus \ldots\oplus H^{0,n-1}(V)
\end{align}
Any algebraic submanifold of $\bP_n$ (a hypersurface for example) is necessarily K\"{a}hler, so we are not imposing any new restriction.

The essence of Griffiths' theorem is that the residue map acts in a very nice way between the order of pole filtration and the Hodge filtration:
\begin{align} \label{filtrationisomorphism}
    \begin{array}{ccccccccc}\cH_1(V) & \subset & \cH_2(V) & \subset & \cdots\cdots & \subset & \cH_n(V) & = & \cH(V) \\  \hcm\downarrow \text{Res} &  & \hcm\downarrow \text{Res} &  &  &  & \hcm\downarrow \text{Res} &  & \hcm\downarrow \text{Res} \\\bF^{n-1,n-1}(V) & \subset & \bF^{n-1,n-2}(V) & \subset & \cdots\cdots & \subset & \bF^{n-1,0}(V) & = & H^{n-1}(V,\bC)\end{array}
\end{align}
We have already confirmed the far left part of the diagram: rational $n$--forms on $\bP_n$ with poles of order 1 on $V$ map to holomorphic $(n-1)$--forms on $V$, i.e. elements of $H^{n-1,0}(V)=\bF^{n-1,n-1}(V)$. The remainder of the proof can be found in \cite{Griffiths}. This is the answer we were looking for to the question: which $(n-1)$--forms on $V$ are the residues of which $n$--forms on $\bP_n-V$? The order of the pole of the form on $\bP_n-V$ (and hence the degree of the numerator) determines which of the Hodge filtrants the residue lies in.

We can say more. It is clear when a form in $\bF^{n-1,n-2}$ is also in $\bF^{n-1,n-1}$, but not so clear yet when a form in $\cH_2(V)$ is also in $\cH_1(V)$. In words: we know from \eqref{reseqn} that the order of the pole of a form can be lowered arbitrarily by adding an appropriate \emph{smooth} form, but when can this decrease be accomplished using only \emph{rational} forms? The key to finding out is to look more closely at rational $n$--forms on $\bP_n-V$. Working in homogeneous coordinates $[x^0,\ldots,x^n]$, one can show that \emph{any} such $n$--form $\varphi$ can be written:\footnote{This is Corollary 2.1 of \cite{Griffiths}. The hat on the $\widehat{\ef{x^i}}$ indicates that it should be left out of the wedge product.}
\begin{align} \label{nform}
    \varphi = \frac{P(x)}{Q(x)^k}\Omega_0, \cm \Omega_0=\sum_{i=0}^n (-1)^i x^i \ef{x^0}\wedge \ldots \widehat{\ef{x^i}}\ldots\wedge\ef{x^n}
\end{align}
where $Q(x)=0$ defines the hypersurface $V$, and $P(x)$ is a homogeneous polynomial obeying $\deg P = k\deg Q - (n+1)$ in order that $\varphi$ is well defined on projective space. It follows that a rational $n$--form in $\bP_n-V$ can be specified by an element of $\bC[x^0,\ldots,x^n]_{kl-(n+1)}$, i.e. a polynomial of degree $kl-(n+1)$ with complex coefficients, where $l = deg Q$. The information we need is contained in the formula:\footnote{Formula 4.5 of \cite{Griffiths}}
\begin{align}
    \frac{\Omega_0}{Q(x)^{k+1}}\sum_{i=0}^n P_i(x)\der{Q(x)}{x^i} = \frac{1}{k} \frac{\Omega_0}{Q(x)^k} \sum_{i=0}^n\der{P_i(x)}{x^i} + \text{exact rational forms}
\end{align}
Thus the order of pole of a form $\varphi=\frac{P(x)}{Q(x)^k}\Omega_0\in H_k(V)$ can be lowered using a rational form iff $P(x) = \sum_{i=0}^nP_i(x)\der{Q(x)}{x^i}$ for some polynomials $P_i(x)$. The ideal generated by $[\partial_0 Q(x),\ldots,\partial_n Q(x)]$ is called the Jacobian ideal of $Q(x)$, and denoted $J(Q)$. So the order of pole can be lowered iff the numerator is in $J(Q)$.

This relates to the Hodge filtration as follows. If we quotient a filtrant (a single group in the filtration) by the filtrant to its left, we find $\bF^{n-1,k}(V)/\bF^{n-1,k+1}(V) = H^{k,n-1-k}(V)$. We have just seen that for the order of pole filtration we have:\footnote{A common alternative notation for the Jacobian ring $\frac{\bC[x^0,\ldots,x^n]_m}{J(Q)}$ is $R^m_Q$.}
\begin{align}
    \frac{\cH_k(V)}{\cH_{k-1}(V)} = \frac{\bC[x^0,\ldots,x^n]_{kl-(n+1)}}{J(Q)}
\end{align}
So the residue map induces an isomorphism:
\begin{align} \label{monomialmap}
    \frac{\bC[x^0,\ldots,x^n]_{kl-(n+1)}}{J(Q)} \to H^{n-k,k-1}(V)
\end{align}

One further note of importance is the following. The image of the map \eqref{monomialmap} in $H^{n-k,k-1}(V)$ is called the \emph{primitive cohomology} of $V$, and denoted $PH^{n-k,k-1}(V)$. If in \eqref{filtrationisomorphism} we replace the Hodge filtrants $\bF^{a,b}(V)$ with their analogs constructed from primitive cohomology groups (denoted $\bF_0^{a,b}(V)$), then the residue maps become isomorphisms.

This is the most we will say about the general properties of the residue map. Later we will consider the simplifications that arise if $V$ has discrete symmetries.

 \subsection{Example: Quintic Calabi--Yau 3--folds}
Consider the case $n=4$, $l=5$. Since $l=n+1$ is precisely the condition $c_1=0$, $V$ is a Calabi--Yau 3--fold. The Hodge filtrants are:
\begin{align}
    \bF^{3,3}_0(V) &= PH^{3,0}(V) \nonumber \\ \nonumber
    \bF^{3,2}_0(V) &= PH^{3,0}(V)\oplus PH^{2,1}(V) \\ \nonumber
    \bF^{3,1}_0(V) &= PH^{3,0}(V)\oplus PH^{2,1}(V) \oplus PH^{1,2}(V) \\ \nonumber
    \bF^{3,0}_0(V) &= PH^{3,0}(V)\oplus PH^{2,1}(V) \oplus PH^{1,2}(V)\oplus PH^{0,3}(V)
\end{align}
The isomorphism with the filtration given by order of pole is:\footnote{For odd dimensional hypersurfaces we have $PH^{p,q}(V)\simeq H^{p,q}(V)$. In the case of Calabi--Yau 3--folds this can be seen as follows: One can define $PH^{p,3-p}(V)$ alternatively as the kernel of the Lefschetz map $L:H^{p,3-p}(V)\to H^5(V)$ defined by $L([\phi]) = [J\wedge \phi]$, where $J$ is a K\"{a}hler form on $V$. For a Calabi--Yau 3--fold with $SU(3)$ holonomy, $b_5=0$, so the kernel of $L$ is the whole of $H^{p,3-p}(V)$.}
\begin{align}
    \begin{array}{ccccccccc}\cH_1(V) & \subset & \cH_2(V) & \subset & \cH_3(V) & \subset & \cH_4(V) & = & \cH(V) \\\hcm\downarrow\text{Res} &  & \hcm\downarrow\text{Res} &  & \hcm\downarrow\text{Res} &  & \hcm\downarrow\text{Res} &  & \hcm\downarrow\text{Res} \\\bF^{3,3}_0(V) & \subset & \bF^{3,2}_0(V) & \subset & \bF^{3,1}_0(V) & \subset & \bF^{3,0}_0(V) & = & H^{3}(V,\bC)\end{array}
\end{align}
Now look for example at Res: $\cH_2(V)\to\bF^{3,2}_0(V)$. Notice that $[\alpha_2]\in\cH_2(V)$ maps to $\bF^{3,3}(V)\subset\bF^{3,2}(V)$ iff $[\alpha_2]=[\alpha_1]$ where $\alpha_1$ has a pole of order 1, i.e. if $\alpha_2 = \alpha_1+\ef{\eta}$. From \cite{Griffiths}, this is equivalent to $P\in J(Q)$, where $\alpha_2 = \tfrac{P}{Q^2}\Omega_0$. We therefore have:
\begin{align} \label{PHjac1}
    \frac{\bC[x_0,\ldots,x_4]_5}{J(Q)} \simeq \frac{\cH_2(V)}{\cH_1(V)} \simeq \frac{\bF^{3,2}_0(V)}{\bF^{3,3}_0(V)} \simeq PH^{2,1}(V)
\end{align}
Similarly, we have:
\begin{align}
    \frac{\bC[x_0,\ldots,x_4]_0}{J(Q)} &\simeq \frac{\cH_1(V)}{\cH_0(V)} \simeq \bF^{3,3}_0(V) \simeq PH^{3,0}(V)   \\
    \frac{\bC[x_0,\ldots,x_4]_{10}}{J(Q)} &\simeq \frac{\cH_3(V)}{\cH_2(V)} \simeq \frac{\bF^{3,1}_0(V)}{\bF^{3,2}_0(V)} \simeq PH^{1,2}(V) \\
    \frac{\bC[x_0,\ldots,x_4]_{15}}{J(Q)} &\simeq \frac{\cH_4(V)}{\cH_3(V)} \simeq \frac{\bF^{3,0}_0(V)}{\bF^{3,1}_0(V)} \simeq PH^{0,3}(V) \label{PHjac4}
\end{align}
The maps are given explicitly by:
\begin{align}
    \frac{\bC[x_0,\ldots,x_4]_{5n}}{J(Q)}\ni[P] \longleftrightarrow (3-n,n) \text{ piece of }\text{Res} \ci{\frac{P}{Q^k}\Omega_0}\in PH^3(V,\bC)
\end{align}
where $k=\frac{\deg P}{5}+1$. This is just the standard and often--used association between $5n^{\text{th}}$ order monomials and elements of $H^{3-n,n}(V)$. For instance, the isomorphism \eqref{PHjac1} ($n=1$) has a familiar interpretation as two different ways of looking at deformations of complex structure: on the one hand as an element of $H^{2,1}(V)$ and on the other as an additional monomial term in the defining equation of the hypersurface $V$.

As an example of the use of the above formalism, we derive a common expression for the unique holomorphic 3--form $\Omega=\text{Res}\ci{\frac{\Omega_0}{Q}}$. Working in the patch $x_0\neq 0$, we can scale the homogeneous coordinates so that $x_0=1$, and therefore $\Omega_0=\ef{x_1}\wedge\ef{x_2}\wedge\ef{x_3}\wedge\ef{x_4}=\ef{z^1}\wedge\ef{z^2}\wedge\ef{z^3}\wedge\ef{z^4}$ where $z^i=x_i/x_0$. Writing $Q$ as $Q(x_0,x_1,x_2,x_3,x_4)$, we define $f = Q(1,z^1,z^2,z^3,z^4)$, and so $\Omega = \text{Res}\ci{\frac{\ef{z^1}\wedge\ef{z^2}\wedge\ef{z^3}\wedge\ef{z^4} }{f}}$. Next we replace the coordinate $z_4$ with $f$, and using $\ef{f} = \frac{\partial f}{\partial z^i}\ef{z^i}$, find:
\begin{align} \label{hol3form}
    \Omega = \text{Res}\ci{\frac{\ef{z^1}\wedge\ef{z^2}\wedge\ef{z^3}\wedge\ef{f} }{f\frac{\partial f}{\partial z^4}}} = \left. \frac{\ef{z^1}\wedge\ef{z^2}\wedge\ef{z^3}}{\frac{\partial f}{\partial z^4}} \right|_V
\end{align}
which can be found in \cite{Witten:1985xc},\cite{Candelas:1987se},\cite{Green:1987mn} for example. It is more difficult to identify forms with monomials of higher order, but explicit expressions are derived in \cite{Candelas:1987se}. Aside from the case of the holomorphic 3--form, the residues are generally not of pure Hodge type, i.e. they are not elements of any single group $H^{p,3-p}(V)$. This conclusion requires modification if $V$ possesses discrete symmetries.

\subsubsection*{Hodge Type of the Non--Holomorphic Forms}
We saw that for a generic quintic in $\bP_4$, the 3--forms corresponding to monomials of order $5n$  live in the $(3-n)$'th Hodge filtrant: $\bF^{3,n} = H^{3,0}\oplus\ldots\oplus H^{3-n,n}$. In the presence of a discrete scaling symmetry, which is also a symmetry of the holomorphic 3--form,\footnote{In the mathematics literature, symmetries that preserve the holomorphic 3--form are called \emph{symplectic automorphisms}.} we can make a stronger statement.

For example, let $[\omega]\in H^{3,0}\oplus H^{2,1}$ be a class corresponding to a 5th order monomial. If $\Omega$ is the holomorphic 3--form, and $\omega^{2,1}$ a $\bar{\partial}$--closed $(2,1)$--form, we can write $[\omega] = c[\Omega] + [\omega^{2,1}]$ for some constant $c\in\bC$. Now consider the integral:
\begin{align}
    \int\omega\wedge\overline{\Omega}=c\int\Omega\wedge\overline{\Omega}
\end{align}
The right hand side is manifestly invariant, so if $[\omega]$ transforms by a nontrivial scaling, then $c=0$. This however is just the same as saying that $[\omega]\in H^{2,1}$. A similar argument shows that the noninvariant 10th order monomials correspond to classes in $H^{2,1}\oplus H^{1,2}$ rather than the full $\bF^{3,1}$.

\subsection{Families of Hypersurfaces and the Period Matrix} \label{periodmatrix}
Say we allow $Q(x)$ and hence $V$ to depend on a parameter $t$ which takes values in a space $T$. If we vary $t$ smoothly from $t_1$ to $t_2$ and do not allow $V$ to become singular along the way, then $V(t_1)$ and $V(t_2)$ are diffeomorphic, but not in general biholomorphic to one another.\footnote{The complex structure we are talking about on $V(t)$ is the one inherited from the embedding in $\bP_n$.} Our interest is in the cohomology of the hypersurface $V$, and in particular in the Hodge decomposition.
\begin{align}
    H^{n-1}(V,\bC) = \bigoplus_{p+q=n-1}H^{p,q}(V)
\end{align}
As we move around in complex structure moduli space, the Hodge decomposition changes because what we mean by a $(p,q)$--form changes. However, one can always define a so--called topological basis of $H^{n-1}(V,\bC)$ that does not change (at least under local deformations). This basis consists of the duals of a basis of topological cycles in $H_{n-1}(V,\bZ)$.\footnote{We make use of the inclusion: $H^{n-1}(V,\bZ)\hookrightarrow H^{n-1}(V,\bC)$} The purpose of this section is to study in concrete terms how the Hodge basis varies with respect to the `anchor' of the topological basis.\footnote{This subject has been formulated in a more abstract fashion, under the name \emph{variations of Hodge structure}. An excellent and readable introduction can be found in \cite{Griffiths2}. }

We can phrase the discussion in terms of a fixed real differentiable manifold $X$ diffeomorphic to $V(t)$, and a basis of $t$--dependent $(n-1)$--forms $\Omega^i_X(t)\in\Lambda^{n-1}(X,\bC)$ where $i=0,\ldots,b_3-1$. We can choose the forms $\Omega^i_X(t)$ to have fixed bidegree $(p,q)$ in the complex structure at point $t$, so that their cohomology classes provide the Hodge decomposition for each $t\in T$. But because residues are generally not of pure Hodge type (see \eqref{filtrationisomorphism}), it makes more sense to work in terms of the Hodge filtration. In other words, we choose forms $\Omega^i_X(t)$ to be elements of $\bF^{n-1,p}(V)$ rather than $H^{n-1-p,p}(V)$.

Next we define the period matrix of $V(t)$ as the integrals of $\Omega^i_X(t)$ over a topological basis of $H_{n-1}(X,\bZ)$. Denoting such a basis $\gamma_i$, $i=1,\ldots,2(n-1)$, the periods are:
\begin{align}
    \int_{\gamma_i}\Omega^j_X(t)
\end{align}
It is these integrals that contain the information about how the Hodge structure varies with $t$. We would therefore like to differentiate them with respect to $t$. It might seem naive to just differentiate under the integral sign, but in fact that is the correct thing to do. Technically we are looking for a connection on the bundle over $T$ whose fibers are $H^{n-1}(V(t),\bC)$ (the so--called  \emph{Hodge bundle}), but this bundle admits a flat connection $\nabla_t$ known as the \emph{Gauss--Manin connection} that can be defined by the property:
\begin{align} \label{GaussManin}
    \nabla_t \int_{\gamma_i}\Omega^j_X(t) = \int_{\gamma_i}\diff{}{t}\Omega^j_X(t)
\end{align}
Notice that by repeatedly differentiating $\Omega^j_X(t)$ we generate a sequence of representatives of classes in $H^{n-1}(X,\bC)$. Since $\dim H^{n-1}(X,\bC)$ is finite (for $X$ compact) we must eventually find that some derivative of $\Omega^j_X(t)$ can be related to lower derivatives up to exact forms, which disappear upon integration. It follows that each column of the period matrix, i.e. the periods $\int_{\gamma_i}\Omega^j_X(t)$ for fixed $j$, obeys a differential equation in the variables $t$, called a Picard--Fuchs equation. By comparing these equations, we will be able to distinguish between different families of Calabi--Yau manifolds.

First though, despite the simplification of \eqref{GaussManin}, we still have a hurdle to overcome. The problem is that in the section \ref{hypersurfacecohomology} we worked with $t$--dependent hypersurfaces in $\bP_n$ rather than an underlying differentiable manifold $X$ with $t$--dependent $(n-1)$--forms. Given a $(n-1)$--cycle (and hence a period) in $V(t_1)$, what is the corresponding $(n-1)$--cycle in $V(t_2)$?

The conclusion we will find is that we should use the tools of section \ref{hypersurfacecohomology} to rewrite the period matrix as integrals of rational forms on the complement of $V$; the reader interested more in the final result than the technical details may want to skip directly to the result, \eqref{diffunderintsign3}.

 We would like to have a precise notion of cycles varying smoothly with $t$. To this end, consider $\pi:X'\to T$ a differentiable proper mapping of differentiable manifolds, with rank $=\dim T$, so that $X_t=\pi^{-1}(t)$ is diffeomorphic to a compact manifold $X$ for any $t$. We are interested in the case where $X_t$ inherits a complex structure from $X'$, and is biholomorphic to $V(t)$. The Ehresmann theorem implies that the fibration $X'\to T$ is locally trivial, so we can think of $X'$ as a fiber bundle with base $T$. We can then specify an \emph{Ehresmann connection} on $X'\to T$, i.e. a decomposition of the tangent space $TX'$ into vertical and horizontal subspaces $T^hX'$ and $T^vX'$ respectively. This defines a notion of parallel transport along a path in $T$. So given a cycle $\gamma(t_1)\in X_{t_1}$ and a smooth path linking $t_1$ and $t_2$, an Ehresmann connection defines a cycle $\gamma(t_2)\in X_{t_2}$. Although $\gamma(t_2)$ depends on the path taken in $T$ as well as the choice of connection, $[\gamma(t_2)]\in H_{n-1}(X_{t_2},\bZ)$ (locally) does not. Moreover, if $\alpha(t):X_t\to X$ is a diffeomorphism, then in $H_{n-1}(X,\bZ)$ we have: $[\alpha(t_1)\gamma(t_1)]=[\alpha(t_2)\gamma(t_2)]$.

To summarize, one can choose a basis of $H_{n-1}(V(t_0),\bZ)$, and parallel transport it using an Ehresmann connection to obtain a (locally) unique \emph{horizontal family of homology classes}. How does this help us? It means we can rewrite the period integrals:
\begin{align} \label{diffunderintsign1}
    \int_{\gamma_i}\Omega^i_X(t) = \int_{\gamma_i(t)}\Omega^i_{V(t)}(t)
\end{align}
where for example $\Omega^0_{V(t)}(t)$ is the holomorphic $(n-1)$--form on $V(t)$, found using the techniques that lead to \eqref{hol3form}. We have transformed the integrand to something we know how to work with, but at the expense of introducing $t$ dependence into the cycle over which we are integrating. How do we then differentiate the periods with respect to $t$? The answer is to represent $\Omega^i_{V(t)}(t)$ as a meromorphic form on $\bP_n$:
\begin{align}
    \int_{\gamma_i(t)}\Omega^i_{V(t)}(t) = \int_{T\ci{\gamma_i(t)}}\frac{P}{Q(t)^k}\Omega_0
\end{align}
which is just to say that $\Omega^i_{V(t)}(t)=\text{Res} \ci{\frac{P\Omega_0}{Q(t)^k}}$. For example, the case $i=0$ is the holomorphic $(n-1)$--form:
\begin{align}
    \int_{\gamma_i(t)}\Omega^0_{V(t)}(t) = \int_{T\ci{\gamma_i(t)}}\frac{\Omega_0}{Q(t)}
\end{align}
For $i\neq 0$, there will be a nontrivial $P$ and a higher power of $Q(t)$ in the denominator.

 Now for $t$ in a sufficiently small neighborhood of $t_0$, $T\ci{\gamma_i(t)}$ is homologous to $T\ci{\gamma_i(t_0)}$ in $H_n(\bP_n-V,\bC)$. For the case of a single parameter, we can then differentiate as follows:
\begin{align} \label{diffunderintsign3}
    \diff{}{t} \int_{T\ci{\gamma_i(t)}}\frac{P\Omega_0}{Q(t)^k} = \diff{}{t}\int_{T\ci{\gamma_i(t_0)}}\frac{P\Omega_0}{Q(t)^k} = -k\int_{T\ci{\gamma_i(t_0)}}\frac{P\Omega_0}{Q(t)^{k+1}}\frac{\ef{Q}}{\ef{t}}
\end{align}
If $r=\dim_{\bC}\ci{\cH_n(V)}=\dim_{\bC}\ci{H^{n-1}(V,\bC)}$, only the first $r-1$ derivatives can be linearly independent. Therefore the periods must satisfy a linear ordinary differential equation of order at most $r$ --- this is a Picard--Fuchs equation.

 \subsection{Picard--Fuchs Equations \`{a} la Griffiths--Dwork} \label{fermatexample}
The tools introduced above provide a systematic, but usually tedious, technique for calculating Picard--Fuchs equations, outlined for example in \cite{CoxKatz}:
\begin{enumerate}
    \item Differentiate the period $r$ times. If $Q(t)$ is linear in $t$, the one finds:
    \begin{align}
        \diff{^r}{t^r} \int_{T\ci{\gamma_i(t)}}\frac{P\Omega_0}{Q(t)^k} = \int_{T\ci{\gamma_i(t)}}\frac{(k+r-1)!}{(k-1)!}\frac{\Omega_0}{Q(t)^{k+r}}P\ci{-\frac{\partial Q}{\partial t}}^r
    \end{align}
    \item Write $P\ci{-\frac{\partial Q}{\partial t}}^r$ explicitly as an element of $J(Q)$, i.e. as $\sum_iA_i(x) \frac{\partial Q}{\partial x_i} $ where $A_i(x)$ are polynomials of degree $(n+1)(r+k-1) - n$.
    \item Use the formula 4.5 from \cite{Griffiths} to reduce the order of the pole:
    \begin{align} \label{reducepoleorder}
        \frac{\Omega_0}{Q^{k+1}} \sum_{i=0}^n A_i\frac{\partial Q}{\partial x_i}  = \frac{1}{k}\frac{\Omega_0}{Q^k}\sum_{i=0}^n\frac{\partial A_i}{\partial x_i}  + \ef{(\cdots)}
    \end{align}
    \item Repeat the above steps until the r$^{\text{th}}$ derivative has been expressed in terms of lower derivatives. This is the required equation.
\end{enumerate}
Appendix \ref{GDexamples} contains two worked out applications of the above steps: the Hesse family of elliptic curves and the Fermat family of quintics in $\bP_4$.

\subsection*{Parameterizing the Moduli Space} \label{symcohom}
Conventionally, the Picard--Fuchs equation satisfied by periods of the holomorphic 3--form on the Fermat quintic or its mirror, is written as in section \ref{fermatexample}:
\begin{align} \tag{\ref{genhyperquintic}}
\sq{\theta^4 - x\ci{\theta+\frac{1}{5}}\ci{\theta+\frac{2}{5}}\ci{\theta+\frac{3}{5}}\ci{\theta+\frac{4}{5}}  }t\cP_1 = 0
\end{align}
Recall that $x=t^{-5}$ and $\theta=x\diff{}{x}$, while $t$ is the parameter of the Fermat family:
\begin{align} \tag{\ref{fermatquintic}}
    Q(t) = \frac{1}{5}\bci{a^5+b^5+c^5+d^5+e^5} - tabcde
\end{align}
For some purposes this is rather convenient; for example, since (\ref{genhyperquintic}) is in generalized hypergeometric form, one can make use of standard results about the monodromy of its solutions.\cite{ohara}

However, the coordinate $x$ is not particularly useful for analysing the equations satisfied by the other periods. To see why, we review the usual argument for the change of variables. It is noted\footnote{See for example section 2.2 of \cite{CoxKatz}.} that the transformation $t\to e^{2\pi i/5} t$ can be undone by a simple change of coordinates: $x_i\to e^{-2\pi i/5} x_i$, where $x_i$ is any of the homogeneous coordinates $[x_0,x_1,x_2,x_3,x_4]=[a,b,c,d,e]$. Since this transformation is holomorphic, it follows that the hypersurfaces specified by $t$ and by $e^{2\pi i/5}t$ are biholomorphic. The natural coordinate on the complex structure moduli space therefore seems to be $t^5$ or $t^{-5}$.

Let us then take $x=t^5$, and consider the following period of a $(2,1)$--form:
\begin{align}
    \int \frac{a^3b^2}{Q(x)^2}\Omega_0 = \int\frac{a^3b^2}{\sq{\frac{1}{5}\ci{a^5+b^5+c^5+d^5+e^5}-x^{1/5}abcde}^2}\Omega_0
\end{align}
Suppose we want to interpret its monodromy around $x=t=0$. We note that circling around $x=0$ corresponds to $t\to e^{2\pi i/5}t$, which can then be undone by $a\to e^{-2\pi i/5}a$. The form as a whole receives a scaling by $e^{2\pi i/5}=e^{-\frac{4}{5}2\pi i/5}$, coming from the $a^3$ in the numerator, as well as the factor of $\ef{a}$ in $\Omega_0$. If on the other hand we decided to absorb the change into a scaling of $b$ instead of $a$, then the form would scale by a different factor. Working in terms of $x$, it is therefore tricky to determine what part of the monodromy of the periods comes from geometric monodromy of the cycles, and what part comes from the scalings of the forms.

This is not an issue for the holomorphic 3--form (or its derivatives), and hence does not arise in discussions of the quintic mirror. In that case, the numerator of the form is some power of $abcde$, so the overall scaling is the same no matter which of the coordinates one chooses to absorb the scaling of $t$. In the case of the holomorphic 3--form, the numerator is $1$, so the only scaling comes from $\Omega_0$ which behaves the same as $abcde$. One can then make the form as a whole \emph{invariant} under the monodromy by including an additional factor of $t$ in the numerator. This explains the (sometimes obscure) appearance of the factor of $t$ in equation (\ref{genhyperquintic}). The monodromy of the periods is then entirely geometric in origin.

For the non--invariant forms it is unclear how to achieve the same outcome, so we work in terms of the original variable $t$, thus ensuring that the monodromy of the Picard--Fuchs equations comes only from the cycles.

\section{Quintic Calabi--Yau 3--Folds along Enhanced Discrete Symmetry Loci} \label{CandelasMethodSec}
As shown by the examples in appendix \ref{GDexamples}, the Griffiths--Dwork method involves some algebraic tedium, particularly at step 2 of the procedure outlined in \ref{fermatexample}. In this section we exploit much simpler technique, similar to that found in \cite{Candelas:2000fq} but one in which our derivation, utilizing the results reviewed in \ref{hypersurfacecohomology}, establishes a different interpretation than that suggested in \cite{Candelas:2000fq}. We then apply it to two families of Calabi--Yau 3--folds.

\subsection{The Diagram Technique for the Fermat Quintic}
\label{DiagramMethod}

We start with formula 4.5 from \cite{Griffiths}, which in our notation is:
\begin{align} \label{basicformula}
    \frac{\Omega_0}{Q(t)^{k+1}}\sum_{i=0}^n A_i\der{Q(t)}{x^i} = \frac{1}{k}\frac{\Omega_0}{Q(t)^k}\sum_{i=0}^n \der{A_i}{x^i} + \text{exact forms}
\end{align}
Recall that $Q(t)$ is the defining equation of the hypersurface, $A_i$ are homogeneous polynomials  in the $[x^i]$, and $\Omega_0=\sum_i (-1)^ix^i\ef{x^0}\wedge\ldots\widehat{\ef{x^i}}\ldots\wedge\ef{x^n}$. We now specialize to the Fermat family of quintics, i.e. $n=4$, and:
\begin{align} \label{fermatquartic}
    Q(t) = \frac{1}{5}\bci{a^5+b^5 + c^5 + d^5 + e^5} - tabcde
\end{align}
where $[a,b,c,d,e]=[x^0,x^1,x^2,x^3,x^4]$ is an alternative notation for the homogeneous coordinates. For this case, we have:
\begin{align} \label{fermatinvrelation}
    \der{Q(t)}{x^i} = (x^i)^4 - tx^0\ldots\widehat{x^i}\ldots x^4
\end{align}
Next we choose $A_i = \delta_{ij}x_i A$ where $A=(x^0)^{v_0}(x^1)^{v_1}(x^2)^{v_2}(x^3)^{v_3}(x^4)^{v_4}=a^{v_0}b^{v_1}c^{v_2}d^{v_3}e^{v_4}$. Griffiths' formula then becomes:
\begin{align}
    \frac{\Omega_0}{Q(t)^{k+1}}A\bsq{(x^i)^5 - tx^0x^1x^2x^3x^4} = \frac{1}{k}\frac{\Omega_0}{Q(t)^k}(1+v_i)A + \text{exact forms}
\end{align}
In order that these forms are well defined on $\bP_4$, we must have:
\begin{align}
    k=\frac{1}{5}\deg A+1 =\frac{1}{5}(v_0+v_1+v_2+v_3+v_4)+1
\end{align}
so we will write $k(v)$ from now on. Integrating over a cycle in $\bP_4-V$ gives:
\begin{align}
    \int\frac{\Omega_0}{Q(t)^{k(v)+1}}A (x^i)^5 = t \int\frac{\Omega_0}{Q(t)^{k(v)+1}}Ax^0x^1x^2x^3x^4 + \frac{(1+v_i)}{k(v)}\int\frac{\Omega_0}{Q(t)^{k(v)}}A
\end{align}
We can write this relation in the form:
\begin{align} \label{basicrel2}
    (v_0,\ldots,v_i+5,\ldots,v_4) = \frac{(1+v_i)}{k(v)}(v_0,v_1,v_2,v_3,v_4) + t(v_0+1,v_1+1,v_2+1,v_3+1,v_4+1)
\end{align}
which uses the following shorthand for the periods:
\begin{align}
    (v_0,v_1,v_2,v_3,v_4) = \int\frac{\Omega_0}{Q(t)^{k(v)}}(x^0)^{v_0}(x^1)^{v_1}(x^2)^{v_2}(x^3)^{v_3}(x^4)^{v_4}
\end{align}
The opportunity to write a differential equation arises because:
\begin{align} \nonumber
    \frac{\mathrm{d}}{\mathrm{d}t} (v_0,v_1,v_2,v_3,v_4) &= \frac{\mathrm{d}}{\mathrm{d}t} \int \frac{\Omega_0}{Q(t)^{k(v)}}a^{v_0}b^{v_1}c^{v_2}d^{v_3}e^{v_4} \\ &= k(v) \int \frac{\Omega_0}{Q(t)^{k(v)+1}}a^{v_0+1}b^{v_1+1}c^{v_2+1}d^{v_3+1}e^{v_4+1} \\ &= k(v)(v_0+1,v_1+1,v_2+1,v_3+1,v_4+1) \nonumber
\end{align}
To organize these relations in a useful way, the authors of \cite{Candelas:2000fq} presented \eqref{basicrel2} as a diagram:
\begin{align} \label{smalldiagram}
    \left.\begin{array}{ccc}(v_0,v_1,v_2,v_3,v_4) & \longrightarrow & (v_0+1,v_1+1,v_2+1,v_3+1,v_4+1) \\\downarrow D_i &  &  \\(v_0,\ldots,v_i+5,\ldots,v_4) &  & \end{array}\right.
\end{align}
One should read this simply as saying that these three periods are linearly related; the subscript on $D_i$ indicating which linear relation is being used. It's also useful to keep in mind that the period on the top right is proportional to the derivative of the period on the top left.  One can then build up larger diagrams, for example:

\small
\begin{align} \nonumber
    \begin{array}{ccccccccccc}
     &   &  (0,0,0,0,0) & \to  &  (1,1,1,1,1)  & \to  & (2,2,2,2,2) & \to  & (3,3,3,3,3) & \to & (4,4,4,4,4) \\
     &   & \downarrow D_0  &   & \downarrow D_0  &   & \downarrow D_0 &   & \downarrow D_0 & & \\
     &   &  (5,0,0,0,0) & \to  &  (6,1,1,1,1)  & \to  & (7,2,2,2,2) & \to  & (8,3,3,3,3) &  &    \\
     &   & \downarrow D_1  &   & \downarrow D_1  &   & \downarrow D_1 &   &  & & \\
     &   &  (5,5,0,0,0) & \to  &  (6,6,1,1,1)  & \to  & (7,7,2,2,2) &   &  &  &   \\
     &   & \downarrow D_2  &   & \downarrow D_2  &   &  &   &  & & \\
     &   &  (5,5,5,0,0) & \to  &  (6,6,1,1,1)  &  &  &   &  &  &     \\
     &   & \downarrow D_3  &   &  &   &  &   &  & &  \\
     (4,4,4,4,-1) & \to & (5,5,5,5,0) &   &   &   &   &   &     &  &    \\
      \downarrow D_4  &   &  &   &  &   &  & & & & \\
      (4,4,4,4,4) &  &  &   &   &   &  &   &   &  &
     \end{array}
\end{align}
\normalsize
Several comments are in order:
\begin{enumerate}
    \item The entry $(4,4,4,4,-1)$ does not correspond to a period. As one can see from \eqref{basicrel2}, this part of the diagram just says that $(4,4,4,4,4)$ is proportional to $(5,5,5,5,0)$.
    \item Only one of the $D_i$ is used in each row, and each $D_i$ is used once.
    \item Working up the diagram using \eqref{basicrel2}, one can write $(4,4,4,4,4)$ at the bottom in terms of the top row of periods. This is then a 4th order differential equation for the period $(0,0,0,0,0)$, i.e. the periods of the holomorphic 3--form.
    \begin{align} \label{pfquarticfermat1}
        \left[  ( t ^5-1)\ndiff{}{ t }{4} + 10 t ^4\ndiff{}{ t }{2} + 25 t ^3\ndiff{}{ t }{2} + 15 t^2 \ndiff{}{ t }{ } +  t  \right] (0,0,0,0) = 0
    \end{align}
    Or, in terms of $\eta =  t \ndiff{}{ t }{}$:
    \begin{align}
        \left[ (\eta+1)^4 - \frac{1}{t^5}\eta(\eta-1)(\eta-2)(\eta-3) \right] (0,0,0,0,0) = 0
    \end{align}
    One can put this in generalized hypergeometric form with a change of variables: $\lambda = t^5$, $\theta= \lambda\ndiff{}{\lambda}{}$:
    \begin{align} \label{pfquarticfermat3}
        \sq{\theta\ci{\theta-\frac{1}{5}}\ci{\theta-\frac{2}{5}}\ci{\theta-\frac{3}{5}} -\lambda \ci{\theta+\frac{1}{5}}^4}(0,0,0,0,0) = 0
    \end{align} This is the equation satisfied by $_4 F_3\sq{\left. \begin{array}{c}\frac{1}{5} , \frac{1}{5} , \frac{1}{5} , \frac{1}{5} \\ \frac{4}{5} , \frac{3}{5} , \frac{2}{5}   \end{array}\right|  t^5 } $
    \item This procedure is rather more convenient than the Griffiths--Dwork approach. \cite{Griffiths,Dwork}
\end{enumerate}

The reason that a 4th order equation appears as opposed to order $204=\dim_{\bC} PH^3(V(t),\bC)=b_3$ can be traced back to the discrete symmetries of $V(t)$ mentioned in the introduction. Recall that the symmetry group is $(S_5 \ltimes (\bZ_5)^4)/\bZ_5$. For the subgroup of scaling symmetries we can take the following for generators:
\begin{align}
    g_1 = (1,0,0,0,4) \cm
    g_2 = (1,0,0,4,0) \cm
    g_3 = (1,0,4,0,0)
\end{align}
where as before the entries indicate powers of a nontrivial 5th root of unity. Notice that the rule
\eqref{smalldiagram} ensures that all periods in a given diagram transform in the same representation of $(\bZ_5)^3$. It follows that as one moves around the base of the family (i.e. as $t$ varies), each 3--form only samples a subspace of $H^3(X,\bC)$ spanned by those 3--forms transforming in the same representation of $(S_5 \ltimes (\bZ_5)^4)/\bZ_5$.

\subsubsection{Interpretation of Equations from the Diagram Technique}
\label{diagraminterpretation}
Equation \eqref{basicrel2} and its diagramatic representation \eqref{smalldiagram} can be found in section 3.1 of \cite{Candelas:2000fq}. However, in this paper a different meaning is attached to the components of the diagram: $(v_0,v_1,v_2,v_3,v_4)$. In particular the authors of \cite{Candelas:2000fq} define:
\begin{align}
  (v_0,v_1,v_2,v_3,v_4) = \frac{1}{2\pi i}\int_{\Gamma} \ef{^5x}\frac{(x^0)^{v_0}(x^1)^{v_1}(x^2)^{v_2}(x^3)^{v_3}(x^4)^{v_4}}{Q(t)^{k(v)+1}}
\end{align}
where $\Gamma$ is a 5--torus in $\bC^5$ whose factors are loops winding around the 5 varieties $\partial_i Q=0$. Equation 3.2 of \cite{Candelas:2000fq} then claims that this is in fact a period of the holomorphic 3--form:
\begin{align}
  (v_0,v_1,v_2,v_3,v_4) = \int_{\gamma_{\vec{v}}}\Omega
\end{align}
where $\gamma_{\vec{v}}$ is a 3--cycle whose homology class corresponds to the element of the Jacobian ideal represented by the monomial $(x^0)^{v_0}(x^1)^{v_1}(x^2)^{v_2}(x^3)^{v_3}(x^4)^{v_4}$. The purpose of the extended introduction in section \ref{hypersurfacecohomology} (and in particular the statement of Griffith's theorems) is to show that this interpretation cannot be correct. It is clear from the definition of the Gauss--Manin connection in equation \eqref{GaussManin} that \emph{all 204 periods of the holomorphic 3--form} obey the 4th order equation \eqref{pfquarticfermat1}. And it is clear from \eqref{filtrationisomorphism} and the derivation above that the equations corresponding to other monomials are not satisfied by different periods of $\Omega$, but rather by \emph{all 204 periods} of other, \emph{non--holomorphic} forms.\footnote{There is a trivial sense in which our interpretation of the diagrams is consistent with that of \cite{Candelas:2000fq}. One can choose a basis of cycles so that all but four periods of the holomorphic 3--form $\Omega$ vanish identically over the entire moduli space. These 200 vanishing periods of $\Omega$ then satisfy the ODEs assigned to them in \cite{Candelas:2000fq}, simply because zero is a solution of \emph{any} linear homogeneous differential equation.}

\subsubsection*{Periods of the Forms Corresponding to 5th Order Monomials}

In general, the quintic monomials map to classes in $\bF^{3,2}=H^{3,0}\oplus H^{2,1}$, but we saw in section \ref{fermatexample} that the symmetries can entail further restrictions. We therefore classify the elements of $\bC[a,b,c,d,e]_5$ by their transformation properties under the $(\bZ_5)^3$ symmetries of the Fermat quintic. Fortunately there is no need to look at each of the 126 monomials separately, because they fall into 5 sets which transform among themselves under permutations of the homogeneous coordinates. See table \ref{Fermatmonomials}.

\begin{table}[h]
\begin{center}
\begin{tabular}{|c|c|c|} \hline Representation Class & Example Monomials & Number of Monomials in Class\\ \hline $\cA$ & $a^5,b^5,c^5,d^5,e^5,abcde$ & 6 \\ $\cB$ & $a^4b,b^2cde$ & 40 \\ $\cC$ & $a^3b^2$ & 20 \\ $\cD$ & $a^3bc$ & 30 \\ $\cE$ & $a^2b^2c$ & 30\\ \hline \end{tabular}
\end{center} \caption{\label{Fermatmonomials} The 126 quintic monomials split into 5 sets under the permutation and scaling symmetries. The number in the right column is the number of example monomials listed supplemented by their permutations. }
\end{table}

As is well known, finding a set of 101 of these 126 that are independent as elements of $\bC[a,b,c,d,e]_5/J(Q)$ is immediate.
From \eqref{fermatinvrelation}, we see that $a^5,b^5,c^5,d^5,e^5$ and $abcde$ are all equivalent, and from
\begin{align}
    b\partial_a Q(t) = a^4b - tb^2cde
\end{align}
we see that we can get rid of half of the monomials in representation class $\cB$. The result is summarized in table \ref{Fermatmonomials2}.
\begin{table}[h]
\begin{center}
\begin{tabular}{|c|c|c|} \hline Representation  & Independent Example & Number of Independent  \\ Class & Monomials  & Monomials in Class   \\ \hline $\cA$ & $abcde$ & 1 \\ $\cB$ & $b^2cde$ & 20 \\ $\cC$ & $a^3b^2$ & 20 \\ $\cD$ & $a^3bc$ & 30 \\ $\cE$ & $a^2b^2c$ & 30\\ \hline \end{tabular}
\end{center} \caption{\label{Fermatmonomials2} After the quotient by the Jacobian ideal $J(Q)=[\partial_i Q(t)]$, there are 101 independent monomials. }
\end{table}
By the argument at the end of section \ref{fermatexample}, if $[m_5]\in\frac{\bC[a,b,c,d,e]_5}{J(Q)}$ is in representations $\cB,\cC,\cD$ or $\cE$, then the classes $\text{Res}\ci{\frac{m_5 \Omega_0}{Q^2}}$ are elements of $H^{2,1}$ rather than $H^{3,0}\oplus H^{2,1}$. There is no such restriction for the single class in representation $\cA$: the derivative of the holomorphic 3--form with respect to $t$.

\subsubsection{Algorithm for Diagram Construction}
\label{diagramalgorithm}
In the previous section as well as in \cite{Candelas:2000fq}, the diagram for the holomorphic 3--form was constructed and utilized in an ad hoc manner. We now present a general algorithm which can be applied straightforwardly to all the forms on several families of hypersurfaces.

\subsubsection*{Representation Class $\cA$: $\cu{abcde}$}

\begin{enumerate}
    \item In this case, we are looking for an equation satisfied by $(1,1,1,1,1)$. First, differentiate with respect to $t$ as many times as is necessary to create the following `staircase diagram':

\small
\begin{align} \label{algorithmstart}
    \begin{array}{ccccccccccc}
    &   &   &   &   &   &   &   & (3,3,3,3,3)   & \to   &  (4,4,4,4,4) \\
    &   &   &   &   &   &   &   & \downarrow D_0    &   &  \\
    &   &   &   &   &   & (7,2,2,2,2)   & \to   & (8,3,3,3,3)   &   & \\
    &   &   &   &   &   & \downarrow D_1    &   &   &   & \\
    &   &   &   & (6,6,1,1,1)   & \to   & (7,7,2,2,2)   &   &   &   & \\
    &   &   &   & \downarrow D_2    &   &   &   &   &   & \\
    &   & (5,5,5,0,0)   & \to   & (6,6,6,1,1)   &   &   &   &   &   & \\
    &   & \downarrow D_3    &   &   &   &   &   &   &   & \\
 (4,4,4,4,-1)   &  \to  & (5,5,5,5,0)   &   &   &   &   &   &   &   & \\
  \downarrow D_4    &   &   &   &   &   &   &   &   &   & \\
 (4,4,4,4,4)    &   &   &   &   &   &   &   &   &   &
    \end{array}
\end{align}
\normalsize
As before, one uses each of the $D_i$'s once, so that the bottom left and top right periods match.

\item Next extend the diagram to the left as far as possible by adding extra mini--diagrams:

\small
\begin{align} \label{Z41RepclassA}
    \begin{array}{ccccccccccc}
    &   &   &   & (1,1,1,1,1)_A & \to   & (2,2,2,2,2)   & \to   & (3,3,3,3,3)   & \to   &  (4,4,4,4,4) \\
    &   &   &   & \downarrow D_0    &   & \downarrow D_0    &   & \downarrow D_0    &   &  \\
    &   & (5,0,0,0,0)_B & \to   & (6,1,1,1,1)   & \to   & (7,2,2,2,2)   & \to   & (8,3,3,3,3)   &   & \\
    &   & \downarrow D_1    &   & \downarrow D_1    &   & \downarrow D_1    &   &   &   & \\
    &   & (5,5,0,0,0)_C & \to   & (6,6,1,1,1)   & \to   & (7,7,2,2,2)   &   &   &   & \\
    &   & \downarrow D_2    &   & \downarrow D_2    &   &   &   &   &   & \\
    &   & (5,5,5,0,0)_D & \to   & (6,6,6,1,1)   &   &   &   &   &   & \\
    &   & \downarrow D_3    &   &   &   &   &   &   &   & \\
 (4,4,4,4,-1)   &  \to  & (5,5,5,5,0)_E &   &   &   &   &   &   &   & \\
  \downarrow D_4    &   &   &   &   &   &   &   &   &   & \\
 (4,4,4,4,4)    &   &   &   &   &   &   &   &   &   &
    \end{array}
\end{align}
\normalsize
The one exception is that we do not add a piece to the left of the period we are interested in: $(1,1,1,1,1)$. The subscripts on the leftmost periods in each row are the symbols we will use to denote them in equations.

\item We now write the relations corresponding to the leftmost mini--diagrams, i.e. a coupled system in $(A,B,C,D,E)$:
\begin{align}
    \frac{1}{24}\eta(\eta-1)(\eta-2) A &=t^4E, \cm  E = \frac{1}{4}(\eta+1)D, \cm  D = \frac{1}{3}(\eta+1)C \\
    C &= \frac{1}{2}(\eta+1)B, \cm \eta B = t(\eta+2)A
\end{align}
where as before $\eta$ is the logarithmic derivative $\diff{}{(\log t)}=t\diff{}{t}$.

\item Finally we manipulate the coupled system to find an equation containing $A$ alone.
\begin{align} \label{algorithmend}
    \bsq{\eta(\eta-1)(\eta-2)(\eta-4) - t^5(\eta+2)^4} A = 0
\end{align}
For this last step relations like $\eta t = (t+1)\eta$ are particularly helpful.

\end{enumerate}

\subsubsection*{Representation Class $\cB$: $\cu{b^2cde}$}
\small
\begin{align} \nonumber
    \begin{array}{ccccccccc}
    &   &   &   &   &   & (0,2,1,1,1)_A & \to   &  (1,3,2,2,2) \\
    &   &   &   &   &   & \downarrow D_4    &   & \downarrow D_4 \\
    &   &   &   &   &   & (0,2,1,1,6)_B & \to   &  (1,3,2,2,7) \\
    &   &   &   &   &   & \downarrow D_3    &   & \\
    &   &   &   & (-1,1,0,5,5)  & \to   & (0,2,1,6,6)_C &   & \\
    &   &   &   & \downarrow D_0    &   &   &   & \\
    &   & (3,0,-1,4,4)  & \to   & (4,1,0,5,5)_D &   &   &   & \\
    &   & \downarrow D_2    &   &   &   &   &   & \\
     (2,-1,3,3,3)   & \to   & (3,0,4,4,4)_E &   &   &   &   &   & \\
     \downarrow D_1 &   &   &   &   &   &   &   & \\
     (2,4,3,3,3)    &   &   &   &   &   &   &   &
    \end{array}
\end{align}
\normalsize
The coupled system is:
\begin{align}
    \frac{1}{6}\eta(\eta-1)A =t^3E, \cm E = tD, \cm D = tC \\
    C = \frac{1}{3}(\eta+2)B, \cm B = \frac{1}{2}(\eta+2)A
\end{align}
The equation for $A$ is:
\begin{align}
    \bsq{\eta(\eta-1) - t^5(\eta+2)^2} A = 0
\end{align}

\subsubsection*{Representation Class $\cC$: $\cu{a^3b^2}$}
\small
\begin{align} \label{FermatRepclassC}
    \begin{array}{ccccccccccc}
    &   &   &   &   &   &   &   & (3,2,0,0,0)_A & \to   &  (4,3,1,1,1) \\
    &   &   &   &   &   &   &   & \downarrow D_4    &   & \downarrow D_4 \\
    &   &   &   &   &   &   &   & (3,2,0,0,5)_B & \to   &  (4,3,1,1,6) \\
    &   &   &   &   &   &   &   & \downarrow D_3    &   & \\
    &   &   &   &   &   & (2,1,-1,4,4)  & \to   & (3,2,0,5,5)_C &   & \\
    &   &   &   &   &   & \downarrow D_2    &   &   &   & \\
    &   & (0,-1,2,2,2)  & \to   & (1,0,3,3,3)_D & \to   & (2,1,4,4,4)   &   &   &   & \\
    &   & \downarrow D_1    &   &   &   &   &   &   &   & \\
 (-1,3,1,1,1)   & \to   & (0,4,2,2,2)_E &   &   &   &   &   &   &   & \\
 \downarrow D_0 &   &   &   &   &   &   &   &   &   & \\
 (4,3,1,1,1)    &   &   &   &   &   &   &   &   &   &
    \end{array}
\end{align}
\normalsize
The coupled system is:
\begin{align}
    \frac{1}{2}\eta A =t^2E, \cm E = tD, \cm \frac{1}{3}\eta D = t^2C \\
    C = \frac{1}{3}(\eta+1)B, \cm B = \frac{1}{2}(\eta+1)A
\end{align}
The equation for $A$ is:
\begin{align}
    \bsq{\eta(\eta-3) - t^5(\eta+1)^2} A = 0
\end{align}

\subsubsection*{Representation Class $\cD$: $\cu{a^3bc}$}
\small
\begin{align} \nonumber
    \begin{array}{ccccccccccc}
    &   &   &   &   &   &   &   & (3,1,1,0,0)_A & \to   &  (4,2,2,1,1) \\
    &   &   &   &   &   &   &   & \downarrow D_4    &   & \downarrow D_4 \\
    &   &   &   &   &   & (2,0,0,-1,4)  & \to   & (3,1,1,0,5)_B & \to   &  (4,2,2,1,6) \\
    &   &   &   &   &   & \downarrow D_3    &   & \downarrow D_3    &   & \\
    &   &   &   &   &   & (2,0,0,4,4)_C & \to   & (3,1,1,5,5)   &   & \\
    &   &   &   &   &   & \downarrow D_2    &   &   &   & \\
    &   &   &   & (1,-1,4,3,3)  & \to   & (2,0,5,4,4)_D &   &   &   & \\
    &   &   &   & \downarrow D_1    &   &   &   &   &   & \\
 (-1,2,2,1,1)   & \to   & (0,3,3,2,2)_E & \to   & (1,4,4,3,3)   &   &   &   &   &   & \\
 \downarrow D_0 &   &   &   &   &   &   &   &   &   & \\
 (4,2,2,1,1)    &   &   &   &   &   &   &   &   &   &
    \end{array}
\end{align}
\normalsize
The coupled system is:
\begin{align}
    \frac{1}{2}\eta A =t^2E, \cm \frac{1}{3}\eta E = t^2D, \cm  D = \frac{1}{3}(\eta+1)C \\
    C = tB, \cm B = \frac{1}{2}(\eta+1)A
\end{align}
The equation for $A$ is:
\begin{align}
    \bsq{\eta(\eta-2) - t^5(\eta+1)(\eta+2)} A = 0
\end{align}

\subsubsection*{Representation Class $\cE$: $\cu{a^2b^2c}$}
\small
\begin{align} \nonumber
    \begin{array}{ccccccccc}
    &   &   &   &   &   & (2,2,1,0,0)_A & \to   &  (3,3,2,1,1) \\
    &   &   &   &   &   & \downarrow D_4    &   & \downarrow D_4 \\
    &   &   &   & (1,1,0,-1,4)  & \to   & (2,2,1,0,5)_B & \to   &  (3,3,2,1,6) \\
    &   &   &   & \downarrow D_3    &   & \downarrow D_3    &   & \\
    &   & (0,0,-1,3,3)  & \to   & (1,1,0,4,4)_C & \to   & (2,2,1,5,5)   &   & \\
    &   & \downarrow D_2    &   & \downarrow D_2    &   &   &   & \\
    &   & (0,0,4,3,3)_D & \to   & (1,1,5,4,4)   &   &   &   & \\
    &   & \downarrow D_1    &   &   &   &   &   & \\
     (-1,4,3,2,2)   & \to   & (0,5,4,3,3)_E &   &   &   &   &   & \\
     \downarrow D_0 &   &   &   &   &   &   &   & \\
     (4,4,3,2,2)    &   &   &   &   &   &   &   &
    \end{array}
\end{align}
\normalsize
The coupled system is:
\begin{align}
    \frac{1}{6}\eta(\eta-1)A =t^3E, \cm E = \frac{1}{3}(\eta+1)D, \cm D = tC \\
    C = tB, \cm B = \frac{1}{2}(\eta+1)A
\end{align}
The equation for $A$ is:
\begin{align}
    \bsq{\eta(\eta-1) - t^5(\eta+3)(\eta+1)} A = 0
\end{align}

\textbox{
\center{\textbf{Summary of Columns of the Period Matrix \\ Corresponding to 5th Order Monomials}}
\begin{align} \nonumber
    \begin{array}{|c|c|c|c|c|} \hline  & \text{Number of Classes} & \text{Operator Annihilating Periods} & \text{Hodge Type} \\
    \hline \cA & 1 & \eta(\eta-1)(\eta-2)(\eta-4) - t^5(\eta+2)^4 & H^{3,0}\oplus H^{2,1} \\
    \cB & 20 & \eta(\eta-1) - t^5(\eta+2)^2 & H^{2,1} \\
    \cC & 20 & \eta(\eta-3) - t^5(\eta+1)^2 & H^{2,1} \\
    \cD & 30 & \eta(\eta-2) - t^5(\eta+1)(\eta+2)  & H^{2,1} \\
    \cE & 30 & \eta(\eta-1) - t^5(\eta+3)(\eta+1) & H^{2,1}\\ \hline\end{array}
\end{align}
}

\subsubsection*{Periods of the Forms Corresponding to 10th Order Monomials}
It looks like an unpleasant task to sift through the ${14\choose 4}=1001$ 10th order monomials, classifying them by their transformations under $(\bZ_5)^3$, and checking for relations in $J(Q)$. So we take a different route.

Our aim is to find a convenient basis of $\bC[a,b,c,d,e]_{10}/J(Q)$. To this end, notice that we can choose a basis of $\bC[a,b,c,d,e]_5/J(Q)$ not containing any homogeneous coordinate raised to the 4th or 5th power. If the basis elements are restricted to being monomials, then the basis is unique:
\begin{align}
    [m_i] = \cu{[abcde],[b^2cde],[a^3b^2],[a^3bc],[a^2b^2c]} + \text{permutations}
\end{align}
Here $i=1,\ldots,101$. In other words, any element of $\bC[a,b,c,d,e]_5/J(Q)$ can be written: $\sum_i\alpha_i[m_i]$ with $\alpha_i\in\bC$. We now define a map denoted $\star$:
\begin{align}
    \star: \frac{\bC[a,b,c,d,e]_5}{J(Q)} &\to \frac{\bC[a,b,c,d,e]_{10}}{J(Q)} \\
    \text{such that}\hcm \star[m] &= \sq{\frac{a^3b^3c^3d^3e^3}{m}}
\end{align}
Note that it is crucial that the basis $[m_i]$ contains no elements like $[a^4b]$ or $[a^5]$ in order that the map is well defined.  We define the action of $\star$ to be linear:
\begin{align}
    \star \bci{\alpha_1[m_1] + \alpha_2 [m_2]} = \alpha_1\star[m_1] + \alpha_2 \star[m_2]
\end{align}
The claim is that if $[m_i]$ is the above basis of $\frac{\bC[a,b,c,d,e]_5}{J(Q)}$, then $\star[m_i]$ is a basis of $\frac{\bC[a,b,c,d,e]_{10}}{J(Q)}$. To prove it, consider the pairing:
\begin{align}
    F: \frac{\bC[a,b,c,d,e]_5}{J(Q)} \times \frac{\bC[a,b,c,d,e]_{10}}{J(Q)} &\to \frac{\bC[a,b,c,d,e]_{15}}{J(Q)} \simeq \bC \\
    \text{given by}\hcm F\sq{\ci{\sum_i\alpha_i[m_i]},\ci{\sum_j\beta_j[\tilde{m}_j]}} &= \sum_{i,j}\alpha_i\beta_j\sq{m_i\tilde{m}_j}
\end{align}
where $[\tilde{m}_i]$ is a basis of $\bC[a,b,c,d,e]_{10}/J(Q)$. The isomorphism with $\bC$ is realized by taking the coefficient of $[a^3b^3c^3d^3e^3]$ in the sum. We'll denote this coefficient $\hat{F}$.

Now we know that:
\begin{align}
    \dim \frac{\bC[a,b,c,d,e]_5}{J(Q)} = \dim \frac{\bC[a,b,c,d,e]_{10}}{J(Q)} = 101
\end{align}
So, $\star[m_i]$ is a basis of basis of $\bC[a,b,c,d,e]_{10}/J(Q)$ iff the 101$\times$101 matrix $\hat{F}\bci{[m_i],\star[m_j]}$ is nondegenerate. But it's not hard to see that $\hat{F}\bci{[m_i],\star[m_j]}=\delta_{ij}$, the $101\times 101$ identity matrix. So $\star[m_i]$ is the dual basis to $[m_i]$. We have therefore found the basis we were looking for.

Before constructing diagrams for the periods corresponding to $\star[m_i]$, it is worth looking at how the $\star$ operator interacts with the discrete symmetries of the Fermat quintic. First some notation: for the $(\bZ_5)^3$ generated by $g_1,g_2$ and $g_3$, we say that a monomial $[m]$ is in the $(n_1,n_2,n_3)$ representation if $g_i[m] = \gamma^{n_i}[m]$ for $i=1,2,3$ where $\gamma$ is a nontrivial 5th root of unity. It is easy to see that if $[m]$ transforms in the representation $(n_1,n_2,n_3)$, then $\star[m]$ transforms in the representation $(5-n,5-n,5-n)$.

We can go a step further, and think of $\star$ as acting on the classes of irreps of $(\bZ_5)^3$ that transform into each other under permutations. (We labeled these $\cA,\cB,\cC,\cD$ and $\cE$.) One finds a simple action:
\begin{align}
    \star \cA = \cA, \cm \star \cB = \cB, \cm \star \cC = \cC, \cm \star \cD = \cD, \cm \star \cE = \cE
\end{align}
For example, representation class $\cC$ includes the monomial $a^3b^2$, with $(n_1,n_2,n_3)=(3,3,3)$. We find $\star[a^3b^2] = [bc^3d^3e^3]$, which has $(n_1,n_2,n_3) = (2,2,2)$. One can check that this representation is also in class $\cC$. Even better, a permutation of $[bc^3d^3e^3]$ already appears in diagram \eqref{FermatRepclassC} for representation class $\cC$, so there is no need to construct a new diagram. We will see that this happens for the other representations as well.

\subsubsection*{Representation Class $\cA$}
The 5th order monomial was $abcde$, with period $(1,1,1,1,1)=A$. Acting with the $\star$ map gives $(2,2,2,2,2) = \frac{1}{2}\diff{A}{t}=\tilde{A}$. All we need to do is write the relations for $A$ in terms of $\tilde{A}$, to get the coupled system:
\begin{align}
    \frac{1}{12}\eta(\eta-1)\tilde{A} = t^3E,\cm E = \frac{1}{4}(\eta+1)D,\cm D = \frac{1}{3}(\eta+1)C \\
    C = \frac{1}{2}(\eta+1)B,\cm \frac{1}{2}\eta(\eta-1)B = t^2(\eta+3)\tilde{A}
\end{align}
which leads to the following equation for $\tilde{A}$:
\begin{align}
    \bsq{\eta(\eta-1)(\eta-3)(\eta-4) -t^5(\eta+3)^4}\tilde{A} = 0
\end{align}

\subsubsection*{Representation Class $\cB$}
The $(2,1)$ period was $(0,2,1,1,1)=A$. The corresponding $(1,2)$ period\footnote{We refer to the periods of forms corresponding to 10th order monomials as $(1,2)$--periods, but one should keep in mind that generally these are integrals of classes contained in $H^{3,0}\oplus H^{2,1} \oplus H^{1,2}$, not just $H^{1,2}$.} is $(3,1,2,2,2) = (1,3,2,2,2)= \frac{1}{2}\diff{A}{t}=\tilde{A}$. Again we write the relations for $A$ in terms of $\tilde{A}$, to get the coupled system:
\begin{align}
    \frac{1}{3}\eta\tilde{A} = t^2E,\cm E = tD,\cm D = tC \\
    C = \frac{1}{3}(\eta+2)B,\cm \eta B = t(\eta+3)\tilde{A}
\end{align}
which leads to the following equation for $\tilde{A}$:
\begin{align}
    \bsq{\eta(\eta-4) -t^5(\eta+3)^2}\tilde{A} = 0
\end{align}

\subsubsection*{Representation Class $\cC$}
The $(2,1)$ period was $(3,2,0,0,0)$. The corresponding $(1,2)$ period is $(0,1,3,3,3) = (1,0,3,3,3)$ which appears in the diagram denoted $D$. We can therefore just use the same coupled system as before to solve for $D$:
\begin{align}
    \bsq{\eta(\eta-2) -t^5(\eta+4)^2}D = 0
\end{align}

\subsubsection*{Representation Class $\cD$}
The $(2,1)$ period was $(3,1,1,0,0)$. The corresponding $(1,2)$ period is $(0,2,2,3,3) = (0,3,3,2,2)$ which appears in the diagram as $E$. So as with representation class $\cC$, we just use the same coupled system as found for the $(2,1)$--forms to solve for $E$:
\begin{align}
    \bsq{\eta(\eta-3) -t^5(\eta+3)(\eta+4)}E = 0
\end{align}

\subsubsection*{Representation Class $\cE$}
The $(2,1)$ period was $(2,2,1,0,0)=A$. The corresponding $(1,2)$ period is $(1,1,2,3,3) = (3,3,2,1,1)= \frac{1}{2}\diff{A}{t}=\tilde{A}$. As with representation classes $\cA$ and $\cB$, we write the relations for $A$ in terms of $\tilde{A}$, to get the coupled system:
\begin{align}
    \frac{1}{3}\eta\tilde{A} = t^2E,\cm E = \frac{1}{3}(\eta+1)D,\cm D = tC \\
    C = tB,\cm \eta B = t(\eta+2)\tilde{A}
\end{align}
which leads to the following equation for $\tilde{A}$:
\begin{align}
    \bsq{\eta(\eta-4) -t^5(\eta+2)(\eta+3)}\tilde{A} = 0
\end{align}

\textbox{
\center{\textbf{Summary of Columns of the Period Matrix \\ Corresponding to 10th Order Monomials}}
\begin{align} \nonumber
    \begin{array}{|c|c|c|c|} \hline & \text{\# Classes} & \text{Operator Annihilating Periods} & \text{Hodge Type}
    \\ \hline \cA & 1 & \eta(\eta-1)(\eta-3)(\eta-4) - t^5(\eta+3)^4 & H^{3,0}\oplus H^{2,1} \oplus H^{1,2}
    \\ \cB & 20 & \eta(\eta-4) - t^5(\eta+3)^2 & H^{2,1}  \oplus H^{1,2}
    \\ \cC & 20 & \eta(\eta-2) - t^5(\eta+4)^2 & H^{2,1}  \oplus H^{1,2}
    \\ \cD & 30 & \eta(\eta-3) - t^5(\eta+3)(\eta+4)  & H^{2,1}  \oplus H^{1,2}
    \\ \cE & 30 & \eta(\eta-4) - t^5(\eta+2)(\eta+3) & H^{2,1} \oplus H^{1,2}
    \\ \hline\end{array}
\end{align}
}

\subsubsection*{Periods of the Class Corresponding to 15th Order Monomials}
The space $\bC[a,b,c,d,e]_{15}/J(Q)$ is 1 dimensional, and we can take the single nonzero basis vector to be the monomial $a^3b^3c^3d^3e^3$. This choice allows us to reuse the diagram for $(1,1,1,1,1)$, now defining $A=(3,3,3,3,3) = \diff{^3\Omega}{t^3}\in H^{3,0}\oplus H^{2,1}\oplus H^{1,2}\oplus H^{0,3}$. The coupled system becomes:
\begin{align}
    \frac{1}{4}\eta A = t^2 E,\cm E = \frac{1}{4}(\eta+1)D, \cm D = \frac{1}{3}(\eta+1)C \\
    C = \frac{1}{2}(\eta+1)B, \cm\frac{1}{6}\eta(\eta-1)(\eta-2)B = t^3(\eta+4)A
\end{align}
The resulting equation is:
\begin{align}
    \bsq{\eta(\eta-2)(\eta-3)(\eta-4) - t^5(\eta+4)^4}A = 0
\end{align}

The algorithm we've given for the diagrammatic method is both systematic and powerful. As our discussion of the Fermat pencil has made clear, the key prerequisite is focusing on a family of varieties each of whose members respects a large discrete symmetry group. Earlier, we emphasized that the quintic moduli space has other, less familiar, loci that respect other, less familiar, discrete symmetries. We now extend the diagrammatic technique to these families, focussing for definiteness on the $\bZ_{41}$ case. The results for the $\bZ_{51}$ family are summarized in appendix \ref{Z51calc}.

 \subsection{The $\bZ_{41}$ Quintic: $Q(t)=\frac{1}{5}\ci{a^4b+b^4c+c^4d+d^4e+e^4a} - t abcde = 0$}
The symmetries of this quintic family are the $\bZ_{41}$ scalings generated by $g=(1,37,16,18,10)$ where the entries now indicate the powers of a nontrivial 41st root of unity multiplying each homogeneous coordinate. There is also a $\bZ_5$ group of cyclic permutations of the homogeneous coordinates generated by $\alpha:(a,b,c,d,e)\to(b,c,d,e,a)$, which is intertwined with the scalings by the relation:
\begin{align}
    \alpha g \alpha^{-1} = g^{10}
\end{align}
 As for the Fermat family we have:
\begin{align}
    \int_{\gamma_i} \Omega^j(t) = \int_{T(\gamma_i)} \frac{a^{v_0}b^{v_1}c^{v_2}d^{v_3}e^{v_4}}{Q(t)^{k(v)}}\Omega_0
\end{align}
But the relation that previously was interpreted diagramatically is now:
\begin{align} \label{Z41derivrel}
    \frac{\Omega_0}{Q(t)^{k(v)+1}}A x^i\der{Q(t)}{x^i} &= \frac{1}{k}\frac{\Omega_0}{Q(t)^{k(v)}}A(1+v_i) + \text{exact forms}
    \\  \label{Z41deriva}
    a\der{Q(t)}{a} &= \frac{4}{5}a^4b+ \frac{1}{5}d^4a -  t  abcde \\
\text{with}\hcm b\der{Q(t)}{b} &= \frac{4}{5}b^4c+ \frac{1}{5}a^4b - t  abcde \\
    c\der{Q(t)}{c} &= \frac{4}{5}c^4d+ \frac{1}{5}b^4c - t  abcde \\  \label{Z41derivd}
    d\der{Q(t)}{d} &= \frac{4}{5} d^4a + \frac{1}{5} c^4d - t  abcde
\end{align}
Equations (\ref{Z41deriva}--\ref{Z41derivd}) now have three terms on the right hand side, in contrast with their counterparts in the $(\bZ_5)^3$ case, so these relations cannot be used to construct diagrams as before.  But one can rectify the problem by taking particular linear combinations:

\begin{align}
    \left(\begin{array}{ccccc}256 & 1 & -4 & 16 & -64 \\-64 & 256 & 1 & -4 & 16 \\16 & -64 & 256 & 1 & -4 \\-4 & 16 & -64 & 256 & 1 \\1 & -4 & 16 & -64 & 256\end{array}\right) \left(\begin{array}{c}a\partial_a Q(t) \\b\partial_b Q(t) \\c\partial_c Q(t) \\d\partial_d Q(t) \\e\partial_e Q(t)\end{array}\right) = 205\left(\begin{array}{c}a^4b - tabcde \\b^4c - tabcde \\c^4d - tabcde \\d^4e - tabcde \\e^4a - tabcde\end{array}\right)
\end{align}
Performing the same manipulations on \eqref{Z41derivrel} and integrating gives:
\begin{align}
    (v_0,\ldots,v_i+4,v_{i+1}+1,\ldots,v_4) = \frac{f(i,\vec{v})}{205k(v)}(v_0,v_1,v_2,v_3,v_4)+t(v_0+1,v_1+1,v_2+1,v_3+1,v_4+1)
\end{align}
where $f(i,\vec{v})$ is the $i$'th component of the column vector:
\begin{align}
    \left(\begin{array}{ccccc}256 & 1 & -4 & 16 & -64\\-64 & 256 & 1 & -4 & 16 \\16 & -64 & 256 & 1 & -4 \\-4 & 16 & -64 & 256 & 0 \\ 1 & -4 & 16 & -64 & 256\end{array}\right)\left(\begin{array}{c}v_0+1 \\ v_1+1 \\v_2+1 \\v_3+1 \\v_4+1\end{array}\right)
\end{align}
and again $k(v) = 1+\sum_iv_i$. The above is encoded in the following diagrams:
\begin{align} \label{Z41diag1}
    \left.\begin{array}{ccc}(v_0,v_1,v_2,v_3,v_4) & \longrightarrow & (v_0+1,v_1+1,v_2+1,v_3+1,v_4+1) \\\downarrow D_0 &  &  \\(v_0+4,v_1+1,v_2,v_3,v_4) &  & \end{array}\right. \end{align} \begin{align}
    \left.\begin{array}{ccc}(v_0,v_1,v_2,v_3,v_4) & \longrightarrow & (v_0+1,v_1+1,v_2+1,v_3+1,v_4+1) \\\downarrow D_1 &  &  \\(v_0,v_1+4,v_2+1,v_3,v_4) &  & \end{array}\right. \end{align} \begin{align}
    \left.\begin{array}{ccc}(v_0,v_1,v_2,v_3,v_4) & \longrightarrow & (v_0+1,v_1+1,v_2+1,v_3+1,v_4+1)  \\\downarrow D_2 &  &  \\(v_0,v_1,v_2+4,v_3+1,v_4) &  & \end{array}\right. \end{align} \begin{align}
    \left.\begin{array}{ccc}(v_0,v_1,v_2,v_3,v_4) & \longrightarrow & (v_0+1,v_1+1,v_2+1,v_3+1,v_4+1)  \\\downarrow D_3 &  &  \\(v_0,v_1,v_2,v_3+4,v_4+1) &  & \end{array}\right. \end{align} \begin{align}
    \left.\begin{array}{ccc}(v_0,v_1,v_2,v_3,v_4) & \longrightarrow & (v_0+1,v_1+1,v_2+1,v_3+1,v_4+1)  \\\downarrow D_4 &  &  \\(v_0+1,v_1,v_2,v_3,v_4+4) &  & \end{array}\right.
\end{align}
The algorithm for finding Picard--Fuchs equations is the same as in the Fermat case except that we can no longer use diagrams with $-1$ appearing in any of the entries of $(v_0,v_1,v_2,v_3,v_4)$.

\subsubsection*{Periods of the Holomorphic 3--Form}
\small
\begin{align} \nonumber
    \begin{array}{ccccccccccc}
     &   &  (0,0,0,0,0)_A & \to  &  (1,1,1,1,1)  & \to  & (2,2,2,2,2) & \to  & (3,3,3,3,3) & \to & (4,4,4,4,4) \\
     &   & \downarrow D_0  &   & \downarrow D_0  &   & \downarrow D_0 &   & \downarrow D_0 & & \\
     &   &  (4,1,0,0,0)_B & \to  &  (5,2,1,1,1)  & \to  & (6,3,2,2,2) & \to  & (7,4,3,3,3) &  &    \\
     &   & \downarrow D_1  &   & \downarrow D_1  &   & \downarrow D_1 &   &  & & \\
     &   &  (4,5,1,0,0)_C & \to  &  (5,6,2,1,1)  & \to  & (6,7,3,2,2) &   &  &  &   \\
     &   & \downarrow D_2  &   & \downarrow D_2  &   &  &   &  & & \\
     &   &  (4,5,5,1,0)_D & \to  &  (5,6,6,2,1)  &  &  &   &  &  &     \\
     &   & \downarrow D_3  &   &  &   &  &   &  & &  \\
     (3,4,4,4,0)_E & \to & (4,5,5,5,1) &   &   &   &   &   &     &  &   \\
      \downarrow D_4  &   &  &   &  &   &  & & & & \\
      (4,4,4,4,4) &  &  &   &   &   &  &   &   &  &
     \end{array}
\end{align}
\normalsize
The coupled system is:
\begin{align}
    \frac{1}{6}\eta(\eta-1)(\eta-2)(\eta-3)A = t^4\eta E,\cm \eta E = t(\eta+1)D \\
    D = \frac{1}{3}(\eta+1)C,\cm C=\frac{1}{2}(\eta+1)B, \cm B=(\eta+1)A
\end{align}
which leads to the equation:
\begin{align}
    \bsq{\eta(\eta-1)(\eta-2)(\eta-3)-t^5(\eta+1)^4}A=0
\end{align}
Notice that the equation for the periods of the holomorphic $3$--form is the same as for the Fermat family.  As indicated in the introduction, this fact can be interpreted as a consequence of mirror symmetry. The Greene--Plesser mirror construction \cite{Greene:1990ud}\cite{Greene:1998yz} involves quotienting the manifold at $t=0$ by the group of scaling symmetries that preserve the holomorphic $3$--form, so the differential forms that descend to the mirror are precisely those that transform trivially.  Now recall from figure \ref{mirrorfig} that the quotient of the Fermat \emph{family} is a 1--parameter familiy in mirror moduli space, varying in complex structure with $t$. The same is true for the quotient of the $\bZ_{41}$ family. But the mirror family of quintics in $\bP_4$ has $h_{2,1}=1$ and $h_{1,1}=101$, so the complex structure moduli space is one dimensional. It follows that the quotients of the Fermat and $\bZ_{41}$ families are isomorphic in terms of their complex structure, differing only in their K\"{a}hler structure.  We therefore expect the invariant periods of both loci to obey the same Picard--Fuchs equations.

\subsubsection*{Periods of the Forms Corresponding to 5th Order Monomials}
As before, elements of $\bC[a,b,c,d,e]_5/J(Q)$ correspond to cohomology classes in $\bF^{3,2}=H^{3,0}\oplus H^{2,1}$. In this case it helps to classify the 126 quintic monomials by their transformation under the $\bZ_{41}$ scaling symmetry generated by $g=(1,37,16,18,10)$. We say a monomial $m$ is in representation $n$ if $g(m) = \gamma^nm$ where $\gamma$ is a nontrivial 41st root of unity. The monomials are listed by  representation in table \ref{Z41monomials}.

\begin{table}[h]
\begin{align}\nonumber
    \begin{array}{|c|c|c|c|c|c|c|c|c|c|c|} \hline
    0 & 1 & 2 & 3 & 4 & 5 & 6 & 7 & 8 & 9 & 10 \\ \hline
    a^4b & e^3bc & b^4d & e^3bd & d^3c^2 & a^5 & d^4c & b^3ad & d^5 & e^5 & d^3ab \\
    b^4c & b^2c^2d & c^3d^2 & a^2c^2e & c^3ab & d^3b^2 & e^3ac & a^2d^2e & b^3e^2 & c^3a^2 & a^2b^2c\\
    c^4d &   &   & b^2d^2c & a^2b^2e & b^3ac & c^2abd & c^2e^2b & e^3ad & a^3be & \\
    d^4e &   &   &   &   & a^2cde &   &   & d^2abc & e^2bcd & \\
    e^4a &   &   &   &   &   &   &   &  & & \\
    abcd &   &   &   &   &   &   &   & & &\\ \hline
 11 & 12    & 13    & 14    & 15    & 16    & 17    & 18    & 19    & 20    & 21    \\ \hline
 a^2c^2d    & a^2b^2d   & c^3be & a^4e  & e^4c  & b^3de & e^4d  & c^3ae & c^4b  &    a^4c   & b^5   \\
 d^2e^2b    & c^2e^2a   & a^2d^2c   & b^3ce & d^3a^2    & d^2e^2a   & a^3bd & a^2e^2b   & d^3be & b^3c^2    & e^3c^2    \\
    &   & b^2e^2a   & e^2acd    & a^3bc &   & d^2bce    &   & b^2ace    & c^2ade    & c^3bd \\
    &   &   &   & c^2bde    &   &   &   &   &   & b^2ade    \\ \hline
 22 & 23    & 24    & 25    & 26    & 27    & 28    & 29    & 30    & 31    & 32    \\ \hline
 a^4d   & a^3e^2    & c^4a  & e^3d^2    & b^4a  & d^4b  & b^2e^2c   & a^3ce & d^3ac & b^3a^2    & d^4a  \\
 e^3b^2 & e^3cd & b^3d^2    & d^3bc & c^3ad & c^3e^2    & c^2d^2a   & b^2d^2a   & a^2c^2b   & a^3de & e^2a^2    \\
b^3cd   & c^2d^2b   & d^3ae & b^2c^2a   & a^2bde    & e^3ab &   & c^2e^2d   & b^2e^2d   & d^2e^2c   & a^2bcd    \\
d^2ace  &   & a^2bce    &   &   & b^2acd    &   &   &   &   &   \\ \hline
 33 & 34    & 35    & 36    & 37    & 38    & 39    & 40    &   &   &   \\ \hline
 c^4e   & a^2d^2b   & b^4e  & e^4b  & a^3cd & a^2e^2c   & c^5   & c^3b^2    &   &   &   \\
d^3e^2  & b^2c^2e   & a^3c^2    & a^3b^2    & c^2d^2e   & b^2d^2e   & a^3d^2    & b^3ac &   &   &   \\
 e^2abc &   & c^3de & b^2cde    &   &   & d^3ce & a^2e^2d   &   &   &   \\
    &   & e^2abd    &   &   &   & c^2abe    &   &   &   &   \\ \hline
    \end{array}
\end{align}
\caption{\label{Z41monomials} The 126 quintic monomials according to their transformation under $\bZ_{41}$. }
\end{table}
The decomposition into representations of $\bZ_{41}$ must be compatible with the permutation symmetry in the sense that cyclic permutation of all the monomials in a given representation should give the monomials in some other representation. We therefore group the representations by their behavior under cyclic permutations in table \ref{Z41cyclicreps}.
\begin{table}[h]
\begin{align}\nonumber
    \begin{array}{|c|c|} \hline \text{Representation Class} & \bZ_{41} \text{ Representations Contained Therein} \\ \hline \cA & 0 \\ \hline \cB1 & 5,8,9,21,39 \\\cB2 & 1,10,16,18,37 \\\cB3 & 2,20,32,33,36 \\\cB4 & 4,23,25,31,40 \\ \hline \cC1 & 15,22,24,27,35 \\\cC2 & 3,7,13,29,30 \\\cC3 & 6,14,17,19,26 \\\cC4 & 11,12,28,34,38 \\ \hline \end{array}
\end{align}
\caption{\label{Z41cyclicreps} The representations of a given row of the right column transform into each other under cyclic permutations of the homogeneous coordinates.}
\end{table}
We now need to take account of the quotient of $\bC[a,b,c,d,e]_5$ by $J(Q)$ to find out how many forms are associated with each representation class. The results are summarized in table \ref{Z41monomialRelations}.
\begin{table}[h]
\begin{align}\nonumber
    \begin{array}{|c|c|c|c|}\hline
     \text{Representation} & & \text{\# Forms per} & \\
     \text{Class} & \text{Relations}  & \text{Representation} & \text{Hodge Type} \\
     \hline \cA & a^4b\sim b^4c\sim c^4d\sim d^4e \sim e^4a \sim abcd  & 1 &H^{3,0}\oplus H^{2,1} \\
    \hline \cB1 & a\partial_bQ = \frac{4}{5}b^3ac + \frac{1}{5}a^5 - ta^2cde  & 3 & H^{2,1} \\
    \cB2 & \text{No relations} & 2 &H^{2,1} \\
    \cB3 & d\partial_cQ = \frac{4}{5}c^3d^2 + \frac{1}{5}b^4d - td^2abe & 2 & H^{2,1}\\
    \cB4 & \text{No relations}  & 3 &H^{2,1} \\
    \hline \cC1 & c\partial_a Q = \frac{4}{5}a^3be+\frac{1}{5}e^4c-tc^2bde  & 3 &H^{2,1}\\
    \cC2 & \text{No relations}  & 3 &H^{2,1} \\
    \cC3 & c\partial_e Q = \frac{4}{5}e^3ac+\frac{1}{5}d^4c-tc^2abd & 2 &H^{2,1} \\
    \cC4 & \text{No relations}  & 2&H^{2,1} \\
    \hline \end{array}
\end{align}
\caption{\label{Z41monomialRelations} The 25 relations among the 126 quintic monomials, considered as elements of $\bC[a,b,c,d,e]_5/J(Q)$. }
\end{table}
In total there are $1+ 5(3+2+2+3+3+3+2+2)=101$ independent monomials as expected.

\subsubsection*{Representation Class $\cA$:  $\cu{abcde}$}

\small
\begin{align} \nonumber
    \begin{array}{ccccccccccc}
    &   &   &   & (1,1,1,1,1)_A &\to    & (2,2,2,2,2)   & \to   & (3,3,3,3,3)   & \to   &  (4,4,4,4,4) \\
    &   &   &   & \downarrow D_0    &   & \downarrow D_0    &   & \downarrow D_0    &   &  \\
    &   & (4,1,0,0,0)_B & \to   & (5,2,1,1,1)   & \to   & (6,3,2,2,2)   & \to   & (7,4,3,3,3)   &   &     \\
    &   &  \downarrow D_1   &   & \downarrow D_1    &   &  \downarrow D_1   &   &   &   & \\
    &   & (4,5,1,0,0)_C & \to   & (5,6,2,1,1)   & \to   & (6,7,3,2,2)   &   &   &   & \\
    &   & \downarrow D_2    &   &  \downarrow D_2   &   &   &   &   &   & \\
    &   & (4,5,5,1,0)_D & \to   & (5,6,6,2,1)   &   &   &   &   &   & \\
    &   &  \downarrow D_3   &   &   &   &   &   &   &   & \\
 (3,4,4,4,0)_E  & \to   & (4,5,5,5,1)   &   &   &   &   &   &   &   & \\
 \downarrow D_4 &   &   &   &   &   &   &   &   &   & \\
 (4,4,4,4,4)    &   &   &   &   &   &   &   &   &   &
    \end{array}
\end{align}
\normalsize
The coupled system is:
\begin{align}
    \frac{1}{6}\eta(\eta-1)(\eta-2) A =t^3\eta E&, \cm \eta E = t(\eta+1)D, \cm D = \frac{1}{3}(\eta+1)C \\
    C &= \frac{1}{2}(\eta+1)B, \cm \eta B = t(\eta+2) A
\end{align}
from which one can find the following equation for the period $A$:
\begin{align}
    \sq{\eta(\eta-1)(\eta-2)(\eta-4) -t^5 (\eta+2)^4}     A &= 0
\end{align}

\subsubsection*{Representation Class $\cB1$: 3 of $\cu{a^5,d^3b^2,b^3ac,a^2cde}$}
In anticipation of using the $\star$ map to find the $(1,2)$--periods, it will be helpful to find a diagram without $a^5$:

\small
\begin{align} \nonumber
    \begin{array}{ccccccccccc}
    &   &   &   &   &   &   &   & (0,2,0,3,0)_A & \to   &  (1,3,1,4,1) \\
    &   &   &   &   &   &   &   & \downarrow D_4    &   & \downarrow D_4 \\
    &   &   &   &   &   &   &   & (1,2,0,3,4)_B & \to   &  (2,3,1,4,5) \\
    &   &   &   &   &   &   &   & \downarrow D_2    &   & \\
    &   &   &   &   &   & (0,1,3,3,3)_C & \to   & (1,2,4,4,4)   &   & \\
    &   &   &   &   &   & \downarrow D_0    &   &   &   & \\
    &   & (2,0,1,1,1)_D & \to   & (3,1,2,2,2)   & \to   & (4,2,3,3,3)   &   &   &   & \\
    &   & \downarrow D_1    &   &   &   &   &   &   &   & \\
 (1,3,1,0,0)_E  & \to   & (2,4,2,1,1)   &   &   &   &   &   &   &   & \\
 \downarrow D_3 &   &   &   &   &   &   &   &   &   & \\
 (1,3,1,4,1)    &   &   &   &   &   &   &   &   &   &
    \end{array}
\end{align}
\normalsize
The coupled system is:
\begin{align}
    \eta A =t\ci{\eta+\frac{37}{41}}E&, \cm \eta E = t\ci{\eta+\frac{18}{41}}D, \cm \frac{1}{2}\eta(\eta-1) D = t^2\ci{\eta+\frac{10}{41}}C \\
    \eta C &= t\ci{\eta+\frac{16}{41}}B, \cm B = \frac{1}{2}\ci{\eta+\frac{1}{41}}A
\end{align}
which results in the following equations for the 3 independent periods:
\begin{align} \nonumber
    \sq{\eta(\eta-1)(\eta-2)(\eta-3)(\eta-4) -t^5 \ci{\eta+\frac{201}{41}}\ci{\eta+\frac{141}{41}}\ci{\eta+\frac{51}{41}}\ci{\eta+\frac{16}{41}}\ci{\eta+\frac{1}{41}}}     A &= 0 \\ \nonumber
    \sq{\eta(\eta-1)(\eta-2)(\eta-3)(\eta-4) -t^5 \ci{\eta+\frac{133}{41}}\ci{\eta+\frac{98}{41}}\ci{\eta+\frac{83}{41}}\ci{\eta+\frac{78}{41}}\ci{\eta+\frac{18}{41}}}     D &= 0 \\
    \sq{\eta(\eta-1)(\eta-2)(\eta-3)(\eta-4) -t^5 \ci{\eta+\frac{182}{41}}\ci{\eta+\frac{92}{41}}\ci{\eta+\frac{57}{41}}\ci{\eta+\frac{42}{41}}\ci{\eta+\frac{37}{41}}}     E &= 0
\end{align}

\subsubsection*{Representation Class $\cB2$:  $\cu{e^3bc,b^2c^2d}$}
\small
\begin{align} \nonumber
    \begin{array}{ccccccccccc}
    &   &   &   &   &   &   &   & (0,1,1,0,3)_A & \to   &  (1,2,2,1,4) \\
    &   &   &   &   &   &   &   & \downarrow D_3    &   & \downarrow D_3 \\
    &   &   &   &   &   &   &   & (0,1,1,4,4)_B & \to   &  (1,2,2,5,5) \\
    &   &   &   &   &   &   &   & \downarrow D_0    &   & \\
    &   &   &   &   &   & (3,1,0,3,3)_C & \to   & (4,2,1,4,4)   &   & \\
    &   &   &   &   &   & \downarrow D_2    &   &   &   & \\
    &   &   &   & (2,0,3,3,2)_D & \to   & (3,1,4,4,3)   &   &   &   & \\
    &   &   &   & \downarrow D_1    &   &   &   &   &   & \\
 (0,2,2,1,0)_E  & \to   & (1,3,3,2,1)   & \to   & (2,4,4,3,2)   &   &   &   &   &   & \\
 \downarrow D_4 &   &   &   &   &   &   &   &   &   & \\
 (1,2,2,1,4)    &   &   &   &   &   &   &   &   &   &
    \end{array}
\end{align}
\normalsize
The coupled system is:
\begin{align}
    \eta A =t\ci{\eta+\frac{33}{41}}E&, \cm \frac{1}{2}\eta(\eta-1)E = t^2\ci{\eta+\frac{20}{41}}D, \cm \eta D = t\ci{\eta+\frac{36}{41}}C \\
    \eta C &= t\ci{\eta+\frac{2}{41}}B, \cm B = \frac{1}{2}\ci{\eta+\frac{32}{41}}A
\end{align}
The equations for the 2 independent periods are:
\begin{align} \nonumber
    \sq{\eta(\eta-1)(\eta-2)(\eta-3)(\eta-4) -t^5 \ci{\eta+\frac{197}{41}}\ci{\eta+\frac{102}{41}}\ci{\eta+\frac{77}{41}}\ci{\eta+\frac{32}{41}}\ci{\eta+\frac{2}{41}}}     A &= 0 \\
    \sq{\eta(\eta-1)(\eta-2)(\eta-3)(\eta-4) -t^5 \ci{\eta+\frac{143}{41}}\ci{\eta+\frac{118}{41}}\ci{\eta+\frac{73}{41}}\ci{\eta+\frac{43}{41}}\ci{\eta+\frac{33}{41}}}     E &= 0
\end{align}

\subsubsection*{Representation Class $\cB3$:  2 of $\cu{b^4d,c^3d^2,d^2abe}$}
\small
\begin{align} \nonumber
    \begin{array}{ccccccccccc}
    &   &   &   &   &   &   &   & (0,4,0,1,0)_A & \to   &  (1,5,1,2,1) \\
    &   &   &   &   &   &   &   & \downarrow D_4    &   & \downarrow D_4 \\
    &   &   &   &   &   &   &   & (1,4,0,1,4)_B & \to   &  (2,5,1,2,5) \\
    &   &   &   &   &   &   &   & \downarrow D_2    &   & \\
    &   &   &   &   &   & (0,3,3,1,3)_C & \to   & (1,4,4,2,4)   &   & \\
    &   &   &   &   &   & \downarrow D_0    &   &   &   & \\
    &   &   &   & (3,3,2,0,2)_D & \to   & (4,4,3,1,3)   &   &   &   & \\
    &   &   &   & \downarrow D_3    &   &   &   &   &   & \\
 (1,1,0,2,1)_E  & \to   & (2,2,1,3,2)   & \to   & (3,3,2,4,3)   &   &   &   &   &   & \\
 \downarrow D_1 &   &   &   &   &   &   &   &   &   & \\
 (1,5,1,2,1)    &   &   &   &   &   &   &   &   &   &
    \end{array}
\end{align}
\normalsize
The coupled system is:
\begin{align}
    \eta A =t\ci{\eta+\frac{81}{41}}E&, \cm  \frac{1}{2}\eta(\eta-1)E = t^2\ci{\eta+\frac{23}{41}}D, \cm \eta D = t\ci{\eta+\frac{4}{41}}C \\
    \eta C &= t\ci{\eta-\frac{10}{41}}B, \cm B = \frac{1}{2}\ci{\eta+\frac{25}{41}}A
\end{align}
The equations for the 2 independent periods are:
\begin{align} \nonumber
    \sq{\eta(\eta-1)(\eta-2)(\eta-3)(\eta-4) -t^5 \ci{\eta+\frac{245}{41}}\ci{\eta+\frac{105}{41}}\ci{\eta+\frac{45}{41}}\ci{\eta+\frac{25}{41}}\ci{\eta-\frac{10}{41}}}     A &= 0 \\
    \sq{\eta(\eta-1)(\eta-2)(\eta-3)(\eta-4) -t^5 \ci{\eta+\frac{146}{41}}\ci{\eta+\frac{86}{41}}\ci{\eta+\frac{81}{41}}\ci{\eta+\frac{66}{41}}\ci{\eta+\frac{31}{41}}}     E &= 0
\end{align}

\subsubsection*{Representation Class $\cB4$: $\cu{d^3c^2,c^3ab,a^2b^2e}$}
\small
\begin{align} \nonumber
    \begin{array}{ccccccccccc}
    &   &   &   &   &   &   &   & (0,0,2,3,0)_A & \to   &  (1,1,3,4,1) \\
    &   &   &   &   &   &   &   & \downarrow D_4    &   & \downarrow D_4 \\
    &   &   &   &   &   &   &   & (1,0,2,3,4)_B & \to   &  (2,1,3,4,5) \\
    &   &   &   &   &   &   &   & \downarrow D_1    &   & \\
    &   &   &   &   &   & (0,3,2,2,3)_C & \to   & (1,4,3,3,4)   &   & \\
    &   &   &   &   &   & \downarrow D_0    &   &   &   & \\
    &   & (2,2,0,0,1)_D & \to   & (3,3,1,1,2)   & \to   & (4,4,2,2,3)   &   &   &   & \\
    &   & \downarrow D_2    &   &   &   &   &   &   &   & \\
 (1,1,3,0,0)_E  & \to   & (2,2,4,1,1)   &   &   &   &   &   &   &   & \\
 \downarrow D_3 &   &   &   &   &   &   &   &   &   & \\
 (1,1,3,4,1)    &   &   &   &   &   &   &   &   &   &
    \end{array}
\end{align}
\normalsize
The coupled system is:
\begin{align}
    \eta A =t\ci{\eta+\frac{5}{41}}E&, \cm \eta E = t\ci{\eta+\frac{21}{41}}D, \cm \frac{1}{2}\eta(\eta-1) D = t^2\ci{\eta+\frac{8}{41}}C \\
    \eta C &= t\ci{\eta+\frac{39}{41}}B, \cm B = \frac{1}{2}\ci{\eta+\frac{9}{41}}A
\end{align}
which results in the following equations for the 3 independent periods:
\begin{align} \nonumber
    \sq{\eta(\eta-1)(\eta-2)(\eta-3)(\eta-4) -t^5 \ci{\eta+\frac{169}{41}}\ci{\eta+\frac{144}{41}}\ci{\eta+\frac{49}{41}}\ci{\eta+\frac{39}{41}}\ci{\eta+\frac{9}{41}}}     A &= 0 \\ \nonumber
    \sq{\eta(\eta-1)(\eta-2)(\eta-3)(\eta-4) -t^5 \ci{\eta+\frac{131}{41}}\ci{\eta+\frac{121}{41}}\ci{\eta+\frac{91}{41}}\ci{\eta+\frac{46}{41}}\ci{\eta+\frac{21}{41}}}     D &= 0 \\
    \sq{\eta(\eta-1)(\eta-2)(\eta-3)(\eta-4) -t^5 \ci{\eta+\frac{185}{41}}\ci{\eta+\frac{90}{41}}\ci{\eta+\frac{80}{41}}\ci{\eta+\frac{50}{41}}\ci{\eta+\frac{5}{41}}}     E &= 0
\end{align}

\subsubsection*{Representation Class $\cC1$:  3 of $\cu{e^4c,d^3a^2,a^3bc,c^2bde}$}
Here it will be convenient to leave out the monomial $e^4c$.
\small
\begin{align} \nonumber
    \begin{array}{ccccccccccc}
    &   &   &   &   &   &   &   & (3,1,1,0,0)_A & \to   &  (4,2,2,1,1) \\
    &   &   &   &   &   &   &   & \downarrow D_3    &   & \downarrow D_3 \\
    &   &   &   &   &   & (2,0,0,3,0)_B & \to   & (3,1,1,4,1)   & \to   &  (4,2,2,5,2) \\
    &   &   &   &   &   & \downarrow D_4    &   & \downarrow D_4    &   & \\
    &   &   &   &   &   & (3,0,0,3,4)_C & \to   & (4,1,1,4,5)   &   & \\
    &   &   &   &   &   & \downarrow D_1    &   &   &   & \\
    &   &   &   & (2,3,0,2,3)_D & \to   & (3,4,1,3,4)   &   &   &   & \\
    &   &   &   & \downarrow D_2    &   &   &   &   &   & \\
 (0,1,2,1,1)_E  & \to   & (1,2,3,2,2)   & \to   & (2,3,4,3,3)   &   &   &   &   &   & \\
 \downarrow D_0 &   &   &   &   &   &   &   &   &   & \\
 (4,2,2,1,1)    &   &   &   &   &   &   &   &   &   &
    \end{array}
\end{align}
\normalsize
The coupled system is:
\begin{align}
    \eta A =t\ci{\eta+\frac{30}{41}}E&, \cm \frac{1}{2}\eta(\eta-1) E = t^2\ci{\eta+\frac{7}{41}}D, \cm \eta D = t\ci{\eta+\frac{13}{41}}C \\
    C &= \frac{1}{2}\ci{\eta+\frac{3}{41}}B, \cm \eta B = t\ci{\eta+\frac{29}{41}}A
\end{align}
The equations for the 3 independent periods are:
\begin{align} \nonumber
    \sq{\eta(\eta-1)(\eta-2)(\eta-3)(\eta-4) -t^5 \ci{\eta+\frac{194}{41}}\ci{\eta+\frac{89}{41}}\ci{\eta+\frac{54}{41}}\ci{\eta+\frac{44}{41}}\ci{\eta+\frac{29}{41}}}     A &= 0 \\ \nonumber
    \sq{\eta(\eta-1)(\eta-2)(\eta-3)(\eta-4) -t^5 \ci{\eta+\frac{193}{41}}\ci{\eta+\frac{153}{41}}\ci{\eta+\frac{48}{41}}\ci{\eta+\frac{13}{41}}\ci{\eta+\frac{3}{41}}}     B &= 0 \\
    \sq{\eta(\eta-1)(\eta-2)(\eta-3)(\eta-4) -t^5 \ci{\eta+\frac{130}{41}}\ci{\eta+\frac{95}{41}}\ci{\eta+\frac{85}{41}}\ci{\eta+\frac{70}{41}}\ci{\eta+\frac{30}{41}}}     E &= 0
\end{align}

\subsubsection*{Representation Class $\cC2$:  $\cu{e^3bd,a^2c^2e,b^2d^2c}$}
\small
\begin{align} \nonumber
    \begin{array}{ccccccccccc}
    &   &   &   &   &   &   &   & (0,1,0,1,3)_A & \to   &  (1,2,1,2,4) \\
    &   &   &   &   &   &   &   & \downarrow D_0    &   & \downarrow D_0 \\
    &   &   &   &   &   &   &   & (4,2,0,1,3)_B & \to   &  (5,3,1,2,4) \\
    &   &   &   &   &   &   &   & \downarrow D_2    &   & \\
    &   &   &   & (2,0,2,0,1)_C & \to   & (3,1,3,1,2)   & \to   & (4,2,4,2,3)   &   & \\
    &   &   &   & \downarrow D_1    &   & \downarrow D_1    &   &   &   & \\
    &   &   &   & (2,4,3,0,1)_D & \to   & (3,5,4,1,2)   &   &   &   & \\
    &   &   &   & \downarrow D_3    &   &   &   &   &   & \\
 (0,2,1,2,0)_E  & \to   & (1,3,2,3,1)   & \to   & (2,4,3,4,2)   &   &   &   &   &   & \\
 \downarrow D_4 &   &   &   &   &   &   &   &   &   & \\
 (1,2,1,2,4)    &   &   &   &   &   &   &   &   &   &
    \end{array}
\end{align}
\normalsize
The coupled system is:
\begin{align}
    \eta A =t\ci{\eta+\frac{17}{41}}E&, \cm \frac{1}{2}\eta(\eta-1)E = t^2\ci{\eta+\frac{14}{41}}D, \cm  D = \frac{1}{2}\ci{\eta+\frac{19}{41}}C \\
    \frac{1}{2}\eta(\eta-1) C &= t^2\ci{\eta+\frac{26}{41}}B, \cm B = \frac{1}{2}\ci{\eta+\frac{6}{41}}A
\end{align}
The equations for the 3 independent periods are:
\begin{align} \nonumber
    \sq{\eta(\eta-1)(\eta-2)(\eta-3)(\eta-4) -t^5 \ci{\eta+\frac{181}{41}}\ci{\eta+\frac{101}{41}}\ci{\eta+\frac{96}{41}}\ci{\eta+\frac{26}{41}}\ci{\eta-\frac{6}{41}}}     A &= 0 \\ \nonumber
    \sq{\eta(\eta-1)(\eta-2)(\eta-3)(\eta-4) -t^5 \ci{\eta+\frac{149}{41}}\ci{\eta+\frac{129}{41}}\ci{\eta+\frac{99}{41}}\ci{\eta+\frac{19}{41}}\ci{\eta+\frac{14}{41}}}     C &= 0 \\
    \sq{\eta(\eta-1)(\eta-2)(\eta-3)(\eta-4) -t^5 \ci{\eta+\frac{142}{41}}\ci{\eta+\frac{137}{41}}\ci{\eta+\frac{67}{41}}\ci{\eta+\frac{47}{41}}\ci{\eta+\frac{17}{41}}}     E &= 0
\end{align}

\subsubsection*{Representation Class $\cC3$:  2 of $\cu{d^4c,e^3ac,c^2abd}$}
\small
\begin{align} \nonumber
    \begin{array}{ccccccccccc}
    &   &   &   &   &   &   &   & (1,0,1,0,3)_A & \to   &  (2,1,2,1,4) \\
    &   &   &   &   &   &   &   & \downarrow D_1    &   & \downarrow D_1 \\
    &   &   &   &   &   &   &   & (1,4,2,0,3)_B & \to   &  (2,5,3,1,4) \\
    &   &   &   &   &   &   &   & \downarrow D_3    &   & \\
    &   &   &   &   &   & (0,3,1,3,3)_C & \to   & (1,4,2,4,4)   &   & \\
    &   &   &   &   &   & \downarrow D_0    &   &   &   & \\
    &   &   &   & (3,3,0,2,2)_D & \to   & (4,4,1,3,3)   &   &   &   & \\
    &   &   &   & \downarrow D_2    &   &   &   &   &   & \\
 (1,1,2,1,0)_E  & \to   & (2,2,3,2,1)   & \to   & (3,3,4,3,2)   &   &   &   &   &   & \\
 \downarrow D_4 &   &   &   &   &   &   &   &   &   & \\
 (2,1,2,1,4)    &   &   &   &   &   &   &   &   &   &
    \end{array}
\end{align}
\normalsize
The coupled system is:
\begin{align}
    \eta A =t\ci{\eta+\frac{34}{41}}E&, \cm \frac{1}{2}\eta(\eta-1)E = t^2\ci{\eta+\frac{11}{41}}D, \cm \eta D = t\ci{\eta+\frac{12}{41}}C \\
    \eta C &= t\ci{\eta+\frac{28}{41}}B, \cm B = \frac{1}{2}\ci{\eta+\frac{38}{41}}A
\end{align}
The equations for the 2 independent periods are:
\begin{align} \nonumber
    \sq{\eta(\eta-1)(\eta-2)(\eta-3)(\eta-4) -t^5 \ci{\eta+\frac{198}{41}}\ci{\eta+\frac{93}{41}}\ci{\eta+\frac{53}{41}}\ci{\eta+\frac{38}{41}}\ci{\eta+\frac{28}{41}}}     A &= 0 \\
    \sq{\eta(\eta-1)(\eta-2)(\eta-3)(\eta-4) -t^5 \ci{\eta+\frac{134}{41}}\ci{\eta+\frac{94}{41}}\ci{\eta+\frac{79}{41}}\ci{\eta+\frac{69}{41}}\ci{\eta+\frac{34}{41}}}     E &= 0
\end{align}

\subsubsection*{Representation Class $\cC4$:  $\cu{a^2c^2d,d^2e^2b}$}
\small
\begin{align} \nonumber
    \begin{array}{ccccccccc}
            &   &   &   &   &   & (2,0,2,1,0)_A & \to   &  (3,1,3,2,1) \\
            &   &   &   &   &   & \downarrow D_1    &   & \downarrow D_1 \\
            &   &   &   &   &   & (2,4,3,1,0)_B & \to   &  (3,5,4,2,1) \\
            &   &   &   &   &   & \downarrow D_4    &   & \\
            &   &   &   & (2,3,2,0,3)_C & \to   & (3,4,3,1,4)   &   & \\
            &   &   &   & \downarrow D_3    &   &   &   & \\
         (0,1,0,2,2)_D  & \to   & (1,2,1,3,3)   & \to   & (2,3,2,4,4)   &   &   &   & \\
         \downarrow D_2 &   & \downarrow D_2    &   &   &   &   &   & \\
         (0,1,4,3,2)_E  & \to   & (1,2,5,4,3)   &   &   &   &   &   & \\
         \downarrow D_0 &   &   &   &   &   &   &   & \\
         (4,2,4,3,2)    &   &   &   &   &   &   &   &
    \end{array}
\end{align}
\normalsize
The coupled system is:
\begin{align}
    \frac{1}{2}\eta(\eta-1) A =t^2\ci{\eta+\frac{22}{41}}E&, \cm E = \frac{1}{2}\ci{\eta+\frac{27}{41}}D, \cm \frac{1}{2}\eta(\eta-1) D = t^2\ci{\eta+\frac{24}{41}}C \\
    \eta C &= t\ci{\eta+\frac{35}{41}}B, \cm B = \frac{1}{2}\ci{\eta+\frac{15}{41}}A
\end{align}
The equations for the 2 independent periods are:
\begin{align} \nonumber
    \sq{\eta(\eta-1)(\eta-2)(\eta-3)(\eta-4) -t^5 \ci{\eta+\frac{150}{41}}\ci{\eta+\frac{145}{41}}\ci{\eta+\frac{65}{41}}\ci{\eta+\frac{35}{41}}\ci{\eta+\frac{15}{41}}}     A &= 0 \\
    \sq{\eta(\eta-1)(\eta-2)(\eta-3)(\eta-4) -t^5 \ci{\eta+\frac{147}{41}}\ci{\eta+\frac{117}{41}}\ci{\eta+\frac{97}{41}}\ci{\eta+\frac{27}{41}}\ci{\eta+\frac{22}{41}}}     D &= 0
\end{align}

In summary, the classes corresponding to representations in $\cB1,\ldots,\cB4,\cC1,\ldots,\cC4$ are of pure Hodge type $H^{2,1}$, and their periods all obey 5th order generalized hypergeometric equations. The periods of the single class in representation $\cA$ (the trivial representation) are of mixed Hodge type, and obey a 4th order generalized hypergeometric equation --- the same as that obeyed by the corresponding periods in the Fermat family.

\subsubsection*{Periods of the Forms Corresponding to 10th Order Monomials}
As for the Fermat quintic, we would like to use the $\star$ map to generate a basis of $\bC[a,b,c,d,e]_{10}/J(Q)$. This requires that we find a basis of $\bC[a,b,c,d,e]_5/J(Q)$ with no monomials containing 4th or 5th powers of any coordinate. One can see by examining the relations in table \ref{Z41monomialRelations} as well as the monomial content of the representation classes,  that such a basis can be found for the $\bZ_{41}$ quintic.

As with the Fermat quintic, $\star$ acts nicely on the representations of the discrete symmetry group. If $[m]$ is in representation $n$ of $\bZ_{41}$, then $\star[m]$ is in representation $41-n$. Here though the $\star$ map acts in a nontrivial way on the representation classes:
\begin{align} \label{Z41star1}
    \star \cA = &\cA,\hcm \star \cB1 = \cB3,\hcm \star \cB2 = \cB4,\hcm \star \cB3 = \cB1,\hcm \star \cB4 = \cB2 \\
    &\star \cC1 = \cC3,\hcm \star \cC2 = \cC4,\hcm \star \cC3 = \cC1,\hcm \star \cC4 = \cC2
\end{align}
For example, the 10th order monomials found by acting with $\star$ on the 5th order monomials of $\cB1$ appear in the diagram for $\cB3$ up to cyclic permutations.

\subsubsection*{Representation Class $\cA$}
The 5th order monomial in class $\cA$ is $abcde$, which maps to $a^2b^2c^2d^2e^2$ under $\star$. We can therefore reuse the diagram \eqref{Z41RepclassA}, but solve for $\diff{A}{t}\propto (2,2,2,2,2)$ rather than $A=(1,1,1,1,1)$. The result is:
\begin{align}
    \bsq{\eta(\eta-1)(\eta-3)(\eta-4) - t^5(\eta+3)^4 }\diff{A}{t} = 0
\end{align}

\subsubsection*{Representation Class $\cB1$}
From \eqref{Z41star1} we see that we should look at the 5th order monomials in $\cB3$: $\cu{c^3d^2, d^2abe}$. Acting with $\star$ gives $\cu{a^3b^3de^3,a^2b^2c^3de^2}$, corresponding to periods $(3,3,0,1,3)$ and $(2,2,3,1,2)$. These are cyclic permutations of (and hence equal to) $(3,1,2,2,2)$ and $(0,1,3,3,3)$ which appear in the $\cB1$ diagram as $C$ and $\frac{1}{2}\diff{D}{t}$ respectively. So all we need to do is write the coupled system in terms of $\diff{D}{t}$ rather than $D$ and then solve for $C$ and $\diff{D}{t}$. The coupled system is:
\begin{align}
    \eta A =t\ci{\eta+\frac{37}{41}}E&, \cm \eta(\eta-1) E = t^2\ci{\eta+\frac{59}{41}}\diff{D}{t}, \cm \frac{1}{2}\eta\diff{D}{t} = t\ci{\eta+\frac{10}{41}}C \\
    \eta C &= t\ci{\eta+\frac{16}{41}}B, \cm B = \frac{1}{2}\ci{\eta+\frac{1}{41}}A
\end{align}
The equations for the 2 independent periods are:
\begin{align} \nonumber
    \sq{\eta(\eta-1)(\eta-2)(\eta-3)(\eta-4) -t^5 \ci{\eta+\frac{174}{41}}\ci{\eta+\frac{139}{41}}\ci{\eta+\frac{124}{41}}\ci{\eta+\frac{119}{41}}\ci{\eta+\frac{59}{41}}}     &\diff{D}{t} = 0 \\ \nonumber
    \sq{\eta(\eta-1)(\eta-2)(\eta-3)(\eta-4) -t^5 \ci{\eta+\frac{180}{41}}\ci{\eta+\frac{165}{41}}\ci{\eta+\frac{160}{41}}\ci{\eta+\frac{100}{41}}\ci{\eta+\frac{10}{41}}}     &C = 0
\end{align}

\subsubsection*{Representation Class $\cB2$}
The 5th order monomials in $\cB4$ are $\cu{d^3c^2, c^3ab,a^2b^2e}$. Acting with $\star$ gives $\cu{a^3b^3ce^3,a^2b^2d^3e^3,abc^3d^3e^2}$, corresponding to periods $(3,3,1,0,3)$, $(2,2,0,3,3)$ and $(1,1,3,3,2)$. These are cyclic permutations of $(3,1,0,3,3)$, $(2,0,3,3,2)$ and $(1,3,3,2,1)$ which appear in the $\cB2$ diagram as $C$, $D$ and $\frac{1}{2}\diff{E}{t}$ respectively. We therefore write the coupled system in terms of $\diff{E}{t}$ rather than $E$ and then solve for $C$, $D$ and $\diff{E}{t}$. The coupled system is:
\begin{align}
    \eta(\eta-1) A = t^2\ci{\eta+\frac{74}{41}}\diff{E}{t}&, \cm \frac{1}{2}\eta\diff{E}{t} = t\ci{\eta+\frac{20}{41}}D, \cm \eta D = t\ci{\eta+\frac{36}{41}}C \\
    \eta C &= t\ci{\eta+\frac{2}{41}}B, \cm B = \frac{1}{2}\ci{\eta+\frac{32}{41}}A
\end{align}
The equations for the 3 independent periods are:
\begin{align} \nonumber
    \sq{\eta(\eta-1)(\eta-2)(\eta-3)(\eta-4) -t^5 \ci{\eta+\frac{184}{41}}\ci{\eta+\frac{159}{41}}\ci{\eta+\frac{114}{41}}\ci{\eta+\frac{84}{41}}\ci{\eta+\frac{74}{41}}}     &\diff{E}{t} = 0 \\ \nonumber
    \sq{\eta(\eta-1)(\eta-2)(\eta-3)(\eta-4) -t^5 \ci{\eta+\frac{196}{41}}\ci{\eta+\frac{166}{41}}\ci{\eta+\frac{156}{41}}\ci{\eta+\frac{61}{41}}\ci{\eta+\frac{36}{41}}}     &C = 0 \\ \nonumber
    \sq{\eta(\eta-1)(\eta-2)(\eta-3)(\eta-4) -t^5 \ci{\eta+\frac{200}{41}}\ci{\eta+\frac{155}{41}}\ci{\eta+\frac{125}{41}}\ci{\eta+\frac{115}{41}}\ci{\eta+\frac{20}{41}}}     &D = 0
\end{align}

\subsubsection*{Representation Class $\cB3$}
The independent 5th order monomials in $\cB1$ can be chosen to be $\cu{d^3b^2, b^3ac,a^2cde}$. Acting with $\star$ gives $\cu{a^3bc^3e^3,a^2c^2d^3e^3,ab^3c^2d^2e^2}$, corresponding to periods $(3,1,3,0,3)$, $(2,0,2,3,3)$ and $(1,3,2,2,2)$. These are cyclic permutations of $(0,3,3,1,3)$, $(3,3,2,0,2)$ and $(2,2,1,3,2)$ which appear in the $\cB3$ diagram as $C$, $D$ and $\frac{1}{2}\diff{E}{t}$ respectively. We therefore write the coupled system in terms of $\diff{E}{t}$ rather than $E$ and then solve for $C$, $D$ and $\diff{E}{t}$. The coupled system is:
\begin{align}
    \eta A =t\ci{\eta+\frac{81}{41}}\diff{E}{t}&, \cm  \frac{1}{2}\eta(\eta-1)\diff{E}{t} = t^2\ci{\eta+\frac{23}{41}}D, \cm \eta D = t\ci{\eta+\frac{4}{41}}C \\
    \eta C &= t\ci{\eta-\frac{10}{41}}B, \cm B = \frac{1}{2}\ci{\eta+\frac{25}{41}}A
\end{align}
The equations for the 2 independent periods are:
\begin{align} \nonumber
    \sq{\eta(\eta-1)(\eta-2)(\eta-3)(\eta-4) -t^5 \ci{\eta+\frac{204}{41}}\ci{\eta+\frac{189}{41}}\ci{\eta+\frac{154}{41}}\ci{\eta+\frac{64}{41}}\ci{\eta+\frac{4}{41}}}     &C = 0 \\ \nonumber
    \sq{\eta(\eta-1)(\eta-2)(\eta-3)(\eta-4) -t^5 \ci{\eta+\frac{168}{41}}\ci{\eta+\frac{163}{41}}\ci{\eta+\frac{148}{41}}\ci{\eta+\frac{113}{41}}\ci{\eta+\frac{23}{41}}}     &D = 0 \\ \nonumber
    \sq{\eta(\eta-1)(\eta-2)(\eta-3)(\eta-4) -t^5 \ci{\eta+\frac{187}{41}}\ci{\eta+\frac{127}{41}}\ci{\eta+\frac{122}{41}}\ci{\eta+\frac{107}{41}}\ci{\eta+\frac{72}{41}}}    &\diff{E}{t} = 0
\end{align}

\subsubsection*{Representation Class $\cB4$}
The 5th order monomials in $\cB2$ are $\cu{e^3bc,b^2c^2d}$. Acting with $\star$ gives $\cu{a^3b^2c^2d^3,a^3bcd^2e^3}$, corresponding to periods $(3,2,2,3,0)$, and $(3,1,1,2,3)$. These are cyclic permutations of $(0,3,2,2,3)$ and $(3,3,1,1,2)$ which appear in the $\cB4$ diagram as $C$, and $\frac{1}{2}\diff{D}{t}$ respectively. We write the coupled system in terms of $\diff{D}{t}$ rather than $D$ and then solve for $C$ and $\diff{D}{t}$. The coupled system is:
\begin{align}
    \eta A =t\ci{\eta+\frac{5}{41}}E&, \cm \eta(\eta-1) E = t^2\ci{\eta+\frac{62}{41}}\diff{D}{t}, \cm \frac{1}{2}\eta\diff{D}{t} = t\ci{\eta+\frac{8}{41}}C \\
    \eta C &= t\ci{\eta+\frac{39}{41}}B, \cm B = \frac{1}{2}\ci{\eta+\frac{9}{41}}A
\end{align}
which results in the following equations for the 3 independent periods:
\begin{align} \nonumber
    \sq{\eta(\eta-1)(\eta-2)(\eta-3)(\eta-4) -t^5 \ci{\eta+\frac{203}{41}}\ci{\eta+\frac{173}{41}}\ci{\eta+\frac{128}{41}}\ci{\eta+\frac{108}{41}}\ci{\eta+\frac{8}{41}}}     C &= 0 \\ \nonumber
    \sq{\eta(\eta-1)(\eta-2)(\eta-3)(\eta-4) -t^5 \ci{\eta+\frac{172}{41}}\ci{\eta+\frac{162}{41}}\ci{\eta+\frac{132}{41}}\ci{\eta+\frac{87}{41}}\ci{\eta+\frac{62}{41}}}     \diff{D}{t} &= 0
\end{align}

\subsubsection*{Representation Class $\cC1$}
The independent 5th order monomials in $\cC3$ can be chosen to be $\cu{e^3ac,c^2abd}$. Acting with $\star$ gives $\cu{a^2b^3c^2d^3,a^2b^2cd^2e^3}$, corresponding to periods $(2,3,2,3,0)$, and $(2,2,1,2,3)$. These are cyclic permutations of $(2,3,0,2,3)$ and $(1,2,3,2,2)$ which appear in the $\cC1$ diagram as $D$, and $\frac{1}{2}\diff{E}{t}$ respectively. We write the coupled system in terms of $\diff{E}{t}$ rather than $E$ and then solve for $D$ and $\diff{E}{t}$. The coupled system is:
\begin{align}
    &\eta(\eta-1) A =t^2\ci{\eta+\frac{71}{41}}\diff{E}{t}, \cm \frac{1}{2}\eta \diff{E}{t} = t\ci{\eta+\frac{7}{41}}D \\ \eta D  = &t\ci{\eta+\frac{13}{41}}C, \cm
    C = \frac{1}{2}\ci{\eta+\frac{3}{41}}B, \cm \eta B = t\ci{\eta+\frac{29}{41}}A
\end{align}
which results in the following equations for the 2 independent periods:
\begin{align} \nonumber
    \sq{\eta(\eta-1)(\eta-2)(\eta-3)(\eta-4) -t^5 \ci{\eta+\frac{177}{41}}\ci{\eta+\frac{167}{41}}\ci{\eta+\frac{152}{41}}\ci{\eta+\frac{112}{41}}\ci{\eta+\frac{7}{41}}}    & D = 0 \\ \nonumber
    \sq{\eta(\eta-1)(\eta-2)(\eta-3)(\eta-4) -t^5 \ci{\eta+\frac{171}{41}}\ci{\eta+\frac{136}{41}}\ci{\eta+\frac{126}{41}}\ci{\eta+\frac{111}{41}}\ci{\eta+\frac{71}{41}}}     &\diff{E}{t} = 0
\end{align}

\subsubsection*{Representation Class $\cC2$}
The 5th order monomials in $\cC4$ are $\cu{a^2c^2d,d^2e^2b}$. Acting with $\star$ gives $\cu{ab^3cd^2e^3,a^3b^2c^3de}$, corresponding to periods $(1,3,1,2,3)$, and $(3,2,3,1,1)$. These are cyclic permutations of $(3,1,3,1,2)$ and $(1,3,2,3,1)$ which appear in the $\cC2$ diagram as $\frac{1}{2}\diff{C}{t}$, and $\frac{1}{2}\diff{E}{t}$ respectively. Rewriting the coupled system in terms of these variables:
\begin{align}
    &\eta(\eta-1) A =t^2\ci{\eta+\frac{58}{41}}\diff{E}{t}, \cm  \frac{1}{2}\eta \diff{E}{t} = t\ci{\eta+\frac{14}{41}}D \\ \eta D = &\frac{1}{2}t\ci{\eta+\frac{60}{41}}\diff{C}{t},\cm
    \frac{1}{2}\eta\diff{C}{t} = t\ci{\eta+\frac{26}{41}}B, \cm B = \frac{1}{2}\ci{\eta+\frac{6}{41}}A
\end{align}
which results in the following equations for the periods:
\begin{align} \nonumber
    \sq{\eta(\eta-1)(\eta-2)(\eta-3)(\eta-4) -t^5 \ci{\eta+\frac{190}{41}}\ci{\eta+\frac{170}{41}}\ci{\eta+\frac{140}{41}}\ci{\eta+\frac{60}{41}}\ci{\eta+\frac{55}{41}}}    \diff{C}{t} &= 0 \\ \nonumber
    \sq{\eta(\eta-1)(\eta-2)(\eta-3)(\eta-4) -t^5 \ci{\eta+\frac{183}{41}}\ci{\eta+\frac{178}{41}}\ci{\eta+\frac{108}{41}}\ci{\eta+\frac{88}{41}}\ci{\eta+\frac{58}{41}}}     \diff{E}{t} &= 0
\end{align}

\subsubsection*{Representation Class $\cC3$}
The independent 5th order monomials in $\cC1$ can be chosen to be $\cu{d^3a^2, a^3bc,c^2bde}$. Acting with $\star$ gives $\cu{ab^3c^3e^3,b^2c^2d^3e^3,a^3b^2cd^2e^2}$, corresponding to periods $(1,3,3,0,3)$, $(0,2,2,3,3)$ and $(3,2,1,2,2)$. These are cyclic permutations of $(0,3,1,3,3)$, $(3,3,0,2,2)$ and $(2,2,3,2,1)$ which appear in the $\cC3$ diagram as $C$, $D$ and $\frac{1}{2}\diff{E}{t}$ respectively. The coupled system is:
\begin{align}
    \eta(\eta-1) A =t\ci{\eta+\frac{75}{41}}\diff{E}{t}&, \cm \frac{1}{2}\eta\diff{E}{t} = t\ci{\eta+\frac{11}{41}}D, \cm \eta D = t\ci{\eta+\frac{12}{41}}C \\
    \eta C &= t\ci{\eta+\frac{28}{41}}B, \cm B = \frac{1}{2}\ci{\eta+\frac{38}{41}}A
\end{align}
The equations for the 3 independent periods are:
\begin{align} \nonumber
    \sq{\eta(\eta-1)(\eta-2)(\eta-3)(\eta-4) -t^5 \ci{\eta+\frac{202}{41}}\ci{\eta+\frac{192}{41}}\ci{\eta+\frac{157}{41}}\ci{\eta+\frac{52}{41}}\ci{\eta+\frac{12}{41}}}     &C = 0 \\ \nonumber
    \sq{\eta(\eta-1)(\eta-2)(\eta-3)(\eta-4) -t^5 \ci{\eta+\frac{176}{41}}\ci{\eta+\frac{161}{41}}\ci{\eta+\frac{151}{41}}\ci{\eta+\frac{116}{41}}\ci{\eta+\frac{11}{41}}}     &D = 0 \\ \nonumber
    \sq{\eta(\eta-1)(\eta-2)(\eta-3)(\eta-4) -t^5 \ci{\eta+\frac{175}{41}}\ci{\eta+\frac{135}{41}}\ci{\eta+\frac{120}{41}}\ci{\eta+\frac{110}{41}}\ci{\eta+\frac{75}{41}}}    &\diff{E}{t} = 0
\end{align}

\subsubsection*{Representation Class $\cC4$}
The 5th order monomials in $\cC2$ are $\cu{e^3bd,a^2c^2e,b^2d^2c}$. Acting with $\star$ gives $\cu{a^3b^2c^3d^2,ab^3cd^3e^2,a^3bc^2de^3}$, corresponding to periods $(3,2,3,2,0)$, $(1,3,1,3,2)$ and $(3,1,2,1,3)$. These are cyclic permutations of $(2,3,2,0,3)$, $(3,1,3,2,1)$ and $(1,2,1,3,3)$ which appear in the $\cC4$ diagram as $C$, $\frac{1}{2}\diff{A}{t}$ and $\frac{1}{2}\diff{D}{t}$ respectively. The coupled system is:
\begin{align}
    \frac{1}{2}\eta\diff{A}{t} =t\ci{\eta+\frac{22}{41}}E&, \cm \eta E = \frac{1}{2}t\ci{\eta+\frac{68}{41}}\diff{D}{t}, \cm \frac{1}{2}\eta \diff{D}{t} = t\ci{\eta+\frac{24}{41}}C \\
    \eta C &= t\ci{\eta+\frac{35}{41}}B, \cm \eta B = \frac{1}{2}t\ci{\eta+\frac{56}{41}}\diff{A}{t}
\end{align}
The equations for the 3 independent periods are:
\begin{align} \nonumber
    \sq{\eta(\eta-1)(\eta-2)(\eta-3)(\eta-4) -t^5 \ci{\eta+\frac{191}{41}}\ci{\eta+\frac{186}{41}}\ci{\eta+\frac{106}{41}}\ci{\eta+\frac{76}{41}}\ci{\eta+\frac{56}{41}}}     &\diff{A}{t} = 0 \\ \nonumber
    \sq{\eta(\eta-1)(\eta-2)(\eta-3)(\eta-4) -t^5 \ci{\eta+\frac{199}{41}}\ci{\eta+\frac{179}{41}}\ci{\eta+\frac{109}{41}}\ci{\eta+\frac{104}{41}}\ci{\eta+\frac{24}{41}}}     &C = 0 \\ \nonumber
    \sq{\eta(\eta-1)(\eta-2)(\eta-3)(\eta-4) -t^5 \ci{\eta+\frac{188}{41}}\ci{\eta+\frac{158}{41}}\ci{\eta+\frac{138}{41}}\ci{\eta+\frac{68}{41}}\ci{\eta+\frac{63}{41}}}    &\diff{D}{t} = 0
\end{align}

\subsubsection*{Periods of the Class Corresponding to 15th Order Monomials}
As for the Fermat quintic, we choose the monomial $a^3b^3c^3d^3e^3$ to represent the single independent class in $\bC[a,b,c,d,e]_{15}/J(Q)$, so we can reuse the diagram for $(1,1,1,1,1)$ and solve for the period $(3,3,3,3,3) = \frac{1}{6}\diff{^2A}{t^2}$. The resulting equation is:
\begin{align}
    \bsq{\eta(\eta-2)(\eta-3)(\eta-4) - t^5(\eta+4)^4}\diff{^2A}{t^2} = 0
\end{align}

\subsection{Decomposition of the Monodromy Representations}\label{monodromyrep}
With the Picard--Fuchs data in hand, we now see what we can learn about the corresponding monodromy representations. Recall from section \ref{hypersurfacecohomology} that the forms we integrate to get periods are single--valued as functions of $t$, i.e. as sections of the Hodge bundle over $\bP_1-F$ where the set $F$ consists of 5th roots of unity and $\infty$. The only source of the monodromy of the solutions to the Picard--Fuchs equations is therefore the geometric monodromy of the cycles.

However, in general the Picard--Fuchs equations contain less information than the monodromy of the cycles. For example, the holomorphic 3--form $\psi$ obeys a 4th order equation. This means that as $t$ varies, $\psi$ moves around in a 4 dimensional space $\Psi\subset H^3(X,\bC)$. If $[\gamma]\in H_3(X,\bC)$ is a class whose dual is in $\Psi$, then $\psi$ will pick up monodromy $[\gamma]\to [\gamma]+[\delta]$ only if $\int_{[\delta]}\psi\neq 0 $, i.e. if $[\delta]$ has a component in the dual of $\Psi$ as well. Another way to say this is that the Picard--Fuchs equations tell us only about particular block diagonal pieces of the monodromy matrices. In particular, we get:
\begin{align}
    \text{Fermat:}\;\;  \left(\begin{array}{ccccc}4\times 4 &   &   &   &   \\  & 2\times 2 &   &   &   \\  &   & 2\times 2 &   &   \\  &   &   & \ldots &   \\  &   &   &   & 2\times 2\end{array}\right), \hcm \bZ_{41}:\;\; \left(\begin{array}{ccccc}4\times 4 &   &   &   &   \\  & 5\times 5 &   &   &   \\  &   & 5\times 5 &   &   \\  &   &   & \ldots &   \\  &   &   &   & 5\times 5\end{array}\right)
\end{align}
In both cases, the single $4\times4$ block is the monodromy representation of the equation \eqref{pfquarticfermat3}. For the Fermat family, the $2\times 2$ blocks correspond to the 2nd order equations satisfied by the 200 other forms. There is a block for each of the 100 other representations of $(\bZ_5)^3$ instantiated by the degree 5 monomials. For the $\bZ_{41}$ family, there is a $5\times5$ block for each of the 40 representations of $\bZ_{41}$.

In general, the non--block diagonal pieces of the monodromy matrices will be nonzero, but the extra symmetry in the Fermat and $\bZ_{41}$ examples provides nongeneric constraints. We now construct a basis in which the action is purely block--diagonal. Start with the 204 forms, at a point $t_0$ such that $t_0^5\neq 1,\infty$:
\begin{align}
    \phi_i = \text{Res}\ci{\frac{P_i}{Q(t_0)^{\frac{1}{5}\text{deg}P_i+1}}\Omega_0 }
\end{align}
where as before $P_i$ are monomials, $Q$ is the polynomial defining the family of hypersurfaces, and $\Omega_0$ is as in equation \eqref{nform}. Let $[\bar{\phi_i}]\in H_3(V(t_0),\bC)$ denote the dual classes to the $\phi_i$, and let $\bar{\phi_i}$ be representative cycles of these classes. We choose the $P_i$ to transform in a representation $g_i$ of the symmetry group, so that the classes transform in the representations $-g_i$. One can then use an Ehresmann connection to generate a family of cycles $\bar{\phi_i}(t)$ in some neighborhood of $t_0$, such that for each $t$, $\bar{\phi_i}(t)$ transforms in the representation $-g_i$.\footnote{Note that in general $ \sq{\bar{\phi_i}(t)}$ is only dual to $\phi_i(t)$ when $t=t_0$.} Since the connection by definition respects the $(\bZ_5)^3$ or $\bZ_{41}$ symmetry, cycles can only mix under monodromy with cycles in the same representation of the symmetry group. This is equivalent to the monodromy representation being block diagonal as above.

\section{Yukawa Couplings of $(2,1)$--Forms}
\label{yukawas}
We stressed in the introduction that the families in table \ref{symmetricfamilies} are distinct from the more familiar Fermat locus \eqref{FermatPencil}. As a first step to seeing how these differences play out in the more detailed properties of the loci, we work out the number of Yukawa couplings constrained to vanish by the discrete symmetry group. This information is also useful for applications to string compactification, since the Yukawa couplings are intimately related to physically measurable constants in the 4d low energy effective theory.

Suppose $\Omega(t)$ is the holomorphic 3--form on a family of Calabi--Yau 3--folds parameterized by $t$. The derivative $\diff{\Omega(t)}{t}{}$ is no longer restricted to $H^{3,0}(X)$, but is instead contained in the second Hodge filtrant:  $\diff{\Omega(t)}{t}\in H^{3,0}(X)\oplus H^{2,1}(X)$.\footnote{In the context of abstract variations of Hodge structure, this property is known as \emph{Griffiths transversality}, and is a useful necessary condition for the variation of Hodge structure to be geometrical in origin. For hypersurfaces in projective space, Griffiths transversality follows from the results of section \ref{hypersurfacecohomology}.} Similarly for the second derivative, we have: $\diff{^2\Omega(t)}{t^2}\in H^{3,0}(X)\oplus H^{2,1}(X) \oplus H^{1,2}(X)$. The following integrals therefore vanish identically:
\begin{align}
    \int \Omega(t)\wedge \diff{\Omega(t)}{t} = \int \Omega(t) \wedge \diff{^2\Omega(t)}{t^2} = 0
\end{align}
But including third derivatives gives a nonzero result:
\begin{align}
    \int \Omega(t) \wedge \diff{^3\Omega(t)}{t^3} \neq 0 \cm \text{for general }t
\end{align}
This is the prototype Yukawa coupling. More generally we can look at the dependence of $\Omega$ over the whole complex structure moduli space (as opposed to just a 1--parameter family). $\Omega$ will then depend on $h_{2,1}$ parameters $t_i$, and the Yukawa couplings are:
\begin{align}
    Y_{ijk}(t_i) = \int \Omega(t_i)\wedge \sq{\diff{}{t_i'}\diff{}{t_j'}\diff{}{t_k'}\Omega(t_i')}_{t_i'=t_i}
\end{align}
Alternatively, with a given normalization for $\Omega(t_i)$, we can interpret the $t_i$'s as different directions in $T_{t_i}\cM\simeq H^{2,1}(X(t_i))$, the tangent space to the complex structure moduli space. The Yukawa couplings are then a map:
\begin{align}
    H^{2,1}(X(t_i))\times H^{2,1}(X(t_i))\times H^{2,1}(X(t_i)) \to \bC
\end{align}
$Y_{ijk}$ is clearly symmetric in its 3 indices, each of which takes $h_{2,1}$ different values. The number of independent Yukawa couplings is therefore:
\begin{align}
    N_{\text{Yukawas}}=\frac{1}{6}h_{2,1}\ci{h_{2,1}+1}\ci{h_{2,1}+2}
\end{align}
For example, quintic hypersurfaces in $\bP_4$ have $h_{2,1}=101$, so  $N_{\text{Yukawas}}=176851$. The technique for performing detailed calculations of Yukawa couplings was presented in \cite{Candelas:1987se}. Here we find the number of $Y_{ijk}$'s that are potentially nonzero in the presence of various discrete symmetries.

For symmetries that preserve $\Omega(t)$ (i.e. projective linear transformations that act trivially on $abcde$), the only Yukawa couplings that are allowed to be finite are those corresponding to 3 monomial deformations of $X$ whose product is invariant. A computer search for such triples (summarized in table \ref{nonzeroyukawas}) yields numbers that approximately satisfy:
\begin{align}
    \text{\# nonzero Yukawas} \simeq \frac{\text{Total \# of Yukawas}}{\text{Ord }G}
\end{align}
\begin{table}
\begin{align}
    \begin{array}{|c|c|c|}\hline \text{Symmetry Group }G & \text{\# Nonzero Yukawas} & (\text{Total \# Yukawas})/(\text{Ord } G) \\ \hline (\bZ_5)^3 & 1431 &1414.8 \\\bZ_{41} & 4321 & 4313.4 \\ \bZ_{51} & 3477 & 3467.7 \\\bZ_5\times\bZ_{13} & 2736 &2720.8 \\ \bZ_3\times\bZ_{13} & 4554 & 4534.6 \\(\bZ_5)^2\times\bZ_3 & 2391 & 2358.0 \\ \hline \end{array}
\end{align}
\caption{Numbers of potentially nonvanishing Yukawa couplings for the six families of quintics in $\bP_4$ listed at the end of section \ref{DiagramMethod}. \label{nonzeroyukawas}}
\end{table}
A relation of this form is somewhat surprising for the following reason. The transformations of the 101 monomials do not exhaust the 125 irreducible representations of $(\bZ_5)^3$, whereas in the $\bZ_{41}$ case there is some monomial transforming in each of the 41 representations. Therefore one might not expect the numbers of nonzero Yukawa couplings for $G=(\bZ_5)^3$ to fit the line defined by the cases with smaller $G$.

It would be interesting to more fully examine the dependence of the number of (potentially) nonzero cubic invariants on the size of the manifold's discrete symmetry group.

\section{Conclusion}

We have investigated some well--known Calabi--Yau 3--folds but focused on unfamiliar loci in their complex structure moduli that give rise to unexpected discrete symmetry groups. With the important role that Calabi--Yau manifolds with enhanced symmetries have played in both the physics and mathematics literatures, there is strong motivation to study these new families.  By carefully deriving a technique apparently similar to that of \cite{Candelas:2000fq} but differing significantly in interpretation, we succeeded in developing a systematic method for computing the Picard--Fuchs equations satisfied by each entry in the full period matrix of along these loci. To illustrate the method, we applied it to the Fermat family \eqref{FermatPencil} as well as the $\bZ_{41}$ quintic hypersurface family (the $\bZ_{51}$ family and a weighted projective space example are handled the appendix).  We then saw how discrete symmetries are reflected in the detailed structure of the geometric monodromy representations. In particular, aside from the 4$\times$4 invariant part the monodromy matrices decompose into block diagonal pieces of different sizes in the different families. Finally we found the number of Yukawa couplings constrained to vanish by the symmetries and noted an intriguing approximate relation between the number of nonzero couplings and the size of the symmetry group.

 The $\bZ_{41}$ and $\bZ_{51}$ families and their cousins in table \ref{symmetricfamilies} are thus a new testing ground for many calculations. For example, as with the Fermat family, computations of periods and Yukawa couplings are more tractable than for a general hypersurface. Such calculations are of interest because in heterotic string compactifications, the Yukawa couplings are eventually nothing but the parameters of the standard model, as well as because of the role periods play in various moduli stabilization schemes. For instance, in any model that purports phenomenological realism, the Yukawas must be able to incorporate the range of observed particle masses, spanning at least 14 orders of magnitude.\footnote{Neutrinos are now known to have a mass approximately $10^{-3}$eV, whereas the Z boson has a mass of $9.1\times 10^{10}$eV.} It would be interesting to know if the moduli space of quintics in $\bP_4$ (which is only a toy example in this context) admits regions with a large enough range of Yukawa couplings, and if so how many flux--stabilized vacua they contain. This could potentially amount to a very severe phenomenological restriction on the `landscape' of vacua which currently plagues attempts to extract TeV scale predictions from string theory.

 A mathematical direction for future work is to use the non--Fermat families to test some of the claims of mirror symmetry. In particular Morrison has constructed \cite{morrison,CoxKatz} a variation of Hodge structure on the even dimensional cohomology of the mirror (the so--called \emph{A--model variation}) in analogy with that coming from the middle--dimensional cohomology of the original manifold (the \emph{B--model variation}). Corresponding to the B--model monodromy action on $H_3(X,\bC)$, there are conjectured automorphisms of the topological K--theory of the mirror.\footnote{One often imagines mirror symmetry exchanging middle cohomology $H^3$ with even cohomology $H^{\text{even}}=H^0\oplus H^2\oplus H^4\oplus H^6$. But the Chern map, which sends an element $(E,F)\in K^0$ to $c(E)/c(F)$ (the quotient of the total Chern classes) is in fact an isomorphism when (as for quintics in $\bP_4$) $H^{\text{even}}(\bZ)$ contains no torsion classes.}\cite{Doran:2005gu} Making this correspondence explicit for the special families considered here should provide new insights into the mathematical structure (quantum cohomology and Gromov--Witten theory) of the A--model on Calabi--Yau threefold hypersurfaces in toric orbifolds\cite{CCIT}.

Our derivation of the corrected version of the technique outlined by Candelas, de la Ossa and Rodriguez--Villegas greatly reduces the computation required to find the Picard--Fuchs equations for a variety of families of Calabi--Yau manifolds with discrete symmetries. The method summarized in section \eqref{CandelasMethodSec} readily extends to the other examples of 3--folds with discrete symmetries, as well as symmetric Calabi--Yau hypersurfaces of other dimensions.\footnote{One might hope to generalize the technique further to hypersurfaces and complete intersections in toric varieties, perhaps with a view to bridge the gap between GKZ systems (which can be derived algorithmically) and true Picard--Fuchs differential equations.\cite{HKTY}}

In \cite{K3forthcoming} we examine 1--parameter families of K3 surfaces. Though the Picard--Fuchs equations can be derived in the same way, the interpretation of the results is more complicated than for 3--folds. The reason is essentially that $H^{n-1,1}$ which controls the deformations of complex structure coincides with $H^{1,1}$ which contains the K\"{a}hler form, as well as information about algebraic cycles. For example, there is an important sublattice of $H^{1,1}\cap H^2(V(t),\bZ)$ known as the \emph{Picard group}, whose classes consist of algebraic cycles. The rank of this group (the \emph{Picard rank}) can jump discontinuously as one deforms the hypersurface, even without passing through singular configurations. Moreover, it has been shown that loci endowed with discrete symmetries are some of the places where such jumps take place.\cite{Nikulin} K3 surfaces also display some extraordinary phenomena that are apparently unrelated to enhancements of Picard rank. An example is the theorem of Oguiso \cite{Oguiso}, that nontrivial projective families (such as quartic hypersurfaces in $\bP_3$) contain dense subsets where the automorphism group is of \emph{infinite} order. Nothing analogous to this occurs in families of 3--folds.

\vspace{1em}

\noindent\large {\bf Acknowledgements:} \normalsize The authors thank Johan de Jong, Brent Doran, and John Morgan for helpful conversations during the course of this work.
BG and SJ gratefully acknowledge the support of DOE grant DE-FG02-92ER40699. SJ acknowledges support from Columbia University ISE and the Pfister Foundation.
C.F.D. is supported in part by a Royalty Research Fund Scholar Award from the Office of Research, University of Washington.

\begin{appendix}

\section{Picard--Fuchs Equations for the $\bZ_{51}$ Quintic}
\label{Z51calc}

The calculation of the Picard--Fuchs equations for the $\bZ_{51}$ quintic:
\begin{align}
    Q(t) = \frac{1}{5}\bci{a^4b+b^4c+c^4d+d^4a+e^5} -tabcde = 0
\end{align}
differs only in detail from that of the $\bZ_{41}$ case. The final results are as follows. The 51 representations of $\bZ_{51}$ group into 14 permutation classes as follows:
\begin{align} \nonumber
    \begin{array}{|ccc|ccc|} \hline \text{Rep. Class} & \text{Reps.} & \# \text{(2,1)--Forms} & \text{Rep. Class} & \text{Reps.} & \# \text{(2,1)--Forms} \\ \hline \cA & 0 & 1 & \cH & 7,10,11,23 & 1 \\\cB & 1,16,38,47 & 1 & \cI & 8,19,26,49 & 2 \\\cC & 2,25,32,43 & 2 & \cJ & 9,15,36,42 & 2 \\\cD & 3,12,39,48 & 2 & \cK & 14,20,22,46 & 2 \\\cE & 4,13,35,50 & 3 & \cL & 17,34 & 2 \\\cF & 5,29,31,37 & 2 & \cM & 18,21,30,33 & 2 \\\cG & 6,24,27,45 & 2 & \cN & 28,40,41,44 & 3\\ \hline \end{array}
\end{align}
The operators annihilating the periods of the (2,1)--forms are then:
\renewcommand{\labelenumi}{$\mathcal{\Alph{enumi}}$:}
\begin{enumerate}
    \item $\eta(\eta-1)(\eta-2)(\eta-4) - t^5(\eta+2)^4$
    \item $\eta(\eta-1)(\eta-2)(\eta-3) - t^5\ci{\eta+\frac{86}{51}}\ci{\eta+\frac{101}{51}}\ci{\eta+\frac{106}{51}}\ci{\eta+\frac{166}{51}}$
    \item $\eta(\eta-1)(\eta-2)(\eta-3) - t^5\ci{\eta+\frac{26}{51}}\ci{\eta+\frac{121}{51}}\ci{\eta+\frac{151}{51}}\ci{\eta+\frac{161}{51}}$ \\ $\eta(\eta-1)(\eta-2)(\eta-3) - t^5\ci{\eta+\frac{19}{51}}\ci{\eta+\frac{49}{51}}\ci{\eta+\frac{59}{51}}\ci{\eta+\frac{179}{51}}$
    \item $\eta(\eta-1)(\eta-2)(\eta-3) - t^5\ci{\eta+\frac{39}{51}}\ci{\eta+\frac{54}{51}}\ci{\eta+\frac{99}{51}}\ci{\eta+\frac{114}{51}}$ \\ $\eta(\eta-1)(\eta-2)(\eta-3) - t^5\ci{\eta+\frac{3}{51}}\ci{\eta+\frac{48}{51}}\ci{\eta+\frac{63}{51}}\ci{\eta+\frac{243}{51}}$
    \item $\eta(\eta-1)(\eta-2)(\eta-3) - t^5\ci{\eta+\frac{1}{51}}\ci{\eta+\frac{16}{51}}\ci{\eta+\frac{191}{51}}\ci{\eta+\frac{251}{51}}$ \\ $\eta(\eta-1)(\eta-2)(\eta-3) - t^5\ci{\eta+\frac{38}{51}}\ci{\eta+\frac{98}{51}}\ci{\eta+\frac{103}{51}}\ci{\eta+\frac{118}{51}}$ \\ $\eta(\eta-1)(\eta-2)(\eta-3) - t^5\ci{\eta+\frac{47}{51}}\ci{\eta+\frac{52}{51}}\ci{\eta+\frac{67}{51}}\ci{\eta+\frac{242}{51}}$
    \item $\eta(\eta-1)(\eta-2)(\eta-3) - t^5\ci{\eta+\frac{46}{51}}\ci{\eta+\frac{71}{51}}\ci{\eta+\frac{116}{51}}\ci{\eta+\frac{226}{51}}$ \\ $\eta(\eta-1)(\eta-2)(\eta-3) - t^5\ci{\eta+\frac{22}{51}}\ci{\eta+\frac{97}{51}}\ci{\eta+\frac{122}{51}}\ci{\eta+\frac{167}{51}}$
    \item $\eta(\eta-1)(\eta-2)(\eta-3) - t^5\ci{\eta+\frac{6}{51}}\ci{\eta+\frac{96}{51}}\ci{\eta+\frac{126}{51}}\ci{\eta+\frac{231}{51}}$ \\ $\eta(\eta-1)(\eta-2)(\eta-3) - t^5\ci{\eta+\frac{27}{51}}\ci{\eta+\frac{57}{51}}\ci{\eta+\frac{147}{51}}\ci{\eta+\frac{177}{51}}$
    \item $\eta(\eta-1)(\eta-2)(\eta-3) - t^5\ci{\eta+\frac{41}{51}}\ci{\eta+\frac{91}{51}}\ci{\eta+\frac{146}{51}}\ci{\eta+\frac{181}{51}}$
    \item $\eta(\eta-1)(\eta-2)(\eta-3) - t^5\ci{\eta+\frac{2}{51}}\ci{\eta+\frac{32}{51}}\ci{\eta+\frac{127}{51}}\ci{\eta+\frac{247}{51}}$ \\ $\eta(\eta-1)(\eta-2)(\eta-3) - t^5\ci{\eta+\frac{43}{51}}\ci{\eta+\frac{53}{51}}\ci{\eta+\frac{83}{51}}\ci{\eta+\frac{178}{51}}$
    \item $\eta(\eta-1)(\eta-2)(\eta-3) - t^5\ci{\eta+\frac{42}{51}}\ci{\eta+\frac{87}{51}}\ci{\eta+\frac{117}{51}}\ci{\eta+\frac{162}{51}}$ \\ $\eta(\eta-1)(\eta-2)(\eta-3) - t^5\ci{\eta+\frac{36}{51}}\ci{\eta+\frac{66}{51}}\ci{\eta+\frac{111}{51}}\ci{\eta+\frac{246}{51}}$
    \item $\eta(\eta-1)(\eta-2)(\eta-3) - t^5\ci{\eta+\frac{37}{51}}\ci{\eta+\frac{82}{51}}\ci{\eta+\frac{107}{51}}\ci{\eta+\frac{182}{51}}$ \\ $\eta(\eta-1)(\eta-2)(\eta-3) - t^5\ci{\eta+\frac{31}{51}}\ci{\eta+\frac{56}{51}}\ci{\eta+\frac{131}{51}}\ci{\eta+\frac{241}{51}}$
    \item $\eta(\eta-1)(\eta-2)(\eta-3) - t^5\ci{\eta+\frac{17}{51}}^2\ci{\eta+\frac{187}{51}}^2$ \\ $\eta(\eta-1)(\eta-2)(\eta-3) - t^5\ci{\eta+\frac{34}{51}}^2\ci{\eta+\frac{119}{51}}^2$
    \item $\eta(\eta-1)(\eta-2)(\eta-3) - t^5\ci{\eta+\frac{21}{51}}\ci{\eta+\frac{81}{51}}\ci{\eta+\frac{171}{51}}\ci{\eta+\frac{186}{51}}$ \\ $\eta(\eta-1)(\eta-2)(\eta-3) - t^5\ci{\eta+\frac{18}{51}}\ci{\eta+\frac{33}{51}}\ci{\eta+\frac{123}{51}}\ci{\eta+\frac{183}{51}}$
    \item $\eta(\eta-1)(\eta-2)(\eta-3) - t^5\ci{\eta+\frac{11}{51}}\ci{\eta+\frac{61}{51}}\ci{\eta+\frac{176}{51}}\ci{\eta+\frac{211}{51}}$ \\ $\eta(\eta-1)(\eta-2)(\eta-3) - t^5\ci{\eta+\frac{23}{51}}\ci{\eta+\frac{58}{51}}\ci{\eta+\frac{113}{51}}\ci{\eta+\frac{163}{51}}$ \\ $\eta(\eta-1)(\eta-2)(\eta-3) - t^5\ci{\eta+\frac{7}{51}}\ci{\eta+\frac{62}{51}}\ci{\eta+\frac{112}{51}}\ci{\eta+\frac{227}{51}}$
\end{enumerate}

\section{Symmetric Hypersurfaces in Weighted Projective Space} \label{weighted}
It is known that any Calabi--Yau 3--fold can be embedded in $\bP_n$ for some sufficiently large $n$, but the case where the embedding is a hypersurface is the exception rather than the rule. More often the embedding can only be realized as an intersection of a large number of hypersurfaces. It is therefore useful to consider other constructions of Calabi--Yau 3--folds. One of the simplest generalizations of a hypersurface in $\bP_n$ is a hypersurface in a weighted projective space: $\ci{\bC_n-\cu{0}}/\sim$ where the equivalence relation $\sim$ is given by:
\begin{align}
    [x_0,\ldots,x_n]\sim[\lambda^{k_0}x_0,\ldots,\lambda^{k_n}]
\end{align}
Here $\lambda$ is any nonzero complex number, and $\cu{k_0,\ldots,k_n}$ are a collection of integers called the weights. This space is denoted $\bW\bP_{[k_0,\ldots,k_n]}$, and one easily sees that ordinary projective space is a special case: $\bP_n = \bW\bP_{[1,\ldots,1]}$.
\end{appendix}
The formulas relating to hypersurfaces in $\bP_n$ generalize straightforwardly to the case of nontrivial weights \cite{Weighted}. As before we have:
\begin{align} \tag{\ref{basicformula}}
    \frac{\Omega_0}{Q(t)^{k+1}}\sum_{i=0}^n A_i\der{Q(t)}{x^i} = \frac{1}{k}\frac{\Omega_0}{Q(t)^k}\sum_{i=0}^n \der{A_i}{x^i} + \text{exact forms}
\end{align}
But the $n$--form $\Omega_0$ is now given by:
\begin{align}
    \Omega_0 = \sum_i (-1)^i k_i x^i \ef{x^0}\wedge\ldots\widehat{\ef{x^i}}\ldots\wedge\ef{x^n}
\end{align}
As an example, consider the collection of weights $[k_0,k_1,k_2,k_3,k_4]=[41,51,52,48,64]$. The condition for a hypersurface to have zero first Chern class is: $\deg Q = \sum_i k_i = 256$, so the Fermat--like Calabi--Yau hypersurface is:
\begin{align}
    Q(t) = \frac{1}{5}\ci{a^5b + b^4c + c^4d + d^4e + e^4} - tabcde = 0
\end{align}
One can check that this hypersurface has $h_{2,1}=1$, and so the only periods to consider are those of the holomorphic 3--form and its derivatives.

As in ordinary projective space, one can find linear combinations of derivatives of $Q$ suitable for constructing diagrams involving 3 periods:
\begin{align}
    \left(\begin{array}{ccccc}256 & 0 & 0 & 0 & 0 \\-64 & 320 & 0 & 0 & 0 \\16 & -80 & 320 & 0 & 0 \\-4 & 20 & -80 & 320 & 0 \\1 & -5 & 20 & -80 & 320\end{array}\right) \left(\begin{array}{c}a\partial_a Q(t) \\b\partial_b Q(t) \\c\partial_c Q(t) \\d\partial_d Q(t) \\e\partial_e Q(t)\end{array}\right) = 256\left(\begin{array}{c}a^5b - tabcde \\b^4c - tabcde \\c^4d - tabcde \\d^4e - tabcde \\e^4 - tabcde\end{array}\right)
\end{align}
and the relations corresponding to diagrams are therefore:
\begin{align}
    (v_0+5,v_1+1,v_2,v_3,v_4) &= \frac{f(0,\vec{v})}{256k(v)}(\vec{v})+t(\vec{v}+\vec{1}) \\
    (v_0,v_1+4,v_2+1,v_3,v_4) &= \frac{f(1,\vec{v})}{256k(v)}(\vec{v})+t(\vec{v}+\vec{1}) \\
    (v_0,v_1,v_2+4,v_3+1,v_4) &= \frac{f(2,\vec{v})}{256k(v)}(\vec{v})+t(\vec{v}+\vec{1}) \\
    (v_0,v_1,v_2,v_3+4,v_4+1) &= \frac{f(3,\vec{v})}{256k(v)}(\vec{v})+t(\vec{v}+\vec{1}) \\
    (v_0,v_1,v_2,v_3,v_4+4) &= \frac{f(4,\vec{v})}{256k(v)}(\vec{v})+t(\vec{v}+\vec{1})
\end{align}
with the coeficients $f(i,\vec{v})$ given by:
\begin{align}
    \left(\begin{array}{c}f(0,\vec{v}) \\f(1,\vec{v}) \\f(2,\vec{v}) \\f(3,\vec{v}) \\f(4,\vec{v})\end{array}\right) = \left(\begin{array}{ccccc}256 & 0 & 0 & 0 & 0 \\-64 & 320 & 0 & 0 & 0 \\16 & -80 & 320 & 0 & 0 \\-4 & 20 & -80 & 320 & 0 \\1 & -5 & 20 & -80 & 320\end{array}\right)\left(\begin{array}{c}v_0+1 \\ v_1+1 \\v_2+1 \\v_3+1 \\v_4+1\end{array}\right)
\end{align}
The diagram for the periods of the holomorphic 3--form is then:
\small
\begin{align} \nonumber
    \begin{array}{ccccccccccc}
     &   &  (0,0,0,0,0)_A & \to  &  (1,1,1,1,1)  & \to  & (2,2,2,2,2) & \to  & (3,3,3,3,3) & \to & (4,4,4,4,4) \\
     &   & \downarrow D_4  &   & \downarrow D_4  &   & \downarrow D_4 &   & \downarrow D_4 & & \\
     &   &  (0,0,0,0,4)_B & \to  &  (1,1,1,1,5)  & \to  & (2,2,2,2,6) & \to  & (3,3,3,3,7) &  &    \\
     &   & \downarrow D_3  &   & \downarrow D_3  &   & \downarrow D_3 &   &  & & \\
     &   &  (0,0,0,4,5)_C & \to  &  (1,1,1,5,6)  & \to  & (2,2,2,6,7) &   &  &  &   \\
     &   & \downarrow D_2  &   & \downarrow D_2  &   &  &   &  & & \\
     &   &  (0,0,4,5,5)_D & \to  &  (1,1,5,6,6)  &  &  &   &  &  &     \\
     &   & \downarrow D_1  &   &  &   &  &   &  & &  \\
     (-1,3,4,4,4) & \to & (0,4,5,5,5)_E &   &   &   &   &   &     &  &    \\
      \downarrow D_0  &   &  &   &  &   &  & & & & \\
      (4,4,4,4,4) &  &  &   &   &   &  &   &   &  &
     \end{array}
\end{align}
\normalsize
The encoded coupled system is:
\begin{align}
    &\frac{1}{4!}\eta(\eta-1)(\eta-2)(\eta-3) A =t^5E, \cm E = \frac{1}{4}\ci{\eta+1}D \\ D &= \frac{1}{3}\ci{\eta+1}C, \cm
    C = \frac{1}{2}\ci{\eta+1}B, \cm B = \ci{\eta+1}A
\end{align}
which results in the following Picard--Fuchs equation:
\begin{align}
    \sq{\eta(\eta-1)(\eta-2)(\eta-3) - t^5(\eta+1)^4} A = 0
\end{align}

\section{Examples of the Griffiths--Dwork Technique} \label{GDexamples}
As examples of the formalism outlined in section \ref{hypersurfacecohomology} we compute some of the Picard--Fuchs equations for a family of elliptic curves and the Fermat family of quintics in $\bP_4$.

\subsection*{Hesse form Cubics in $\bP_2$} 
Cubic hypersurfaces in $\bP_2$ are elliptic curves, and hence admit a single holomorphic 1--form. We have the following correspondence in general:\footnote{For the rest of this section we will abbreviate $H^{p,q}(V(t))$ and $\bF^{p,q}(V(t))$ with $H^{p,q}$ and $\bF^{p,q}$. }
\begin{align} \nonumber
    \begin{array}{ccccc} \Omega\in H^{1,0}=\bF^{1,1} & \to & [1]\in \bC[a,b,c]_0/[\partial_i Q] & \to & \cP_1 = \int\frac{1}{Q}\Omega_0 \\ \diff{\Omega}{t} \in H^{1,0}\oplus H^{0,1}=\bF^{1,0} & \to & [abc]\in \bC[a,b,c]_3/[\partial_i Q] & \to & \cP_2= \int\frac{abc}{Q}\Omega_0\end{array}
\end{align}
Here $[a,b,c]$ are homogeneous coordinates, $\cP$ is generic notation for a column of the period matrix, and the polynomial $Q(t)$ defining the hypersurface is:
\begin{align} \label{Hessecubic}
    Q(t) = \frac{1}{3}\bci{a^3+b^3+c^3} - tabc  = 0
\end{align}

\subsubsection*{Holomorphic 1--Form}
Differentiating a period of the holomorphic 1--form twice gives: $\cP''_1=\int \frac{2\Omega_0}{Q^3}(abc)^2$, where $\cP_1$ is a period, and the prime denotes differentiation with respect to $t$. As an element of the Jacobian ideal, we find:
\begin{align}
    (1-t^3)(abc)^2 = t^2a^2bc\frac{\partial Q}{\partial a} + ta^3b\frac{\partial Q}{\partial b}+a^2b^2\frac{\partial Q}{\partial c}
\end{align}
Applying \eqref{reducepoleorder} results in:
\begin{align} \label{hessecalc1}
    (1-t^3)\diff{^2\cP_1}{t^2} = 2\int \frac{\Omega_0}{Q^2}t^2abc + \int\frac{\Omega_0}{Q^2}ta^3
\end{align}
Now writing $a^3 = tabc + a\frac{\partial Q}{\partial a}$, we find:
\begin{align}
    \int\frac{\Omega_0}{Q^2}ta^3 = \int\frac{\Omega_0}{Q^2}t^2abc + t\int\frac{\Omega_0}{Q^2}a\frac{\partial Q}{\partial a} = t^2\diff{\cP_1}{t} + t\cP_1
\end{align}
Substituting back into \eqref{hessecalc1} gives the Picard--Fuchs equation:
\begin{align}
    (t^3-1)\diff{^2\cP_1}{t^2} + 3t^2 \diff{\cP_1}{t} + t\cP_1 = 0
\end{align}
In terms of the logarithmic derivative: $\eta=t\diff{}{t}$, this becomes:
\begin{align}
    \sq{(\eta+1)^2 - \frac{1}{t^3}\eta(\eta-1)}\cP_1 = 0
\end{align}
Making the substitution $x=t^{-3}$ and $\theta=x\tfrac{\ef{}}{\ef{x}}$ results in an equation in standard hypergeometric form:
\begin{align} \label{Hesseholohyper}
    \sq{\theta^2 - x\ci{\theta+\frac{1}{3}}  \ci{\theta+\frac{2}{3}}  } t\cP_1 = 0
\end{align}
Since nothing in the reasoning above depends on the choice of cycle integrated over, we conclude that both of the $b_1=2$ integrals of the holomorphic 1--form (one column of the period matrix $\int_{\gamma_i}\Omega_X^j$) obey the equation \eqref{Hesseholohyper}.

\subsubsection*{Mixed 1--Form}
The Griffiths--Dwork technique doesn't work so well for the periods  of the derivative of the holomorphic 1--form:
\begin{align}
    \cP_2 = \int \frac{abc}{Q(t)^2}\Omega_0 = \diff{}{t}\int \frac{\Omega_0}{Q(t)}
\end{align}
The problem is that we only have to differentiate once in order that the numerator of the integrand is in the ideal $[\partial_i Q(t)]$, so one might guess that the $\cP_2$ obeys a 1st order equation. Indeed a 1st order equation can be derived, but it contains $\cP_1$ as well, so it is not a Picard--Fuchs equation. To eliminate $\cP_1$ one must differentiate $\cP_2$ a second time. In the method introduced in section \ref{DiagramMethod}, this happens automatically. The end result is:
\begin{align}
    (t^3-1)\diff{^2\cP_2}{t^2} + \ci{5t^2+\frac{1}{t}}\diff{\cP_2}{t} + 4t\cP_2 = 0
\end{align}
Or, with $x=t^3$ and $\theta=x\diff{}{x}$, in hypergeometric form:
\begin{align}
    \sq{\theta\ci{\theta-\frac{2}{3}} - x\ci{\theta+\frac{2}{3}}^2   } \cP_2 = 0
\end{align}
Again, since nothing depends on the cycle integrated over, both periods obey the above equation. The Hesse form cubic is the simplest possible example; in general computational techniques are required to do the algebra.

 \subsection*{Fermat form Quintics in $\bP_4$}
Smooth quintic hypersurfaces in $\bP_4$ are Calabi--Yau 3--folds with $b_3 = 204$. The results of section \ref{hypersurfacecohomology} give the following correspondence:
\begin{align} \nonumber
    \begin{array}{ccccc} \Omega \in H^{3,0}& \to & [1]\in \frac{\bC[a,b,c,d,e]_0}{[\partial_i Q]} & \to & \cP_1 = \int\frac{1}{Q}\Omega_0 \\
    \omega_{\alpha}\in H^{3,0}\oplus H^{2,1} & \to & [M_{\alpha}]\in \frac{\bC[a,b,c,d,e]_5}{[\partial_i Q]} & \to & \cP_{\alpha} = \int\frac{M_{\alpha}}{Q}\Omega_0  \\
    \omega'_{\zeta} \in H^{3,0}\oplus H^{2,1} \oplus H^{1,2}& \to & [M_{\zeta}]\in \frac{\bC[a,b,c,d,e]_{10}}{[\partial_i Q]} & \to & \cP_{\zeta} = \int\frac{M_{\zeta}}{Q}\Omega_0  \\
    \ndiff{\Omega}{t}{3}\in H^{3,0}\oplus H^{2,1} \oplus H^{1,2}\oplus H^{0,3} & \to & [a^3b^3c^3d^3e^3]\in \frac{\bC[a,b,c,d,e]_{15}}{[\partial_i Q]} & \to & \cP_{204}= \int\frac{a^3b^3c^3d^3e^3}{Q}\Omega_0\end{array}
\end{align}

Again $[a,b,c,d,e]$ are homogeneous coordinates, and $Q(t)$ is the polynomial defining the hypersurface. The indices $\alpha$ and $\zeta$ have the ranges $\cu{2,\ldots,102}$ and $\cu{103,\ldots,203}$ respectively. If we now specialize to the Fermat family of quintic hypersurfaces:
\begin{align} \label{fermatquintic}
    Q(t) = \frac{1}{5}\bci{a^5+b^5+c^5+d^5+e^5} - tabcde
\end{align}
then there are two simplifications. For a general quintic, each period satisfies a 204th order differential equation. In computational terms, one expects to have to differentiate periods 204 times before the numerator in the integrand lies in the Jacobian ideal $J(Q)$. For \eqref{fermatquintic} the order of the equations is reduced to 4 or less. The other simplification is that we can say more about the Hodge type of the forms than merely which Hodge filtrant they are in. 

\subsubsection*{Periods of the Holomorphic 3--form}
One finds (by Gr\"{o}bner basis techniques for example) that it is sufficient to differentiate the periods of the holomorphic 3--form just 4 times.
\begin{align}
    \diff{^4\cP_1}{t^4} &= \int \frac{(4!) \Omega_0}{Q^5}(abcde)^4 \\ \nonumber
    (1-t^5)(abcde)^4 &= \bsq{t^4a^4(bcde)^3}\partial_a Q + \bsq{t^3a^7b^3(cde)^2}\partial_b Q + \bsq{t^2(ab)^6c^2de}\partial_c Q \\&\hcm+ \bsq{t(abc)^5d}\partial_d Q + \bsq{(abcd)^4}\partial_e Q
\end{align}
Then proceeding in the same way as with the Hesse cubic, one finds the equation:
\begin{align} \label{pfeqquintic}
    (t^5-1)\diff{^4\cP_1}{t^4} + 10t^4\diff{^3\cP_1}{t^3} + 25t^3\diff{^2\cP_1}{t^2}+15t^2\diff{\cP_1}{t}+t\cP_1 =0
\end{align}
Substituting $x=t^{-5}$ and $\theta=x\diff{}{x}$ gives an equation in generalized hypergeometric form:
\begin{align} \label{genhyperquintic}
    \sq{\theta^4 - x\ci{\theta+\frac{1}{5}}\ci{\theta+\frac{2}{5}}\ci{\theta+\frac{3}{5}}\ci{\theta+\frac{4}{5}}  }t\cP_1 = 0
\end{align}
Indeed this is how the Picard--Fuchs equation for the invariant periods is most often presented. From hereon though we will not make such changes of variables, but rather work directly in terms of the variable $t$, and the logarithmic derivative $\eta=\diff{}{t}$. The reasons for this choice are summarized at the end of section \ref{symcohom}.

As before, nothing depends on which cycle is integrated over, so all 204 integrals of $\Omega_X^0$  obey the 4th order equation \eqref{pfeqquintic}. This fact is related to the symmetries of the Fermat locus in section \ref{DiagramMethod}.

\subsubsection*{Periods of the Other Forms}
In a similar way, one can pick a basis of the rest of $\bC[a,b,c,d,e]/J(Q)$ and work out the equations satisfied by each of the $203$ other forms. It is easier to do this with the techniques introduced in  section \ref{DiagramMethod}. In particular we take advantage of the symmetries of the Fermat--form quintic with greater ease.

\section{Geometric Mondromy} \label{geometricmonodromy}
In section \ref{hypersurfacecohomology} we explored the connections between two descriptions of hypersurfaces; on the one hand as objects embedded in projective space (via the order of pole filtration), and on the other as complex manifolds (via the Hodge filtration). As already alluded to, a great deal of information about the relation between these two points of view is contained in the period matrix:
\begin{align}
    H_{n-1}\bci{V(t),\bZ}\times H^{n-1}\bci{V(t)&,C} \longrightarrow \bC \\
    [\gamma_i(t)], \;[\Omega_j(t)] \longrightarrow \Pi_{ij}(t) =\int_{\gamma_i(t)}&\Omega_j(t)
\end{align}
The $t$ dependence in $[\gamma_i(t)]$ is locally trivial, but if one follows the homology classes $[\gamma_i(t)]$ around a path enclosing a singular hypersurface, one finds that the class at the finish is not the same as at the beginning.

\begin{figure}[h]
\begin{center} 
\includegraphics[scale=0.9]{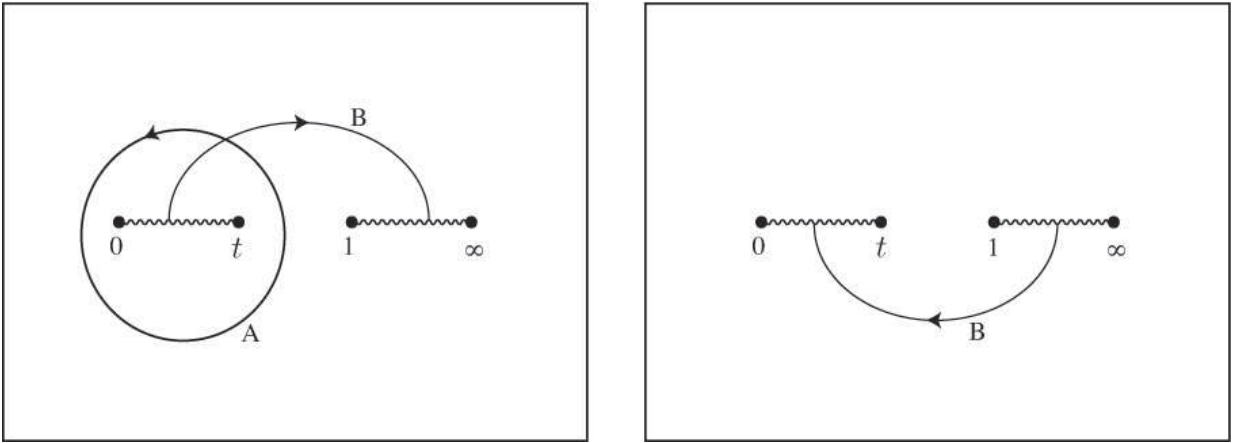}
\vspace{-1em} \begin{minipage}[h]{5in}\caption{\label{riemann1} \small The two sheets of the Riemann surface of $y^2=x(x-1)(x- t)$, which topologically glue together to make a torus. The $A$ and $B$ cycles are also shown.}\end{minipage}\end{center}\
\end{figure}

\begin{figure}[h]
\begin{center} 
\includegraphics[scale=1]{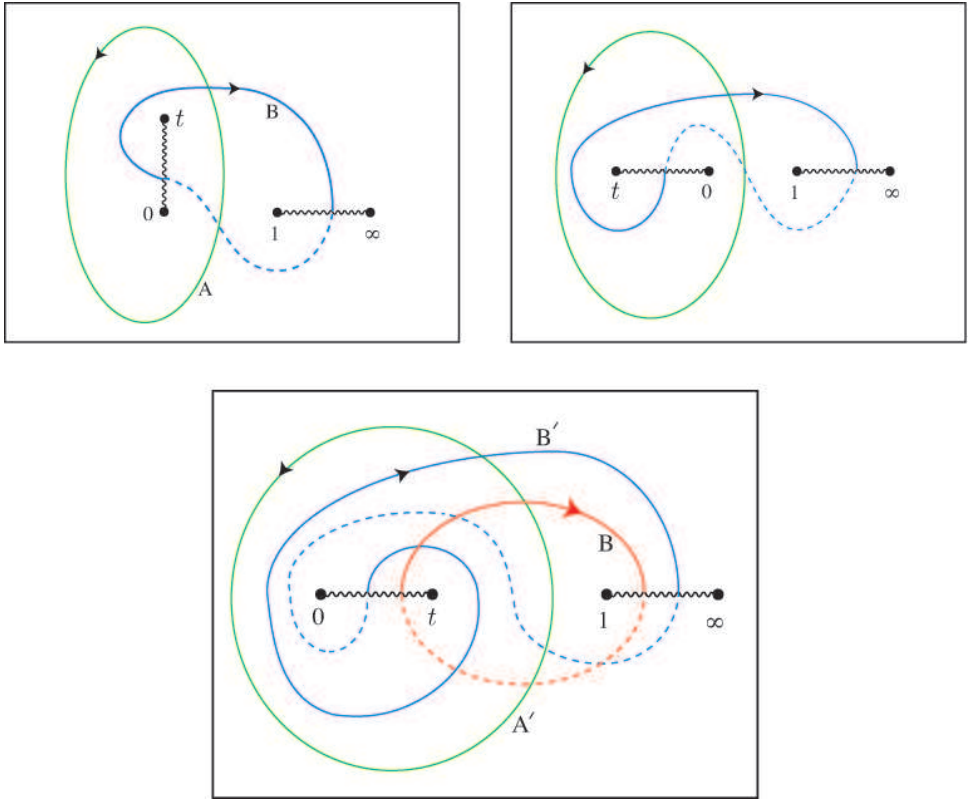}
\vspace{-1em} \begin{minipage}[h]{5in}\caption{\label{riemann2} \small The $A$ and $B$ cycles (in green and blue respectively) after $ t$ rotates by $\pi/2$, $\pi$ and $2\pi$. The dashed lines are on the 2nd sheet.}\end{minipage}\end{center}\
\end{figure}

Though the reader may be familiar with monodromy of in general, for the purpose of interpreting Picard--Fuchs equations, it is useful to have in mind a simple example of geometric monodromy acting on the homology groups of a manifold. As is often the case, elliptic curves provide a beautifully concrete case study.\footnote{The following argument and diagrams are adapted from \cite{PMPD}.}  Consider for example the Riemann surface of the function:\footnote{This set of curves parametrized by $t$ is called the \emph{Legendre family}.}
\begin{align} \label{ec}
    y^2 = x(x-1)(x- t)
\end{align}
which is singular at $t=0,1$ and $\infty$. There are 2 sheets, and we choose the branch cuts to be as in figure \ref{riemann1}.
As defined pictorially, the intersection matrix is given by:
\begin{align} \label{intersectionform}
    \left(\begin{array}{cc}A\cap A & A\cap B \\B\cap A & B\cap B\end{array}\right) = \left(\begin{array}{cc}0 & -1 \\1 & 0\end{array}\right)
\end{align}
But now imagine that $ t$ executes a small circle around $0$. The consequences for the cycles $A$ and $B$ are shown in figure \ref{riemann2}.
In particular, the bottom diagram shows both the old $B$ cycle (in red) and the new $B'$ (in blue). One can then read off the intersections:
\begin{align}
    A' \cap A = 0, \cm &A' \cap B = -1 \\
    B' \cap A = 1, \cm &B' \cap B = 2
\end{align}
From this it follows that the homology classes of the primed cycles are related to those of the unprimed cycles in the following way:
\begin{align}
    \left(\begin{array}{c} \sq{A'} \\  \sq{B'} \end{array}\right)  = \left(\begin{array}{cc}1 & 0 \\-2 &  1\end{array}\right)   \left(\begin{array}{c} \sq{A} \\ \sq{B} \end{array}\right)
\end{align}
In general, it is easy to see that moving cycles around $0,1,\infty$ gives a map:
\begin{align}
    \pi_1\bci{\bP_1-\cu{0,1,\infty}} \to Sp(2,\bZ)
\end{align}
called the \emph{monodromy representation}, or the \emph{monodromy action}. The image is $Sp(2,\bZ)$ rather than a more general matrix because the intersection form \eqref{intersectionform} must be preserved.  An important consequence of nontrivial monodromy is that the period matrix is a multivalued function of $t$. To distinguish the monodromy of cycles from the monodromy of anything else (hypergeometric functions say), we call the former \emph{geometric monodromy}.

The families of Calabi--Yau 3--folds we consider in section \ref{CandelasMethodSec} have a somewhat different singularity structure from the elliptic curves \eqref{ec}. They have the singularities of the Fermat form quintic \eqref{fermatquintic}.\footnote{One of them \emph{is} the Fermat form quintic. The other is the $\bZ_{41}$--symmetric example mentioned in the introduction.} Rather than at $t=0,1,\infty$, the singular hypersurfaces are at $t^5=1$ and $t=\infty$.

\end{spacing}
\end{document}